\documentclass[onecolumn,pre,floats,aps,amsmath,amssymb,nofootinbib]{revtex4-2}
\usepackage{comment}
\usepackage[utf8]{inputenc}
\usepackage{mathtools}
\usepackage{bbm}
\usepackage{cancel} 
\usepackage{dcolumn}
\usepackage{bm}
\usepackage{blindtext}
 \usepackage{relsize}
\usepackage{MnSymbol}
\usepackage{xcolor}
\usepackage{hyperref}
\usepackage{graphicx}
\usepackage{caption}
\usepackage{amsthm}
\usepackage{subfig}
\usepackage{bm}
\usepackage{verbatim}
\usepackage{microtype}
\usepackage{amsmath}
\usepackage{adjustbox}
\usepackage{siunitx}
\usepackage{physics}
\usepackage{amssymb}
\usepackage{multirow}
\usepackage{slashed}
\newcommand{\be}{\begin{equation}}
\newcommand{\ee}{\end{equation}}
\newcommand{\beq}{\begin{equation}}
\newcommand{\eeq}{\end{equation}}
\newcommand{\bea}{\begin{eqnarray}}
\newcommand{\eea}{\end{eqnarray}}
\newcommand{\msbar}{\overline{\footnotesize\textrm{MS}}}

\newcommand{\bs}{\boldsymbol}

\newcommand{\Romatre}{Dipartimento di Fisica, Universit\`a  Roma Tre and INFN, Sezione di Roma Tre,\\ Via della Vasca Navale 84, I-00146 Rome, Italy}
\newcommand{\RomatreINFN}{Istituto Nazionale di Fisica Nucleare, Sezione di Roma Tre,\\ Via della Vasca Navale 84, I-00146 Rome, Italy}
\newcommand{\Romadue}{Dipartimento di Fisica and INFN, Universit\`a di Roma ``Tor Vergata",\\ Via della Ricerca Scientifica 1, I-00133 Roma, Italy}
\newcommand{\LaSapienza}{Physics Department and INFN Sezione di Roma La Sapienza,\\ Piazzale Aldo Moro 5, 00185 Roma, Italy}
\newcommand{\soton}{Department of Physics and Astronomy, University of Southampton,\\ Southampton SO17 1BJ, UK}

\begin{document}
\title{ The $B_{s}\to \mu^{+}\mu^{-}\gamma$ decay rate at large $q^{2}$ from lattice QCD} 

\author{R.\,Frezzotti}\affiliation{\Romadue} 
\author{G.\,Gagliardi}\affiliation{\RomatreINFN}
\author{V.\,Lubicz}\affiliation{\Romatre} 
\author{G.\,Martinelli}\affiliation{\LaSapienza}
\author{C.T.\,Sachrajda}\affiliation{\soton}
\author{F.\,Sanfilippo}\affiliation{\RomatreINFN}
\author{S.\,Simula}\affiliation{\RomatreINFN}
\author{N.\,Tantalo}\affiliation{\Romadue}
\date{\today}
\begin{abstract}
We determine, by means of lattice QCD calculations, the local form factors describing the $B_{s}\to \mu^{+}\mu^{-}\gamma$ decay. For this analysis we make use of the gauge configurations produced by the ETM Collaboration with $N_{f}=2+1+1$ flavour of Wilson-Clover twisted-mass fermions at maximal twist. To obtain the $B_{s}$ meson form-factors, we perform simulations for several heavy-strange meson masses $m_{H_{s}}$ in the range $m_{H_{s}} \in [ m_{D_{s}}, 2 m_{D_{s}} ]$, and extrapolate to the physical $B_{s}$ meson point $m_{B_{s}}\simeq 5.367~{\rm GeV}$ making use of the HQET scaling laws. We cover the region of large di-muon invariant masses $\sqrt{q^{2}} > 4.16\,{\rm GeV}$, and use our results to determine the branching fraction for $B_{s}\to \mu^{+}\mu^{-}\gamma$, which has been recently measured by LHCb in the region $\sqrt{q^{2}} > 4.9\,{\rm GeV}$. The largest contribution to the uncertainty in the partial branching fractions at values of $\sqrt{q^{2}} < 4.8\,{\rm GeV}$ is now due to resonance and other long-distance effects, including those from "charming penguins", which we estimate by summing over the contributions from the $J_P=1^-$ charmonium resonances.
\end{abstract}

\maketitle

\section{Introduction}

The flavour-changing-neutral current (FCNC) transition $B_s\to\mu^+\mu^-\gamma$, being strongly suppressed in the Standard Model (SM), represents an ideal channel to look for signals of New Physics (NP).
Although there is an additional factor of $\alpha_\mathrm{em}$ in the amplitude for this process compared to that for the widely-studied $B_s\to\mu^+\mu^-$ decay, the presence of the final state energetic photon removes the helicity suppression making the rates for the two processes approximately comparable. 
The LHCb Collaboration has recently searched for signals of this process\,\cite{LHCb:2021awg, LHCb:2021vsc} but found no significant events resulting in an upper limit for the branching ratio of $\mathcal{B}(B_{s} \to \mu^{+}\mu^{-}\gamma) < 2.0\times 10^{-9} $ for photons $\gamma$ emitted by the quarks\footnote{The final-state radiation (FSR) contribution, in which the photon is emitted from a final-state muon, dominates at small photon energies and has been subtracted in Ref.\,\cite{LHCb:2021awg}. The interference between FSR and ISR is instead found to be negligible. }  (the so-called initial-state radiation contribution, or ISR) and for di-muon inviariant masses $\sqrt{q^{2}} > 4.9\,{\rm GeV}$. Future measurements will be able to reduce the experimental uncertainties and cover a larger portion of the phase space reaching lower values of $q^{2}$. 
On the other hand, a first-principles theoretical prediction of the $B_{s} \to \mu^{+}\mu^{-}\gamma$ decay rate is currently missing.
While the leading hadronic effects in the $B_s\to\ell^+\ell^-$ ($\ell=e,\mu,\tau$) decay amplitude depend only on the $B_s$-meson decay constant $f_{B_s}$, which is known to sub-percent precision from lattice computations, the determination of the amplitude for the $B_s\to\mu^+\mu^-\gamma$ decay is much more complex. 
In this case the non-perturbative hadronic effects depend not only on local form factors, but also on resonance contributions.
Existing estimates of the rate are based on light-cone sum rules (LCSR)\,\cite{Janowski:2021yvz}, on model/effective-theory calculations, such as the relativistic dispersion approach based on the constituent-quark picture\,\cite{Kozachuk:2017mdk} and, more recently, on the use of existing lattice QCD results for the radiative leptonic form factors of the $D_{s}$ meson to estimate some of the $B_{s}\to \mu^{+}\mu^{-}\gamma$ transition form factors assuming vector-meson dominance (VMD)\,\cite{Guadagnoli:2023zym}.

The aim of this paper is to provide a first-principles determination, using lattice QCD, of the local form factors $F_{V}, F_{A}, F_{TV}, F_{TA}$ and $\bar{F}_{T}$, which represent the only non-perturbative QCD input in the determination of the $\bar{B}_{s}\to \mu^{+}\mu^{-}\gamma$ transition matrix elements\,\footnote{The text and diagrams here and below correspond to the decay of the $\bar{B}_s$ meson, which contains a valence $b$-quark.}
$\langle \gamma(\varepsilon)\,\mu^{+}\mu^{-} | \mathcal{O}_{7,9,10} | \bar{B}_{s}\rangle $, where $\varepsilon$ is the photon's polarization vector, and the $\mathcal{O}_{i}$ are the standard operators appearing in the effective weak Hamiltonian $H_{\rm eff}^{b\to s}$ describing the FCNC $b\to s$ transition and are defined in Eq.\,(\ref{eq:O_def})  below. 
We explore the region of large di-muon invariant masses $\sqrt{q^{2}} > 4.16~{\rm GeV}$. In this region, the impact of the contributions from the operators $\mathcal{O}_{1-6,8}$ (which are neglected at present) stemming from the four-quark operators and from the chromomagnetic penguin operator in $H_{\rm eff}^{b\to s}$ is expected to be modest~\cite{Guadagnoli:2017quo}, and the rate can be reliably computed from the knowledge of the local form factors only. 
As an estimate of the systematic error induced by this approximation, we employ a phenomenological description of the charming-penguin contribution, illustrated in Figure\,\ref{fig:penguin} below, which is expected to be among the largest of the contributions we have neglected because of the presence of broad charmonium resonances which are near or within the region of $q^{2}$ we consider. 
Whilst we find that the differential branching fractions themselves are dominated by the form factors (in particular by $F_V$), the dominant uncertainty 
for $\sqrt{q^2}<4.8\,\mathrm{GeV}$ is that due to charming penguin contributions (see Figure\,\ref{fig:total_branching}) and therefore in order to improve the precision and to be able to reach lower values of $q^2$ the development of a rigorous treatment of the contributions from $\mathcal{O}_{1-6,8}$ will be necessary. 

For this calculation we employ the same set of gauge configurations  which have recently been used in our work on the radiative leptonic form factors of the $D_{s}$ meson\,\cite{Frezzotti:2023ygt}. The configurations have been generated by the Extended Twisted Mass Collaboration (ETMC) with $N_{f}=2+1+1$ flavors of Wilson-clover twisted-mass fermions at maximal twist, and sea-quark masses tuned very close to their physical values for all quark flavours. The ensembles correspond to four values of the lattice spacing $a$ in the range $[0.056, 0.09]\,{\rm fm}$. 

Our strategy for obtaining results for the physical $\bar{B}_s$ meson, 
is to perform simulations at a series of unphysical (lighter) heavy-strange pseudoscalar mesons $\bar{H}_{s}$, consisting of a heavy quark ($h$) and a strange anti-quark ($\bar{s}$), with $m_{H_{s}} \in [ m_{D_{s}}, 2 m_{D_{s}} ]$. We then use heavy quark effective theory (HQET) relations to guide the extrapolation of the results to the physical $\bar{B}_{s}$ meson. 
For each heavy-quark mass we   
evaluate the form factors at four different values of the energy of the photon $E_{\gamma}$ (as measured in the rest frame of the decaying meson), which we keep fixed in units of the heavy-strange meson mass $m_{H_{s}}$. The values are given by $x_\gamma\equiv 2E_{\gamma}/m_{H_{s}} = 0.1, 0.2, 0.3$ and $0.4$, and for $m_{H_{s}}= m_{B_{s}}$ this corresponds to $q^{2} > (4.16\,{\rm GeV})^{2}$. Our main result is the calculation of $\mathcal{B}_{\rm SD}(x_{\gamma}^{\rm cut})$, which is the ISR contribution (to which we refer to in the paper as the structure-dependent contribution) to the branching fraction for $q^{2} > m^{2}_{B_{s}}(1 - x_{\gamma}^{\rm cut})$, and is given in Table~\ref{tab:branching}.  Our results at $x_{\gamma}^{\rm cut} = 0.166$ (i.e. $q^{2} > (4.9~{\rm GeV})^{2}$) is
\begin{align}
\mathcal{B}_{\rm SD}(x_{\gamma}^{\rm cut}=0.166) = 6.9(9) \times 10^{-11}~,
\end{align}
which is well within the current upper-bound set by the LHCb, $\mathcal{B}(x_{\gamma}^{\rm cut}=0.166) < 2.0\times 10^{-9}$. 
Anticipating that future experiments will be able to access values of $q^2$ below $(4.9\,\mathrm{GeV})^2$, we present in Table\,\ref{tab:branching} the partial branching fractions corresponding to values of the lower cut-off $\sqrt{q^2_{\mathrm{cut}}}=m_{B_s}\sqrt{1-x_\gamma^{\mathrm{cut}}}$ from 4.1\,GeV to 5.2\,GeV in steps of 0.1\,GeV. We find that the partial branching fractions in the $q^{2}$-region we explored is dominated by the contribution of the vector form factor $F_{V}$; the combined contribution of all other local form factors $F_{A}, F_{TV}, F_{TA}, \bar{F}_{T}$ is of the order of $\mathcal{O}(10\%)$.
   
The plan for the remainder of this paper is as follows. In Section\,\ref{sec:eff_weak} we briefly recall the definition of the local form factors in terms of matrix elements of the operators in the effective weak Hamiltonian. 
In Section\,\ref{sec:FF_num} we explain our strategy for the determination of the local form factors $F_{V}$, $F_{A}$, $F_{TV}$ and $F_{TA}$, and present the results of the continuum extrapolation for each value of the simulated heavy-strange meson mass. 
We also discuss the heavy-quark scaling relations, which are then used to extrapolate the results to the mass of the physical $B_{s}$ meson. 
In Section\,\ref{sec:FbarT} we present our strategy for evaluating the local form factor $\bar{F}_{T}$, 
whose lattice determination is complicated by the problem of the analytic continuation to Euclidean space-time of the relevant Minkowski correlation functions. We tackle this problem using the spectral reconstruction technique developed in Ref.\,\cite{Frezzotti:2023nun}. In Section\,\ref{sec:rate} we provide our determination of the differential cross-section for $B_{s}\to \mu^{+}\mu^{-}\gamma$ as well as the total differential rate for different $q^{2}$ intervals. We then compare our results with existing estimates as well as with the LHCb measurement~\cite{LHCb:2021vsc,LHCb:2021awg} corresponding to the interval $q^{2} > (4.9~{\rm GeV})^{2}$. 
Finally, in Section\,\ref{sec:conclusions} we present our conclusions and outlook for future improvements.

\section{The effective weak Hamiltonian and local form factors}
\label{sec:eff_weak}
The low-energy effective weak Hamiltonian describing the $b\to s$ transition, neglecting doubly Cabibbo-suppressed contributions, is given 
 by~\cite{Beneke:2020fot}
\begin{align}
\label{weak_hamiltonian}
\mathcal{H}_{\rm eff}^{b\to s} = 2\sqrt{2}G_{F} V_{tb}V_{ts}^{*}\left[ \sum_{i=1,2} C_{i}(\mu)\mathcal{O}_{i}^{c} + \sum_{i=3}^{6} C_{i}(\mu)\mathcal{O}_{i} +\frac{\alpha_{\rm em}}{4\pi}\sum_{i=7}^{10}C_{i}(\mu)\mathcal{O}_{i} \right]~,
\end{align}
where $G_{F}$ is the Fermi constant, $C_{i}$ are the Wilson coefficients and $\mathcal{O}_{i}$ are local operators renormalized at the scale $\mu$. The latter are given by ($P_{L(R)} = (1\mp\gamma^{5})/2$)
\begin{align}
\label{eq:O_def}
\mathcal{O}_{1}^{c} &= \left(\bar{s}_{i}\gamma^{\mu}P_{L}c_{j}\right)( \bar{c}_{j}\gamma^{\mu}P_{L}b_{i})~, \qquad\qquad \mathcal{O}_{2}^{c} = \left(\bar{s}\gamma^{\mu}P_{L}c\right)( \bar{c}\gamma^{\mu}P_{L}b)~,~\\[8pt]
\mathcal{O}_{7} &= -\frac{m_b}{e}\bar{s}\sigma^{\mu\nu}F_{\mu\nu}P_{R}b~, \qquad\qquad\,\,\,\,\,\,\,\,\,\,\,\,\,\,\,\,  \mathcal{O}_{8} = -\frac{g_{s}m_b}{4\pi\alpha_{\rm em}}\bar{s}\sigma^{\mu\nu}G_{\mu\nu}P_{R}b~,\\[8pt]
\mathcal{O}_{9} &= \left(\bar{s}\gamma^{\mu}P_{L}b\right) (\bar{\mu}\gamma_{\mu} \mu)~, \qquad\qquad\qquad\, \mathcal{O}_{10} = \left(\bar{s}\gamma^{\mu}P_{L}b\right) (\bar{\mu}\gamma_{\mu}\gamma^{5} \mu)~
\end{align}
while the operators $\mathcal{O}_{3-6}$ are the QCD penguins. In the previous equations $i,j$ are color indices, while $F_{\mu\nu}$ and $G_{\mu\nu}$ are the electromagnetic and gluonic field strength tensor, respectively. In the following, for the CKM matrix elements we use the PDG values $|V_{tb}| =  1.014(29)$ and $|V_{ts}|= 4.15(9)\times 10^{-2}$\,\cite{ParticleDataGroup:2020ssz}. Our conventions for the gamma matrices are
\begin{align}
\gamma^{5} = i\gamma^{0}\gamma^{1}\gamma^{2}\gamma^{3}\, , \qquad \sigma^{\mu\nu} = \frac{i}{2}[\gamma^{\mu},\gamma^{\nu}]~,
\end{align}
while for the Levi-Civita tensor we adopt the convention $\varepsilon^{0123}=-1$. The transition amplitude for the decay of the $\bar{B}_{s}$ meson is given by
\begin{align}
\mathcal{A}[\bar{B}_{s}\to \mu^{+}\mu^{-}\gamma ] = \langle \gamma(\bs{k},\varepsilon) \mu^{+}(p_{1})\mu^{-}(p_{2}) | \, -\mathcal{H
}_{\rm eff}^{b\to s} \, | \bar{B}_{s}(\bs{p}) \rangle_{\rm QCD+QED} ~,
\end{align}
where $k=(E_{\gamma}=|\bs{k}|,\bs{k})$ and $p=(E_{B_{s}}, \bs{p})$ are the momenta of the photon and $\bar{B}_{s}$ meson respectively, $\varepsilon$ is the photon's polarization vector, and $p_{1}$ and $p_{2}$ the momenta of the $\mu^{+}$ and $\mu^{-}$ respectively. The di-muon four-momentum is then $q=p_{1}+p_{2}= p-k$. The amplitude $\mathcal{A}$ is then expanded to leading non-vanishing order $(\mathcal{O}(\alpha_{\rm em}^{3/2}))$ in the electromagnetic coupling $\alpha_{\rm em}$, and can be expressed as~\cite{Beneke:2020fot}
\begin{align}
\label{eq:ampl_exp}
\mathcal{A}[\bar{B}_{s}\to \mu^{+}\mu^{-}\gamma ] = -e\frac{\alpha_{\rm em}}{\sqrt{2}\pi}V_{tb}V_{ts}^{*}\varepsilon_{\mu}^{*}\left[  \sum_{i=1}^{9} C_{i}H_{i}^{\mu\nu}L_{V\nu} + C_{10}\left( H_{10}^{\mu\nu}L_{A\nu}  -\frac{i}{2}f_{B_{s}}L_{A}^{\mu\nu}p_{\nu}\right)   \right]~
\end{align}
where we have defined 
\begin{align}
L_{V}^{\nu} = \bar{u}(p_{2})\gamma^{\nu}v(p_{1})~, \qquad\qquad L_{A}^{\nu} = \bar{u}(p_{2})\gamma^{\nu}\gamma^{5}v(p_{1}) ~.
\end{align}
The last term in Eq.~(\ref{eq:ampl_exp}), which only depends on the leptonic tensor $L_{A}^{\mu\nu}$~\cite{Beneke:2020fot} and on the decay constant $f_{B_{s}}$ of the $B_{s}$ meson, corresponds to the final-state radiation (FSR) contribution (to which we refer in the following as to the point-like contribution). The non-perturbative contribution to the structure-dependent part of the amplitude is instead encoded in the hadronic tensors $H^{\mu\nu}_{i}$, which can be grouped into three different categories: the contribution from the semileptonic operators $\mathcal{O}_{9-10}$, the contribution from the photon penguin operator $\mathcal{O}_{7}$, and finally the contributions from the four-fermion operators $\mathcal{O}_{1-6}$ and from the chromomagnetic penguin operator $\mathcal{O}_{8}$. 

\begin{figure}[t]
\includegraphics[scale=0.6]{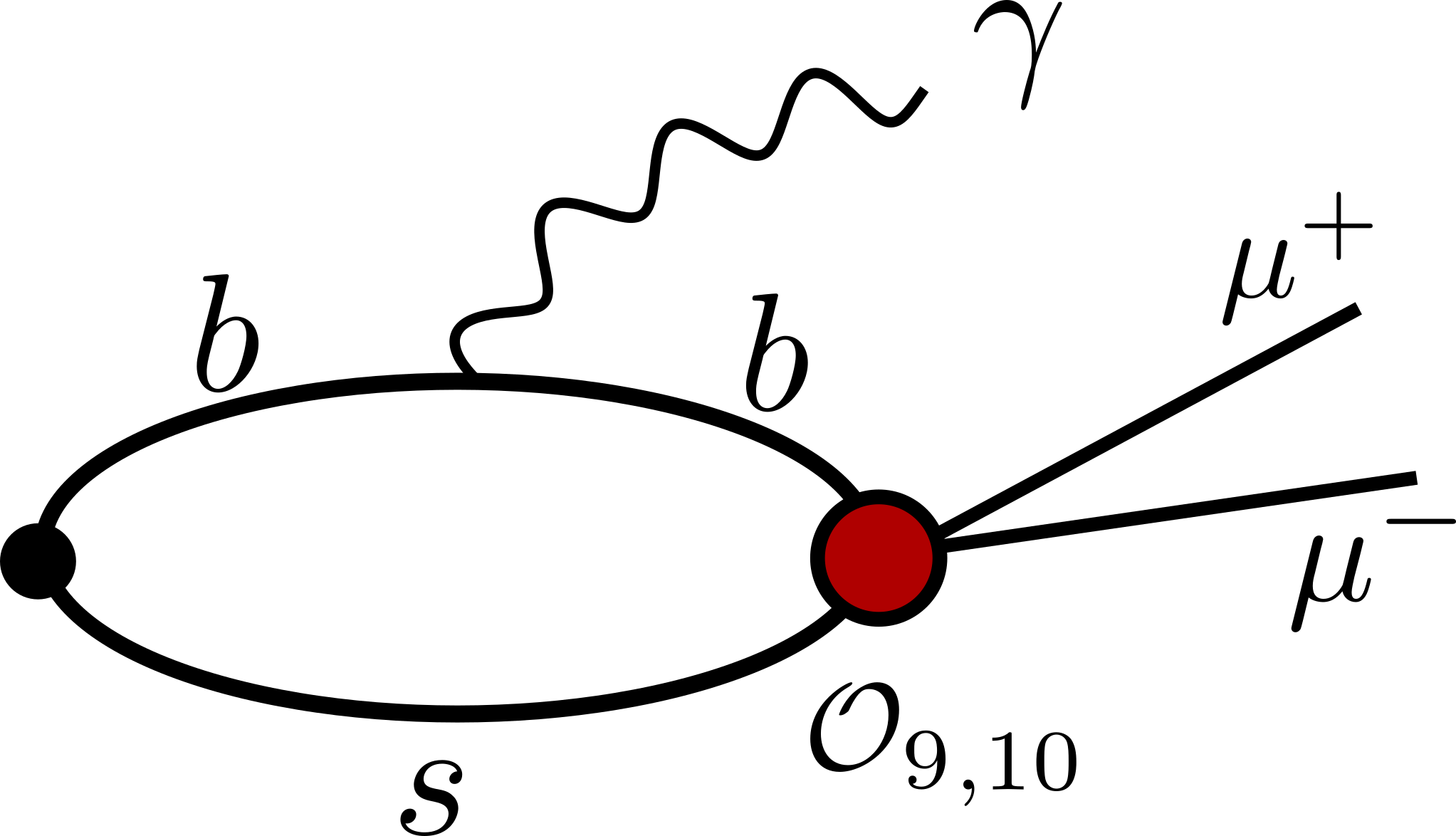}$\qquad\qquad\qquad\qquad$
\includegraphics[scale=0.6]{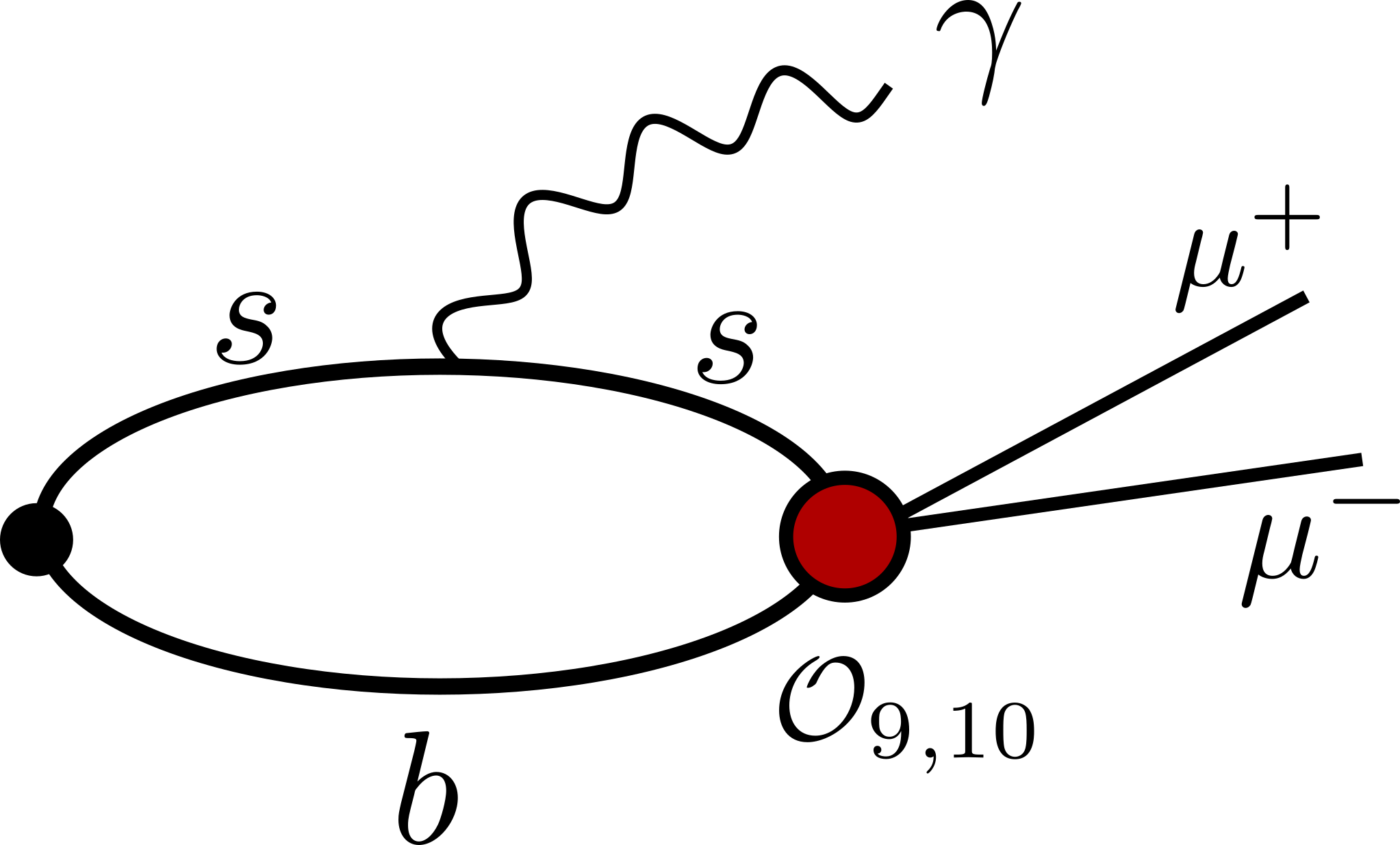}
\caption{\small\it Graphical representation of the contribution to the $\bar{B}_{s}\to \mu^{+}\mu^{-}\gamma$ decay amplitude from the semileptonic operators $\mathcal{O}_{9}$ and $\mathcal{O}_{10}$.}
\label{fig:semileptonic}
\end{figure}

The contributions from the semileptonic operators are depicted graphically in Figure\,\ref{fig:semileptonic}.
For these contributions the real photon $\gamma$ is emitted directly from one of the two quarks. The corresponding tensors $H^{\mu\nu}_{9-10}$ are given by
\begin{align}
\label{eq:FV_FA}
H_{9}^{\mu\nu}(p,k) = H_{10}^{\mu\nu}(p,k) &= i\int d^{4}y ~ e^{iky} ~ {\rm \hat{T}} \langle 0 | \left[\bar{s}\gamma^{\nu}P_{L}b\right](0) J_{{\rm em}}^{\mu}(y) | \bar{B}_{s}(\bs{p}) \rangle \nonumber \\[8pt]
&= -i\left[ g^{\mu\nu}(k\cdot q) -q^{\mu}k^{\nu}\right]\frac{F_{A}}{2m_{B_{s}}} +\varepsilon^{\mu\nu\rho\sigma}k_{\rho}q_{\sigma}\frac{F_{V}}{2m_{B_{s}}}~,
\end{align}
where $J_{\rm em}^{\mu}$ is the e.m. current, and $\hat{T}$ represents ``time-ordered". The two tensors are parameterized by vector ($F_{V}$) and axial ($F_{A}$) form factors, which are scalar functions of the single invariant of the process, namely the di-muon inviarant mass $q^{2}= (p-k)^{2}$. In the following, as in our previous papers, we present the form factors as functions of the dimensionless variable
\begin{align}
x_{\gamma} = \frac{2\,p\cdot k}{m_{B_{s}}^{2}} = 1 - \frac{q^{2}}{m_{B_{s}}^{2}}~,\qquad\qquad  0\leq x_{\gamma} \leq 1 - \frac{4m^{2}_{\mu}}{m_{B_{s}}^{2}}\,,
\end{align}
which in the decaying meson rest frame reduces to $x_{\gamma} = 2E_{\gamma}/m_{B_{s}}$, where $E_{\gamma}$ is the energy of the emitted photon.

\begin{figure}[t]
\includegraphics[scale=0.6]{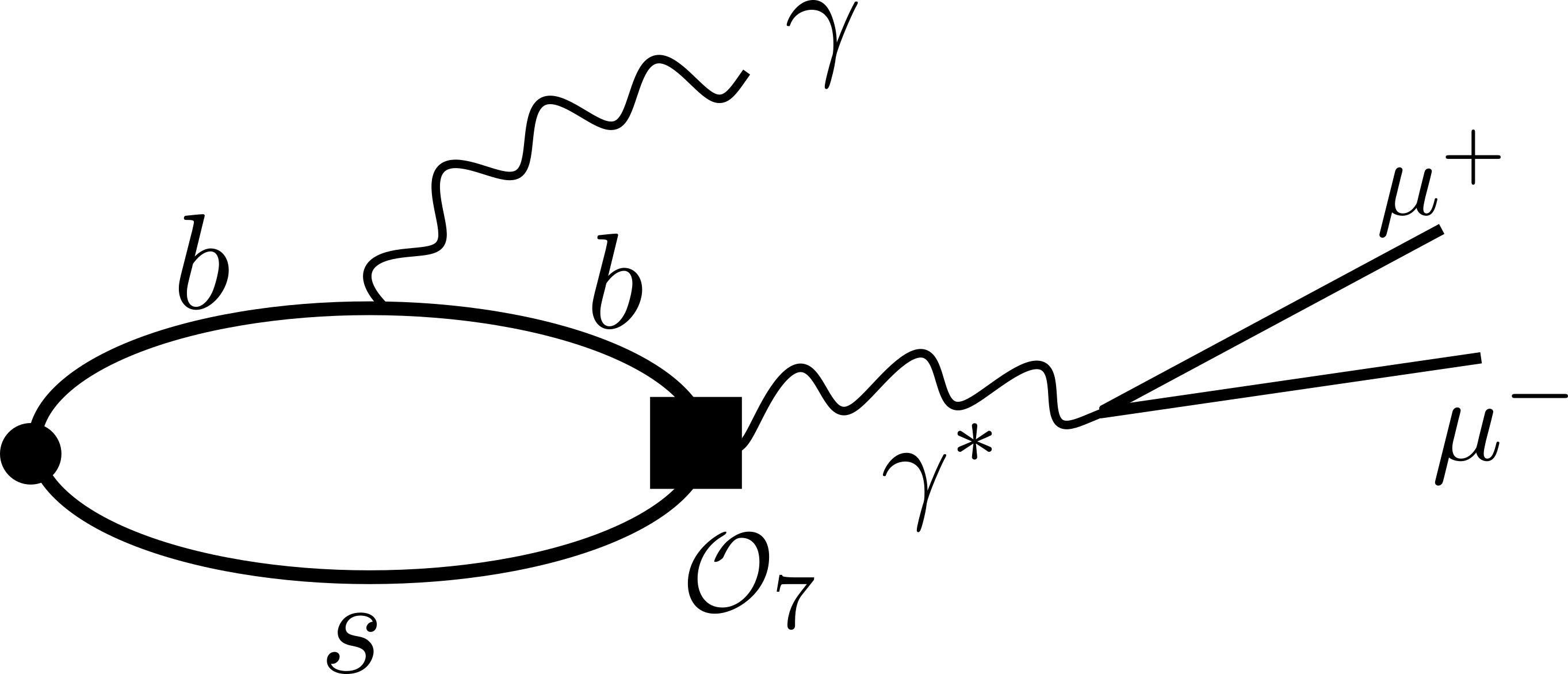}
$\qquad\qquad\qquad\qquad$
\includegraphics[scale=0.6]{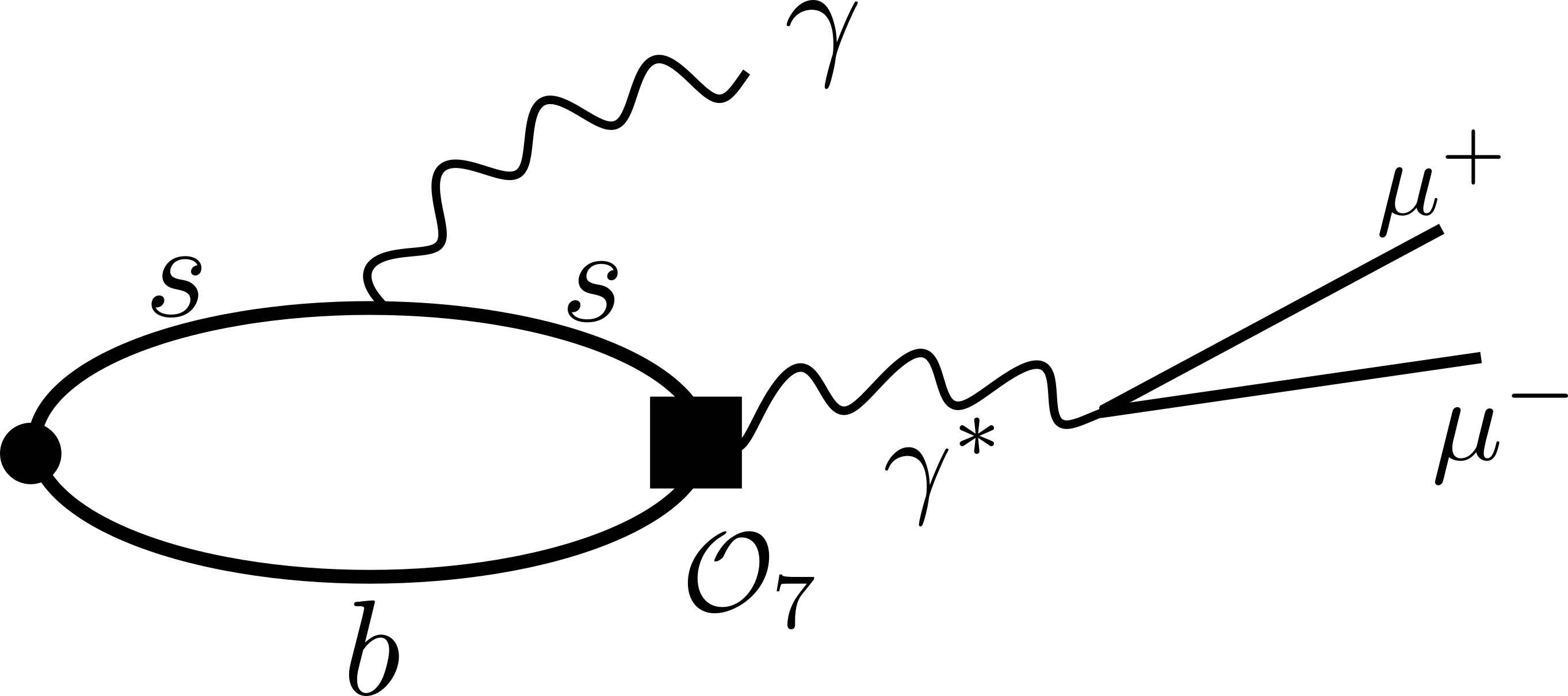}
\caption{\small\it Graphical representation of the contribution to the $\bar{B}_{s}\to \mu^{+}\mu^{-}\gamma$ decay amplitude from the photon penguin operator $\mathcal{O}_{7}$ in which the final-state photon is emitted directly from one of the valence quarks.}
\label{fig:ph_penguin1}
\end{figure}

\begin{figure}[t]
\includegraphics[scale=0.6]{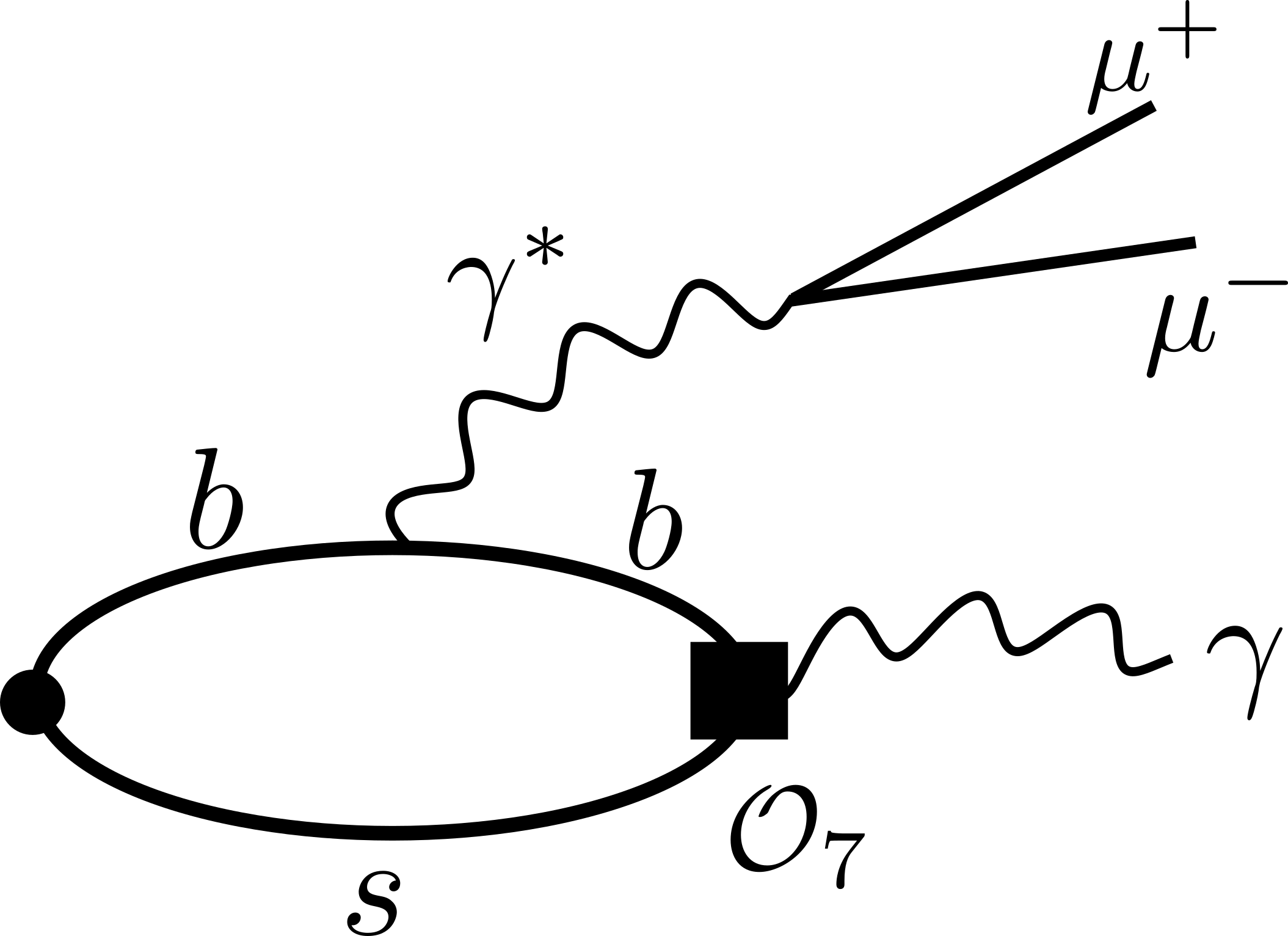}
$\qquad\qquad\qquad\qquad\qquad\qquad\qquad$
\includegraphics[scale=0.6]{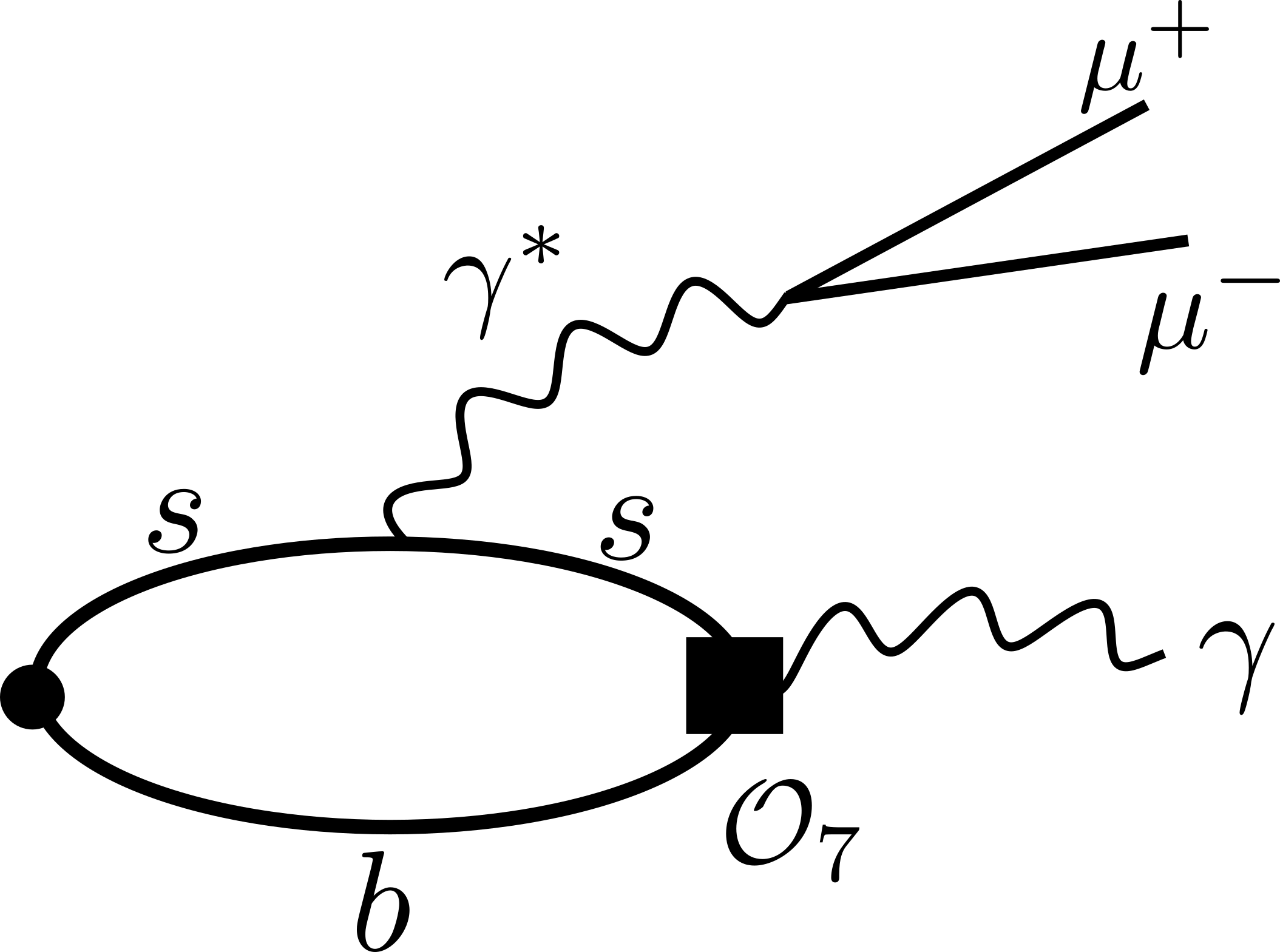}
\caption{\small\it Graphical representation of the contribution to the $\bar{B}_{s}\to \mu^{+}\mu^{-}\gamma$ amplitude from the photon penguin operator $\mathcal{O}_{7}$ in which  the final-state photon is emitted from the penguin operator.}
\label{fig:ph_penguin2}
\end{figure}

The contribution from the photon penguin operator $\mathcal{O}_{7}$ is illustrated in Figures\,\ref{fig:ph_penguin1} and \ref{fig:ph_penguin2}. 
In this case there are two types of contribution: those in which the final-state real photon is emitted by the valence quarks (Figure\,\ref{fig:ph_penguin1}), and those in which the real photon is emitted by the penguin vertex (Figure~\ref{fig:ph_penguin2}). We indicate by $H_{7A}^{\mu\nu}$ and $H_{7B}^{\mu\nu}$ the hadronic tensor corresponding to the first and second contribution respectively, with $H_{7}^{\mu\nu}=H_{7A}^{\mu\nu}+H_{7B}^{\mu\nu}$. The hadronic tensor $H_{7A}^{\mu\nu}$ is given by
\begin{align}
\label{eq:ph_penguin_T}
H_{7A}^{\mu\nu}(p,k) &= i\frac{2m_{b}}{q^{2}}\int d^{4}y ~ e^{iky} ~ {\rm \hat{T}} \langle 0 | \left[-i\bar{s}\sigma^{\nu\rho}q_{\rho}P_{R}b\right](0) J_{{\rm em}}^{\mu}(y) | \bar{B}_{s}(\bs{p})  \rangle \nonumber \\[8pt]
&= -i\left[ g^{\mu\nu}(k\cdot q) -q^{\mu}k^{\nu}\right]\frac{F_{TA}m_{b}}{q^{2}} +\varepsilon^{\mu\nu\rho\sigma}k_{\rho}q_{\sigma}\frac{F_{TV}m_{b}}{q^{2}}
\end{align}
where the two tensor form factors $F_{TV}$ and $F_{TA}$ are again scalar functions of $x_{\gamma}$. Exploiting the relation $\gamma^{5}\sigma^{\mu \nu}= -i\varepsilon^{\mu\nu\rho\sigma}\sigma_{\rho\sigma}/2$ one can show that the two tensor form factors obey the kinematical constraint $F_{TV}(1) = F_{TA}(1)$ (see also Ref.~\cite{Kozachuk:2017mdk}). The hadronic tensor $H^{\mu\nu}_{7B}$, corresponding to the emission of the real photon from the FCNC vertex is instead given by
\begin{align}
H_{7B}^{\mu\nu}(p,k) &= i\frac{2m_{b}}{q^{2}}\int d^{4}y ~ e^{iqy} ~ {\rm \hat{T}} \langle 0 | \left[-i\bar{s}\sigma^{\mu\rho}k_{\rho}P_{R}b\right](0) J_{{\rm em}}^{\nu}(y) | \bar{B}_{s}(\bs{p})  \rangle \nonumber \\[8pt]
&= -i\left[ g^{\mu\nu}(k\cdot q) -q^{\mu}k^{\nu}\right]\frac{\bar{F}_{TA}m_{b}}{q^{2}} +\varepsilon^{\mu\nu\rho\sigma}k_{\rho}q_{\sigma}\frac{\bar{F}_{TV}m_{b}}{q^{2}}
\end{align}
In this case, as discussed in~Ref.\,\cite{Kozachuk:2017mdk}, the two form factors obey $\bar{F}_{TV}(x_{\gamma})=\bar{F}_{TA}(x_{\gamma}) \equiv \bar{F}_{T}(x_{\gamma})$.\footnote{Again this can be shown making use of the relation $\gamma^{5}\sigma^{\mu\nu}= -i\varepsilon^{\mu\nu\rho\sigma}\sigma_{\rho\sigma}/2$.} Moreover at $x_{\gamma}=1$, i.e. at $q^{2}=k^{2}=0$ one has
\begin{align}\label{eq:F1}
F_{TV}(1) = F_{TA}(1) = \bar{F}_{T}(1)~.
\end{align}
The form factor $\bar{F}_{T}(x_{\gamma})$ is the most difficult to determine on the lattice. When the virtual photon $\gamma^{*}$ is emitted by a valence strange quark, the presence of intermediate $J^{P}= 1^{-}$ $s\bar{s}$ resonance states forbids the analytic continuation to Euclidean spacetime of the relevant Minkowskian correlation functions needed to evaluate $\bar{F}_{T}(x_{\gamma})$. In this case, in order to evaluate the form factor $\bar{F}_{T}$,  we rely on the spectral density reconstruction technique developed in Ref.\,\cite{Frezzotti:2023nun}. It is the form factors $F_{V}, F_A, F_{TV}, F_{TA}$ and $\bar{F}_{T}$ which we evaluate from first principles via lattice QCD simulations. In the following we sometimes refer to them as local form factors.

\begin{figure}
    \centering
    \includegraphics[scale=0.65]{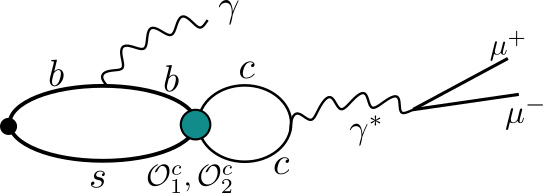}
    \caption{\small\it Graphical representation of the contribution to the $\bar{B}_{s}\to \mu^{+}\mu^{-}\gamma$ amplitude from the four-quarks operators $\mathcal{O}_{1}^{c}$ and $\mathcal{O}_{2}^{c}$ with the virtual photon $\gamma^{*}$ emitted by the charm loop (the corresponding 
    diagram with the real photon emitted from the strange valence quark is not shown). }
    \label{fig:penguin}
\end{figure}

The remaining contributions to the amplitude $\mathcal{A}[\bar{B}_{s}\to \mu^{+}\mu^{-}\gamma]$ are those corresponding to the four-quark operators and to the chromomagnetic penguin operator. The corresponding hadronic tensors $H^{\mu\nu}_{i=1-6,8}$ are given by
\begin{align}
\label{eq:H_penguins}
H^{\mu\nu}_{i=1-6,8}(p,k) = \frac{(4\pi)^{2}}{q^{2}}\int d^{4}y~d^{4}x~ e^{iky} e^{iqx}\, {\rm{\hat{T}}}\langle 0 | J_{\rm em}^{\mu}(y) J_{\rm em}^{\nu}(x) \mathcal{O}_{i}(0) | \bar{B}_{s}(\bs{p}) \rangle ~. 
\end{align}
In the high-$q^{2}$ region which we consider, as discussed in Ref.~\cite{Guadagnoli:2017quo}, the contribution of the presently neglected terms from $\mathcal{O}_{i=1-6,8}$ is expected to be small, since they are of higher-order in the $1/m_{b}$ expansion. Among them, one of the most important contribution is that of the charming-penguin diagram depicted in Figure~\ref{fig:penguin}, due to the presence of broad charmonium resonance contributions, which are near or within the region of $q^{2}$ we have explored. 

To take into account this contribution, we follow Refs.~\cite{Kozachuk:2017mdk, Guadagnoli:2017quo, Guadagnoli:2023zym} and include the charming-penguin diagram in Figure~\ref{fig:penguin} as a $q^{2}$-dependent shift of the Wilson coefficient $C_{9}$, namely
\begin{align}
C_{9} \to C_{9}^{\rm eff}(q^{2}) = C_{9}+ \Delta C_{9}(q^{2})~,
\end{align}
where $\Delta C_{9}(q^{2})$ can be phenomenologically modelled as a sum over the contributions from all the  $J^{P}= 1^{-}$ charmonium resonances~\cite{Kruger:1996cv, Guadagnoli:2023zym, Kozachuk:2017mdk}
\begin{align}
\label{eq:delta_C9V}
\Delta C_{9}(q^{2}) = -\frac{9 \pi}{\alpha_{\rm em}^2} \, (C_{1}+\frac{C_{2}}{3}) \,  \sum_V |k_{V}| e^{i\delta_{V}} 
 \frac{m_V \, B(V \to \mu^+ \mu^-) \, \Gamma_V}{q^2 - m_V^2 + i m_V \Gamma_V}\,,
\end{align}
where $\Gamma_{V}$ is the total decay width of the resonance $V$, $m_{V}$ its mass, and $B(V\to \mu^{+}\mu^{-})$ the branching fraction for the decay into a di-muon. 
The coefficient $k_{V}$ and the phase shift $\delta_{V}$ take into account deviations from the factorization approximation, which corresponds to $\delta_{V}=|k_{V}|-1=0$. 
The values of (some of) the parameters entering Eq.\,(\ref{eq:delta_C9V}), for the low-lying resonances, can be taken from experiments, but clearly this introduces a systematic error in our prediction. In the evaluation of the rate, we will use Eq.~(\ref{eq:delta_C9V}), and estimate the associated systematic error in a conservative way by varying the input parameters over a sufficiently broad range. The conservative systematic we associate to the missing charming-penguin diagram in Figure~\ref{fig:penguin} is expected to be sufficiently large to cover the uncertainty of all the other missing contributions from Eq.~\ref{eq:H_penguins}. In the future, in order to remove this source of systematic uncertainty and reduce the theory error, it will be extremely important to evaluate $H^{\mu\nu}_{i=1-6,8}$ on the lattice, in particular the charming-penguin contribution of Figure~\ref{fig:penguin}, which we plan to do. We now turn to the discussion of the calculation of the local form factors $F_{V}, F_{A}, F_{TV}, F_{TA}$ and $\bar{F}_{T}$. 
\section{The local form factors $F_{V}, F_{A}, F_{TV}$ and $F_{TA}$}
\label{sec:FF_num}
As illustrated in the previous section, the form factors $F_{W}(x_{\gamma})$, $W=\{ V, A, TV, TA\}$, can be computed from QCD matrix elements involving the e.m. and the following currents
\begin{align}
J^{\nu}_{A} =  Z_{V}\bar s \gamma^\nu\gamma_5 b , \qquad   J^{\nu}_{V} =  Z_{A}\bar s \gamma^\nu b , \qquad 
J^{\nu}_{TA} =  -iZ_{T}(\mu)\bar s \sigma^{\nu\rho}\gamma_5 b~\frac{q_{\rho}}{m_{B_{s}}} , \qquad   J^{\nu}_{TV} =  -iZ_{T}(\mu)\bar s \sigma^{\nu\rho} b~\frac{q_{\rho}}{m_{B_{s}}}~,
\end{align}
where, as already stated, $q^{\nu} = p^{\nu}-k^{\nu}$ is the four-momentum of the charged muon pair. In the previous equation we have introduced the scheme- and scale-dependent renormalization constant (RC) $Z_{T}(\mu)$ of the tensor current, and the (finite) RCs of the axial and vector currents that in twisted-mass QCD are chirally rotated with respect to the ones of standard Wilson fermions. 
From now on, we work in the rest frame of the decaying meson and thus set $p=(m_{B_{s}}, \bs{0})$. 
In terms of the hadronic tensors
\begin{align}
\label{hadronic_tensor_Ds}
 H^{\mu\nu}_{W}(p,k) \equiv i\int d^{4}y ~ e^{iky} ~ {\rm \hat{T}} \langle 0 | J_{W}^{\nu}(0) J_{{\rm em}}^{\mu}(y) | \bar{B}_{s}(\bs{0}) \rangle~,\qquad W=\{V,A,TV,TA\}~,
\end{align} 
and recalling the definitions given in Eqs.~(\ref{eq:FV_FA})-(\ref{eq:ph_penguin_T}), one has that
\begin{align}
H^{\mu\nu}_{A}(p,k) &= i\left[(k\cdot q)g^{\mu\nu} - q^{\mu}k^{\nu}\right]\frac{F_{A}}{m_{B_{s}}}~, \qquad\qquad
H^{\mu\nu}_{V}(p,k) = \epsilon^{\mu\nu\rho\sigma}k_{\rho}p_{\sigma}\frac{F_{V}}{m_{B_{s}}} \nonumber \\[10pt]
H^{\mu\nu}_{TA}(p,k) &= -i\left[(k\cdot q)g^{\mu\nu} - q^{\mu}k^{\nu}\right]\frac{F_{TA}}{m_{B_{s}}}~, \qquad\quad\,
H^{\mu\nu}_{TV}(p,k) = \epsilon^{\mu\nu\rho\sigma}k_{\rho}p_{\sigma}\frac{F_{TV}}{m_{B_{s}}} \,.
\end{align}
In Section\,III and Appendix\,B of Ref.\,\cite{Desiderio:2020oej} we show in detail that for the emission of a real photon, the hadronic tensor $H_{W}^{\mu\nu}$ can be extracted for all values of $x_{\gamma}$ from the Euclidean three-point correlation function:
\begin{equation}\label{eq:Cmunudef}
B^{\mu\nu}_W(t,\bs{k}) =a\sum_{t_y=0}^{T} a^{3}\sum_{\bs{y}}a^{3}\sum_{\bs{x}}\left( \theta(T/2-t_{y}) + \theta(t_{y}-T/2)e^{-E_{\gamma}T}\right) ~e^{\hspace{1pt}t_{y}E_\gamma-i\bs{k}\cdot\bs{y}}~
\bra 0\hat{\mathrm{T}}\,[J^\nu_W(t,\bs{0})J_\mathrm{em}^\mu(t_{y}, \bs{y})\phi^\dagger_{B_s}(0,\bs{x})]\ket 0\,,
\end{equation}
where $T$ is the temporal extent of the lattice\,\footnote{$T$ is not to be confused with $\hat{\mathrm{T}}$ which represents ``time-ordered".}, $a$ is the lattice spacing, and $\phi^\dagger_{B_s}$ is an interpolating operator with the quantum numbers to create the $\bar{B}_s$ meson which, as in Ref.\,\cite{Frezzotti:2023ygt}, we smear using Gaussian smearing. 
For the electromagnetic current $J^\mu_{\rm em}$ we use the exactly-conserved point-split lattice operator
\begin{align}
\label{eq:1PS_em}
J_{\mathrm{em}}^{\mu}(x) = \sum_{f} J_{f}^{\mu} = -\sum_{f} q_{f} \left\{
\bar \psi_f(x)\frac{ir_{f}\gamma_5-\gamma^\mu}{2}\, U^\mu(x)\psi_f(x+\hat \mu)
-
\bar \psi_f(x+\hat \mu)\frac{ ir_{f}\gamma_5+\gamma^\mu}{2}U^\mu(x)^\dagger \psi_f(x) \right\}\,~.
\end{align} 
In the forward half of the lattice $0\ll t \ll T/2$ one has 
\begin{equation}
R^{\mu\nu}_W(t,\bs{k}) 
\equiv \frac{2m_{B_{s}}}{e^{-t(m_{B_{s}}-E_\gamma)}\, \langle \bar{B}_{s}(\bs{0})|\phi^{\dagger}_{B_s}(0) | 0 \rangle } \, B^{\mu \nu}_W(t, \bs{k})
=
H^{\mu\nu}_W(p,k) + \cdots ~,
\label{eq:Rinf}
\end{equation}
where the ellipsis indicates terms that vanish exponentially in the large $t$ limit. 
 Eq.\,(\ref{eq:Cmunudef}) is valid for $t < T/2$, however, as explained in Appendix B of Ref.\,\cite{Desiderio:2020oej}, $H^{\mu\nu}_W(p,k)$ can also be obtained from the backward half of the lattice $T/2\ll t\ll T$ exploiting time-reversal symmetry. 
  \begin{figure}[t]
    \begin{center}
    \includegraphics[width=0.42\textwidth]{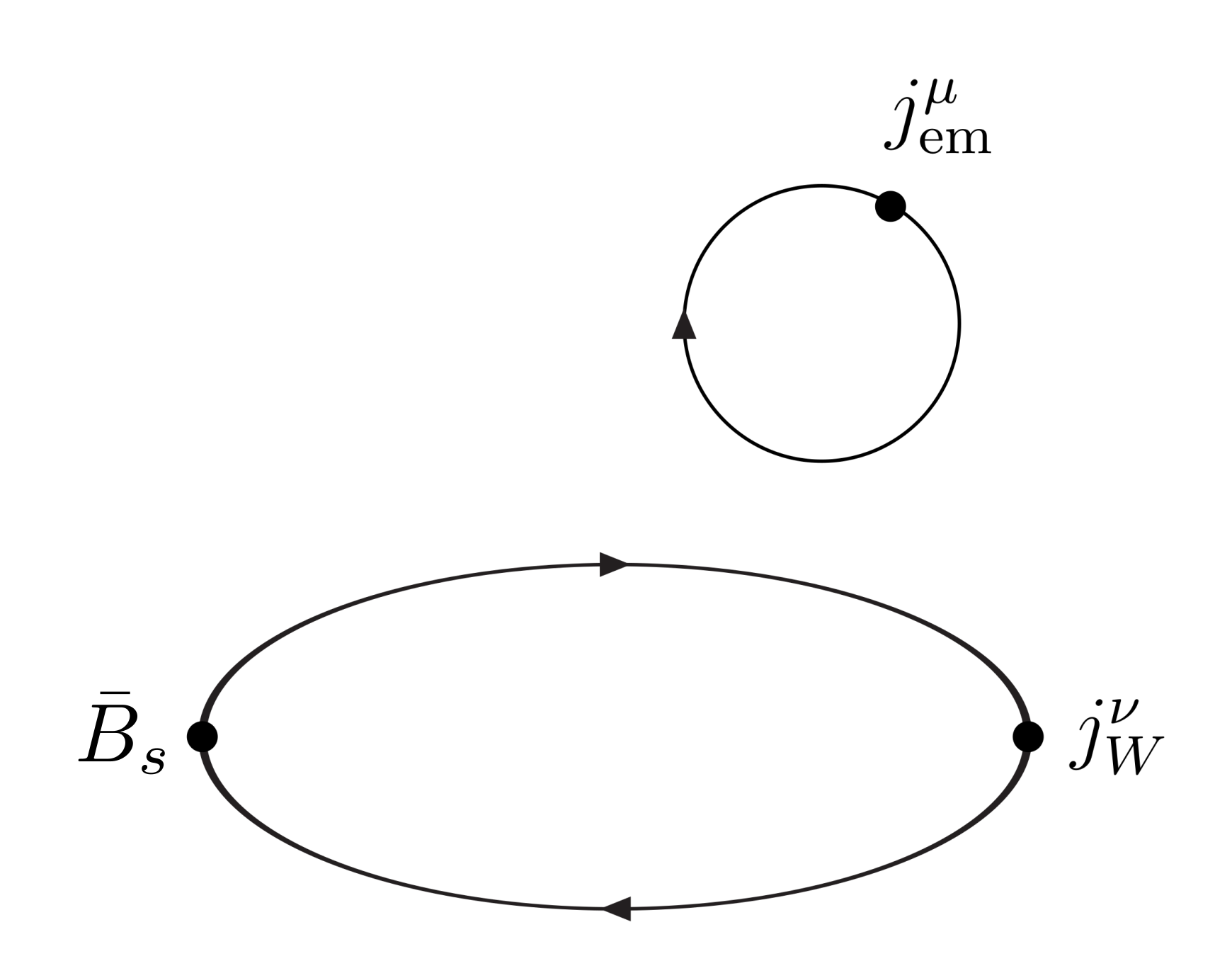}
    \hspace{0.1cm}
    \includegraphics[width=0.42\textwidth]{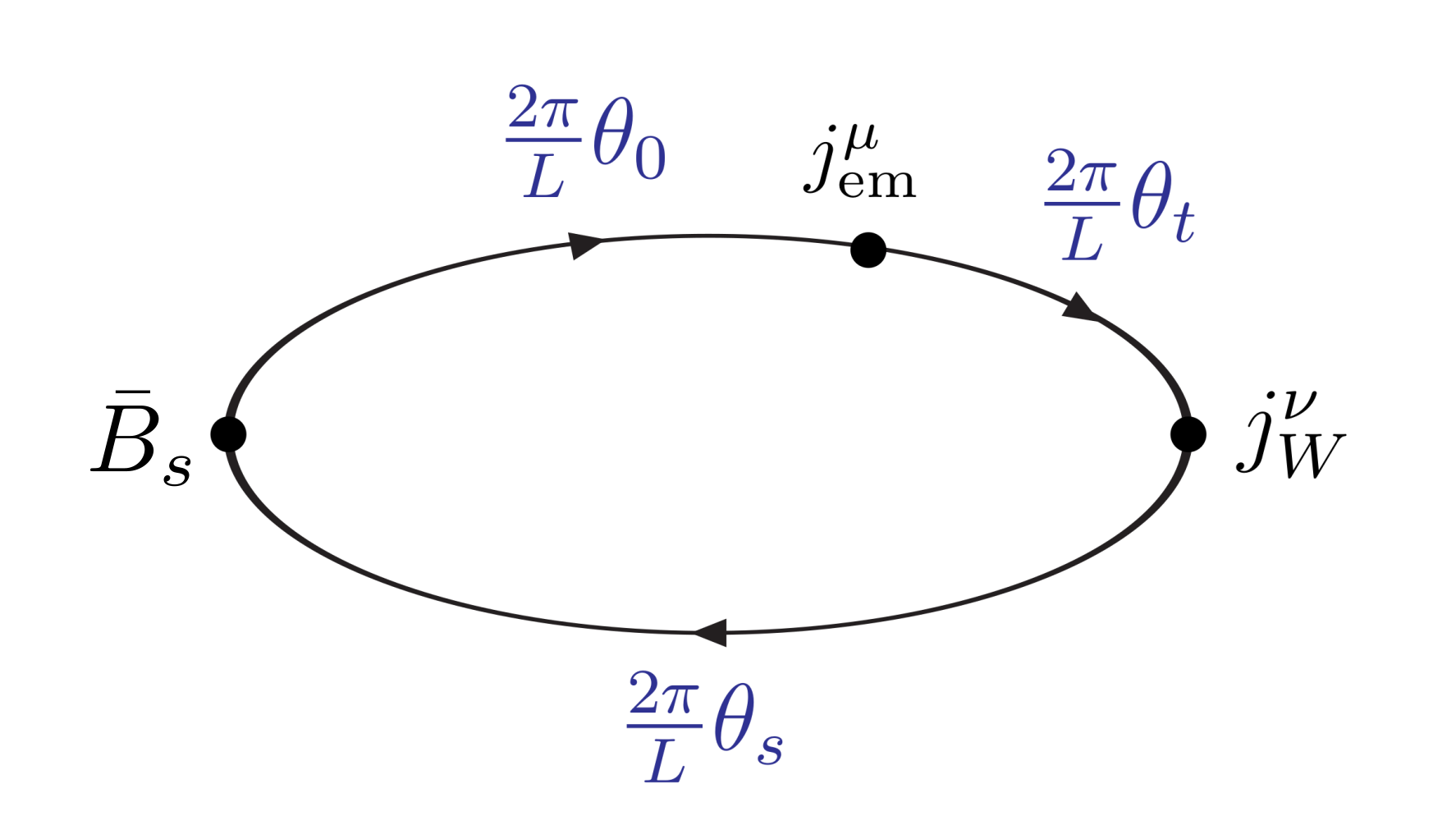}
    \end{center}
    \caption{\small\it The diagram on the left represents the quark-line disconnected contributions to the correlation function $B^{\mu\nu}_{W}$ in which the photon is emitted by a sea quark. In our numerical simulations we work in the electroquenched approximation and neglect such diagrams. The diagram on the right represents the quark-line connected contributions and illustrates our choice of the spatial boundary conditions, which allow us to set arbitrary values for the meson and photon spatial momenta.
    The spatial momenta of the valence quarks in terms of the twisting angles are as indicated. Each diagram implicitly includes all orders in QCD.
    \hspace*{\fill}
    \label{fig:Feyn_conn_disc}}
    \end{figure}
    The Wick contractions of the correlation function in Eq.\,(\ref{eq:Cmunudef}) give rise to two distinct topologies of Feynman diagrams, namely to quark-line connected and quark-line disconnected diagrams; these are illustrated in Figure\,\ref{fig:Feyn_conn_disc}. In the disconnected diagrams the photon is emitted from a sea quark. This contribution vanishes in the $\rm{SU}(3)$-symmetric limit and when loop of charmed and heavier quarks are omitted, and is neglected in the present study; this is the so-called electroquenched approximation. 
We focus instead on the calculation of the dominant, quark-connected contributions for which  only the strange- and bottom-quark components of the electromagnetic current $J_{\rm em}^{\mu}$ contribute.

As explained in Ref.\,\cite{Desiderio:2020oej}, it is possible to use twisted boundary conditions to assign arbitrary values to momenta of the photon and $\bar{B}_s$-meson, 
$\bs{k}$ and $\bs{p}$ respectively, at the price of violations of unitarity which vanish exponentially with the lattice extent $L$
\,\cite{Sachrajda:2004mi,Flynn:2007ess,deDivitiis:2004kq}. 
This is achieved by treating the two quark propagators 
beginning or ending at $y$, i.e. the point at which the 
electromagnetic current is inserted in the right-hand diagram of 
Figure\,\ref{fig:Feyn_conn_disc}, as corresponding to two distinct quark fields $\psi^{0}, \psi^{t}$ having the same mass and quantum number, but satisfying different spatial boundary conditions.
Defining $\psi^{s}$ to be the \textit{spectator} quark-field in the right-hand diagram of Figure\,\ref{fig:Feyn_conn_disc}, we set the spatial boundary  conditions of the three quark fields $\psi^{0},\psi^{t},\psi^{s}$ as follows:
\begin{align}
\psi^{r}(x+\bs{n}L) = \exp{(2\pi i\bs{n}\cdot \bs{\theta}_{r})}\psi^{r}(x)\;,\qquad r=\{ 0, t, s \}~,
\end{align}
where $\bs{\theta}_{\{0,t,s\}}$ are arbitrary spatial-vectors of angles, in terms of which the photon and meson lattice momenta can be written as
\begin{align}
\bs{p} = \frac{2}{a}\sin\left(\frac{a\pi}{L}\left( \bs{\theta}_0-\bs{\theta}_s\right)\right)\;,
\qquad
\bs{k} = \frac{2}{a}\sin\left(\frac{a\pi}{L}\left( \bs{\theta}_0-\bs{\theta}_t\right)\right)\;,
\label{eq:momenta}    
\end{align}
We choose the photon momentum to be in the $z$-direction, $\bs{k}=(0,0,k_z)$, and set
\begin{align}
\bs{\theta}_{0}= \bs{\theta}_{s} = \bs{0}\;,\qquad \bs{\theta_t} = (0,0,\theta_t).
\end{align}
With such a choice of kinematics, the form factors can be obtained from the large time behaviour, $0\ll t \ll T/2$, of the following estimators
\begin{align}
\label{eq:RV}
R_{V}(t,\bs{k}) &\equiv  \frac{1}{2k_{z}}\left( R_{V}^{12}(t,\bs{k}) - R_{V}^{21}(t,\bs{k})\right) ~  \xrightarrow[0\ll t \ll T/2]{} ~ F_{V}(x_{\gamma})\, ,  \\[10pt]
\label{eq:RA}
R_{A}(t, \bs{k}) &\equiv  \frac{i}{2E_{\gamma}}\left( R_{A}^{11}(t,\bs{k}) + R_{A}^{22}(t,\bs{k})\right)  ~ \xrightarrow[0\ll t \ll T/2]{} ~ F_{A}(x_{\gamma})\,,\\[10pt]
\label{eq:RTV}
R_{TV}(t,\bs{k}) &\equiv  \frac{1}{2k_{z}}\left( R_{TV}^{12}(t,\bs{k}) - R_{TV}^{21}(t,\bs{k})\right) ~  \xrightarrow[0\ll t \ll T/2]{} ~ F_{TV}(x_{\gamma})\, ,  \\[10pt]
\label{eq:RTA}
R_{TA}(t, \bs{k}) &\equiv  -\frac{i}{2E_{\gamma}}\left( R_{TA}^{11}(t,\bs{k}) + R_{TA}^{22}(t,\bs{k})\right)  ~ \xrightarrow[0\ll t \ll T/2]{} ~ F_{TA}(x_{\gamma})\,.
\end{align}
For each form factor it is useful to distinguish the two contributions due to the emission of the real photon from the bottom and strange quarks (left and right diagrams in Figures\,\ref{fig:semileptonic} and \ref{fig:ph_penguin1}). 
We denote the two contributions by $F_{W}^{b}$ and $F_{W}^{s}$ for $W=\{ V, A, TV, TA\}$. 
They are simply obtained by setting respectively the electric charges $q_{s}=0$ and $q_{b}=0$ in all the previous formulae. 
A minor complication arises in the axial channel $W=A$ due to the presence of a point-like contribution, proportional to $q_{b}f_{B_{s}}$ and $-q_{s}f_{B_{s}}$ respectively in $F_{A}^{b}$ and $F_{A}^{s}$, which then cancels in the sum of the two contributions due to $q_{b}=q_{s}$. This point-like contribution, which is always present in the radiative leptonic decays of charged pseudoscalar mesons~\cite{Desiderio:2020oej}, can however, be easily removed by calculating the following zero-momentum-subtracted estimator
\begin{align}
R_{A}^{(s,b)}(t, \bs{k}) \equiv \frac{i}{2E_{\gamma}}\left( R_{A}^{11,(s,b)}(t,\bs{k}) - R_{A}^{11,(s,b)}(t,\bs{0})  + R_{A}^{22,(s,b)}(t,\bs{k}) - R_{A}^{22,(s,b)}(t,\bs{0})\right)  ~ \xrightarrow[0\ll t \ll T/2]{} ~ F_{A}^{(s,b)}(x_{\gamma})~.
\end{align}
We refer the reader to Ref.\,\cite{Desiderio:2020oej} for more details on the removal of the point-like contribution.

\subsection{Numerical results for $F_{V}, F_{A}, F_{TV}, F_{TA}$}
\label{subsec:formfactorsnumerical}
\begin{figure}[t]
    \centering
    \includegraphics[scale=0.525]{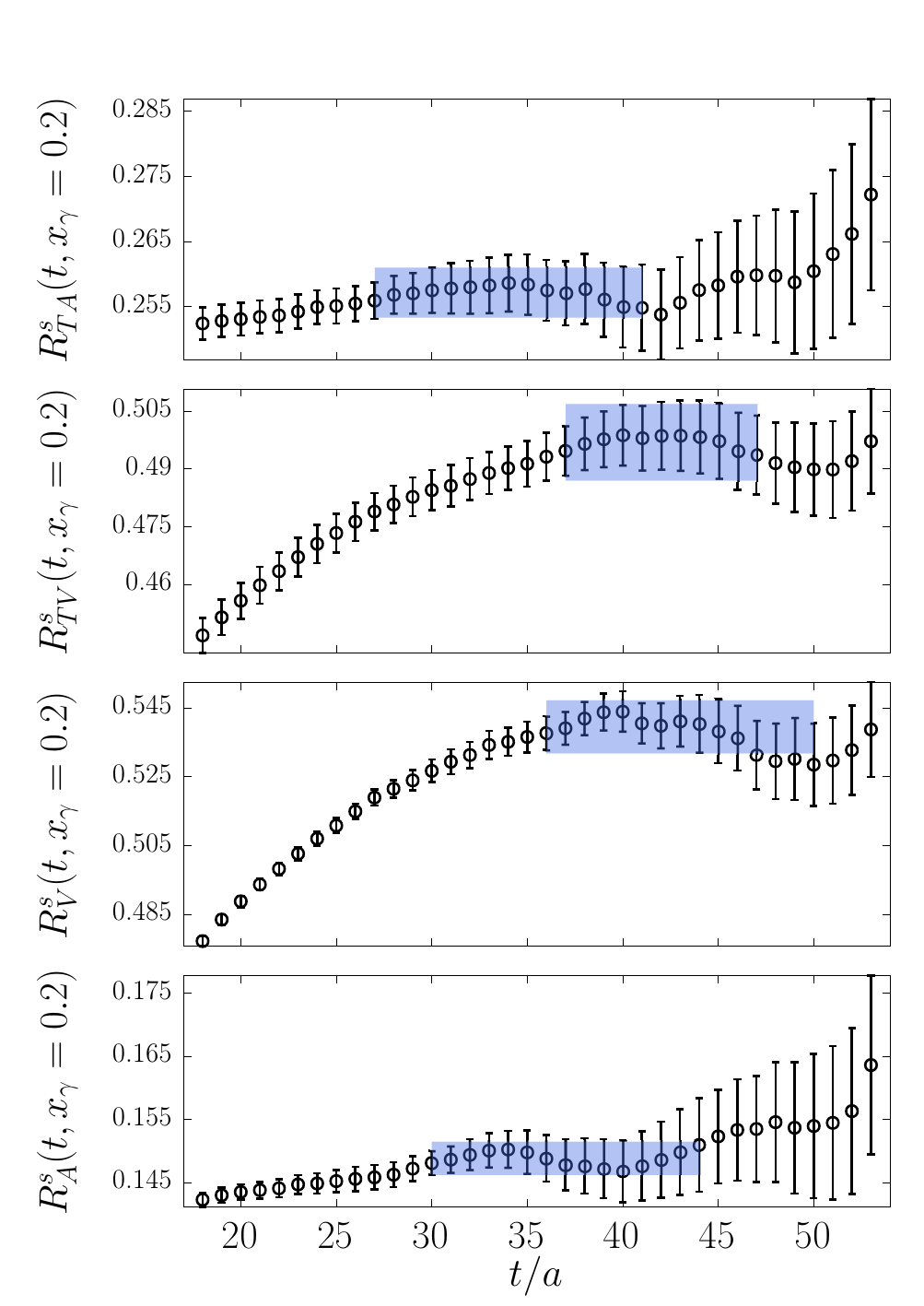}
    \includegraphics[scale=0.525]{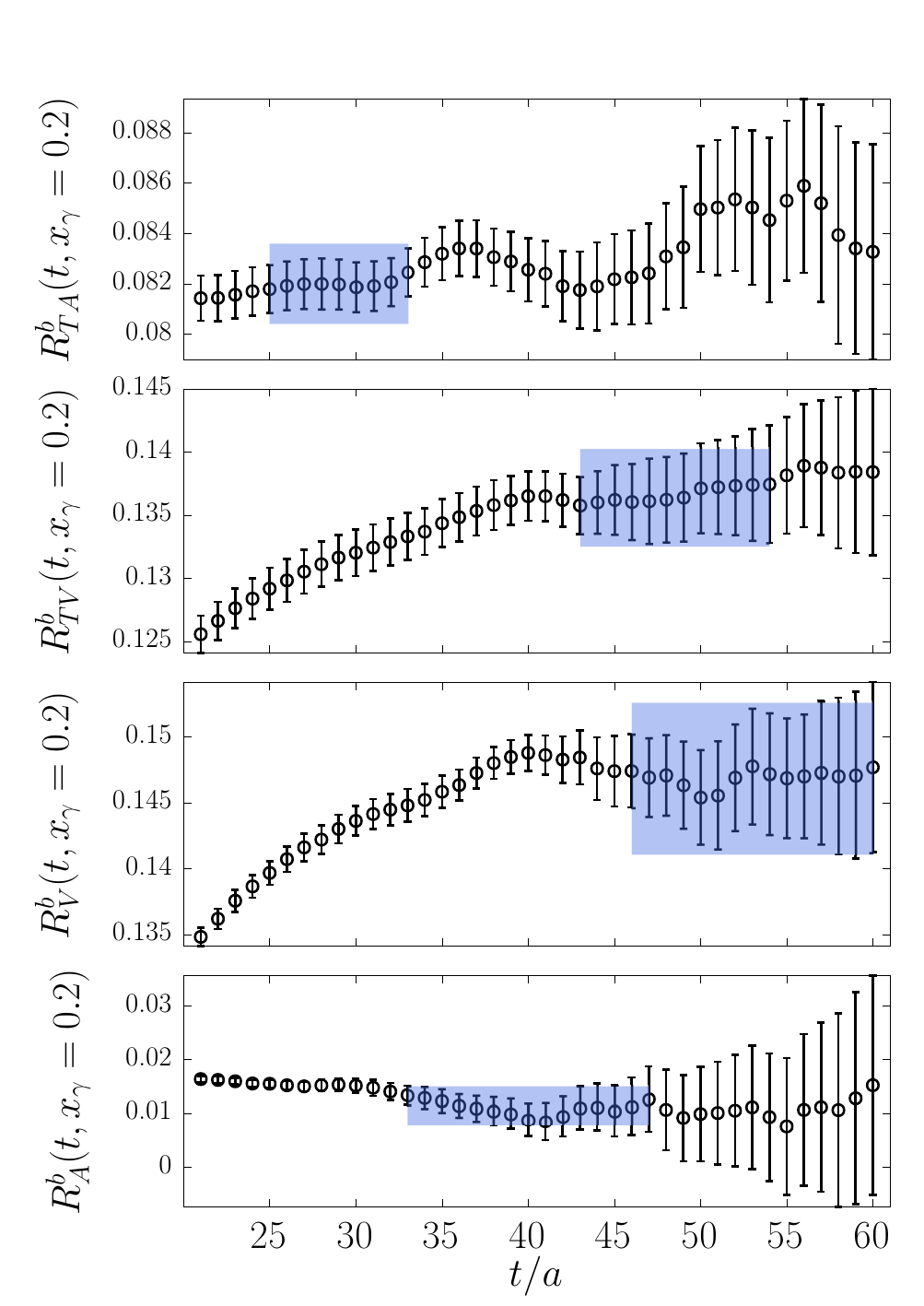}
    \caption{\small\it The estimators $R_{W}^{s}(t, x_{\gamma})$ (left) and $R_{W}^{b}(t,x_{\gamma})$ (right) for $W=\{V,A,TV,TA\}$. The data corresponds to the D96 ensemble at $x_{\gamma}=0.2$ and for $m_{h}\simeq 2m_{c}$. The blue bands show our estimates of the form factors from a constant fit in the region where the estimators $R_{W}^{(s,b)}(t,x_{\gamma})$ display a plateau.}
    \label{fig:plat_D64}
\end{figure}

We now turn to the discussion of our numerical results for $F_{V}, F_{A}, F_{TV}$ and $F_{TA}$. They have been obtained using the gauge field configurations generated by the Extended Twisted Mass Collaboration (ETMC) employing the Iwasaki gluon action~\cite{Iwasaki:1985we} and $N_{f}=2+1+1$ flavours of Wilson-Clover twisted-mass fermions at maximal twist~\cite{Frezzotti:2000nk}. This framework guarantees the automatic $\mathcal{O}(a)$ improvement of parity-even observables~\cite{Frezzotti:2003ni,Frezzotti:2004wz}. A detailed description of the ETMC ensembles can be found in Refs.~\cite{ExtendedTwistedMass:2021gbo,ExtendedTwistedMass:2021qui,Alexandrou:2022amy,Alexandrou:2018egz}, and we also refer to Ref.\,\cite{Frezzotti:2023ygt} for additional information on the tuning of the sea and valence quark masses.
In Table\,\ref{tab:simudetails} we present the parameters of the ETMC ensembles that have been used in the present computation, while in Table\,\ref{tab:renormalization} we collect the relevant RCs used to renormalize the vector, axial, and tensor currents.
\begin{table}
\begin{center}
    \begin{tabular}{||c||c|c|c|c|c||c||c||}
    \hline
    ~~~ ensemble ~~~ & ~~~ $\beta$ ~~~ & ~~~ $V/a^{4}$ ~~~ & ~~~ $a$\,(fm) ~~~ & ~~~ $a\mu_{\ell}$ ~~~ & ~ $m_{\pi}$\,(MeV) ~ & ~ $L$ (fm) ~ &  ~ $N_{g}$  ~ \\
  \hline
  A48 & $1.726$ & $48^{3}\cdot 128$ & $0.09075~(54)$ & $0.00120$ & $174.5~(1.1)$ & $4.36$ & $109$  \\
  
  B64 & $1.778$ & $64^{3}\cdot 128$ & $0.07957~(13)$ & $0.00072$ & $140.2~(0.2)$ & $5.09$   & $400$ \\
  
  C80 & $1.836$ & $80^{3}\cdot 160$ & $0.06821~(13)$ & $0.00060$ & $136.7~(0.2)$ & $5.46$   & $72$ \\
  
  D96 & $1.900$ & $96^{3}\cdot 192$ & $0.05692~(12)$ & $0.00054$ & $140.8~(0.2)$ & $5.46$   & $100$ \\
  \hline
    \end{tabular}
\end{center}
\caption{\it \small Parameters of the ETMC ensembles used in this work. We present the light-quark bare mass, $a \mu_\ell = a \mu_u = a \mu_d$, the lattice spacing $a$,  the pion mass $m_\pi$, the lattice size $L$ 
and the number of gauge configurations $N_{g}$ that have been used for each ensemble. The values of the lattice spacing are determined as explained in Appendix B of Ref.~\cite{Alexandrou:2022amy} using the value $f_\pi^{isoQCD} = 130.4(2)$\,{\rm MeV}  for the pion decay constant.}
\label{tab:simudetails}
\end{table}
\begin{center}
\begin{table}[]
    \centering
    \begin{tabular}{||c||c|c|c||}
    \hline
    ensemble & $Z_{V}$ & $Z_{A}$ & $Z_{T}(\msbar, 5~{\rm GeV})$ \\
    \hline
    ~ A48 ~ & ~ $0.68700~(15)$ ~ & ~ $0.7284~(18)$ ~ & ~ $0.7541~(98)$  \\
    ~ B64 ~ & ~ $0.706379~(24)$ ~ & ~ $0.74294~(24)$ ~ & ~ $0.7735~(93)$  \\
    ~ C80 ~ & ~ $0.725404~(19)$ ~ & ~ $0.75830~(16)$ ~ & ~ $0.7928~(85)$ \\
    ~ D96 ~ & ~ $0.744108~(12)$ ~ & ~ $0.77395~(12)$ ~ & ~ $0.8141~(74)$ \\
    \hline
    \end{tabular}
    \caption{\it \small The values of the vector ($Z_{V}$), axial ($Z_{A}$), and tensor ($Z_{T}$) renormalization constants, for the ETMC ensembles of Table\,\ref{tab:simudetails}.
      $Z_{T}$ values were kindly provided to us by the ETMC, and are from a preliminary analysis~\cite{ExtendedTwistedMass_RCs}. For the present work, we increased their uncertainties by a factor of 3. The scale-independent renormalization constants $Z_{V}$ and $Z_{A}$ have been determined in Ref.~\cite{Alexandrou:2022amy} using Ward-identity methods.}
    \label{tab:renormalization} 
\end{table}
\end{center}
The presently available lattice spacings are not small enough to perform simulations at the physical bottom quark mass. 
For this reason our strategy to reach the physical $B_{s}$ meson mass, is to perform simulations for a series of heavy-strange quark masses, and then extrapolate to the physical point using heavy-quark effective theory (HQET) scaling relations, to be discussed in the next sections. 
For each of the ensembles of Table\,\ref{tab:simudetails}, we have performed simulations at five different values of $m_{H_s}$, the mass of the lightest pseudoscalar meson composed of a valence heavy quark of mass $m_h$ and a strange antiquark with mass $m_s$.
The five values correspond to the following five $m_h/m_c$ ratios ($m_c$ is the mass of the charm quark determined by the condition $m_{\eta_{c}}= 2.984(4) ~{\rm GeV}$, see Refs.~\cite{Frezzotti:2023nun, Alexandrou:2022amy}):
\begin{align}
\label{eq:mass_list}
\frac{m_{h}}{m_{c}} \simeq 1,~1.5,~2,~2.5,~3~.
\end{align}
Such values of the heavy quark masses $m_{h}$ give rise to heavy-strange meson masses $m_{H_{s}}$ in the range $m_{H_{s}}/m_{D_{s}} \mathlarger{\mathlarger{\in}}\, [1, 2]$.  
For each ensemble and heavy quark mass $m_{h}$, we evaluate 
the Euclidean three point function $B_{W}^{\mu\nu}(t;k,p)$ at four evenly-spaced values of the dimensionless variable $x_{\gamma}$:
\begin{align}
\label{eq:xg_val}
x_{\gamma} = \frac{2E_{\gamma}}{m_{H_{s}}} =  0.1, 0.2, 0.3, 0.4~. 
\end{align}
For an illustration of the quality of the plateaus, we present
in Figure\,\ref{fig:plat_D64} the estimators $R^{(s,b)}_{W}(t,x_{\gamma})\equiv R_{W}^{(s,b)}(t, (0,0, k_{z}(x_{\gamma}))$, $W=\{V,A,TV, TA\}$,  obtained at $x_{\gamma}=0.2$ on the finest lattice spacing ensemble (D96) for $m_{h}/m_{c} \simeq 2$.
In each figure the blue band shows our estimate of the corresponding form factor, obtained from a constant fit in the region where the estimators $R_{W}^{(s,b)}(t,x_{\gamma})$ display a plateau. The band already includes the systematic error due to the choice of the fit-interval, which is estimated by performing a second fit shifting the fit-interval forward in time by an amount $\Delta T= 0.4, 0.35, 0.30, 0.27,0.25\,{\rm fm}$, respectively for $m_{h}/m_{c} \simeq 1, 1.5, 2, 2.5, 3$, and then adding the difference between the central values obtained in the two different fits as a systematic error\footnote{The choice of $\Delta T$, for each value of $m_{h}$, has been adjusted so that this is small enough to avoid the region of large times where the signal-to-noise ratio of the estimator $R_{W}^{s,b}$ is very small, and at the same time large enough to provide a reasonable estimate of the systematics due to the choice of the fit interval.}. For the tensor form factors the results are obtained using the preliminary values of $Z_{T}(\mu)$ in the $\msbar$ scheme at $\mu=5~{\rm GeV}$, provided to us by the ETMC~\cite{ExtendedTwistedMass_RCs}. 

The ensembles of Table\,\ref{tab:simudetails} all correspond to lattices with a spatial  extent in the range $L \simeq 4.4$\,-\,$5.4\,{\rm fm}$. 
These volumes are expected to be large enough for the finite size effects (FSEs) on the form factors to be small. For the smallest heavy quark mass considered, $m_{h}=m_{c}$, and for the form factors $F_{A}$ and $F_{V}$, this has been explicitly checked in Ref.~\cite{Frezzotti:2023ygt} using an additional ensemble, the B96, which has a large spatial extent $L$ of more than $7.5~{\rm fm}$. 
Here, using the B96 ensemble, we have checked that FSEs are very small (at the level of our statistical uncertainty or smaller)  also  for the tensorial form factors $F_{TV}$ and $F_{TA}$, and we therefore consider our results on the ensembles listed in Table\,\ref{tab:simudetails} as infinite-volume quantities.  

Next we consider the cut-off effects. For each value of $x_{\gamma}$ and $m_{H_{s}}$, the extrapolation to the continuum limit is performed using the following Ansatz
\begin{align}
\label{eq:fit_ansatz}
F_{W}^{(b,s)}(x_{\gamma}, m_{H_{s}}, a) = F_{W}^{(b,s)}(x_{\gamma}, m_{H_{s}})\left( 1 + D_{W}^{(b,s)}(x_{\gamma}, m_{H_{s}})\, a^{2} \right)~,\qquad W= \{ V, A, TV, TA \}~,
\end{align}
where $F_{W}^{(b,s)}(x_{\gamma}, m_{H_{s}})$ and $D_{W}^{(b,s)}(x_{\gamma}, m_{H_{s}})$ are fit parameters which depend on $x_{\gamma}$ and $m_{H_{s}}$, and are different for the four channels $W = \{ V, A, TV, TA\}$ and for the two contributions $F_{W}^{s}$ and $F_{W}^{b}$. 
We estimate the systematic uncertainty due to the continuum-limit extrapolation by performing two different linear extrapolations: in the first one we include the full dataset, and in the second one we remove the measurements on the ensemble with the largest lattice spacing (A48). The two results are combined as follows: let $f_A$ and $f_B$ represent generically the continuum values of $F_W(x_\gamma)$, for a given $W=\{V,A,TV,TA\}$ and $x_\gamma \in \{ 0.1, \ldots, 0.4\}$, obtained respectively from the linear fit by including or omitting the result at the coarsest lattice spacing.
 We determine the final central value $\bar{f}$ through a weighted average of the form
\begin{align}
\label{eq:weight_ave_1}
\bar{f} = w_{A}~f_{A}~+~w_{B}~f_{B},\qquad w_{A}+w_{B}=1~.
\end{align}
Our estimate of the systematic error, which is added (linearly to be conservative) to the statistical uncertainty, is then obtained using
\begin{align}
\label{eq:weight_ave_2}
\sigma_{\mathrm{syst}}^{2} =  \sum_{i=A,B} w_{i}~( f_{i} - \bar{f})^{2}~.      
\end{align}
The weights $w_{i}$, with  $i=\{A,B\}$, are chosen according to the Akaike Information Criterion\,\cite{Akaike} (AIC), namely 
\begin{align}
 \label{eq:AIC}
    w_{i} \propto e^{- \left(\chi_{i}^2 + 2 N^{(i)}_{\rm{pars}} -  N^{(i)}_{\rm{data}}\right) / 2} ~ , ~
\end{align}
where $\chi_{i}^{2}$ is the total $\chi^{2}$ obtained in the $i$-th fit, and $N_{\rm{pars}}^{(i)}$ and $N_{\rm{meas}}^{(i)}$ are the corresponding number of fit parameters and measurements. 

\begin{figure}
    \centering
    \includegraphics[scale=0.6]{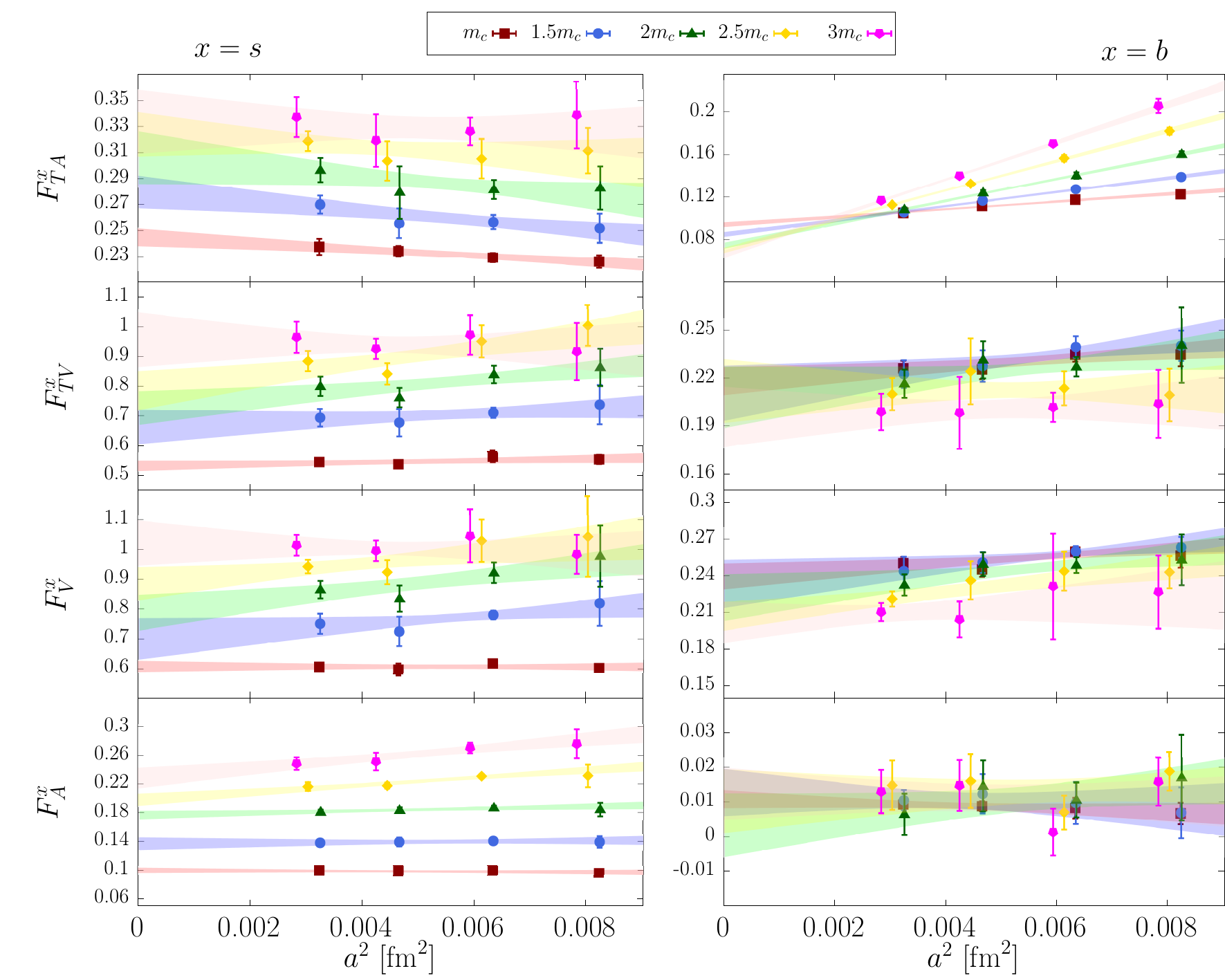}
    \caption{\small\it Continuum limit extrapolation of the lattice data for $F_{W}^{s}$ (left) and $F_{W}^{b}$ (right) for $x_{\gamma}=0.1$.  The transparent bands correspond to the best-fit function obtained in the linear $a^{2}$ fit employing the full dataset. In the panels, the different colors correspond to different values of the heavy quark mass $m_{h}$. }
    \label{fig:cont_extr_ixg_1}
\end{figure}

\begin{figure}
  \centering
    \includegraphics[scale=0.6]{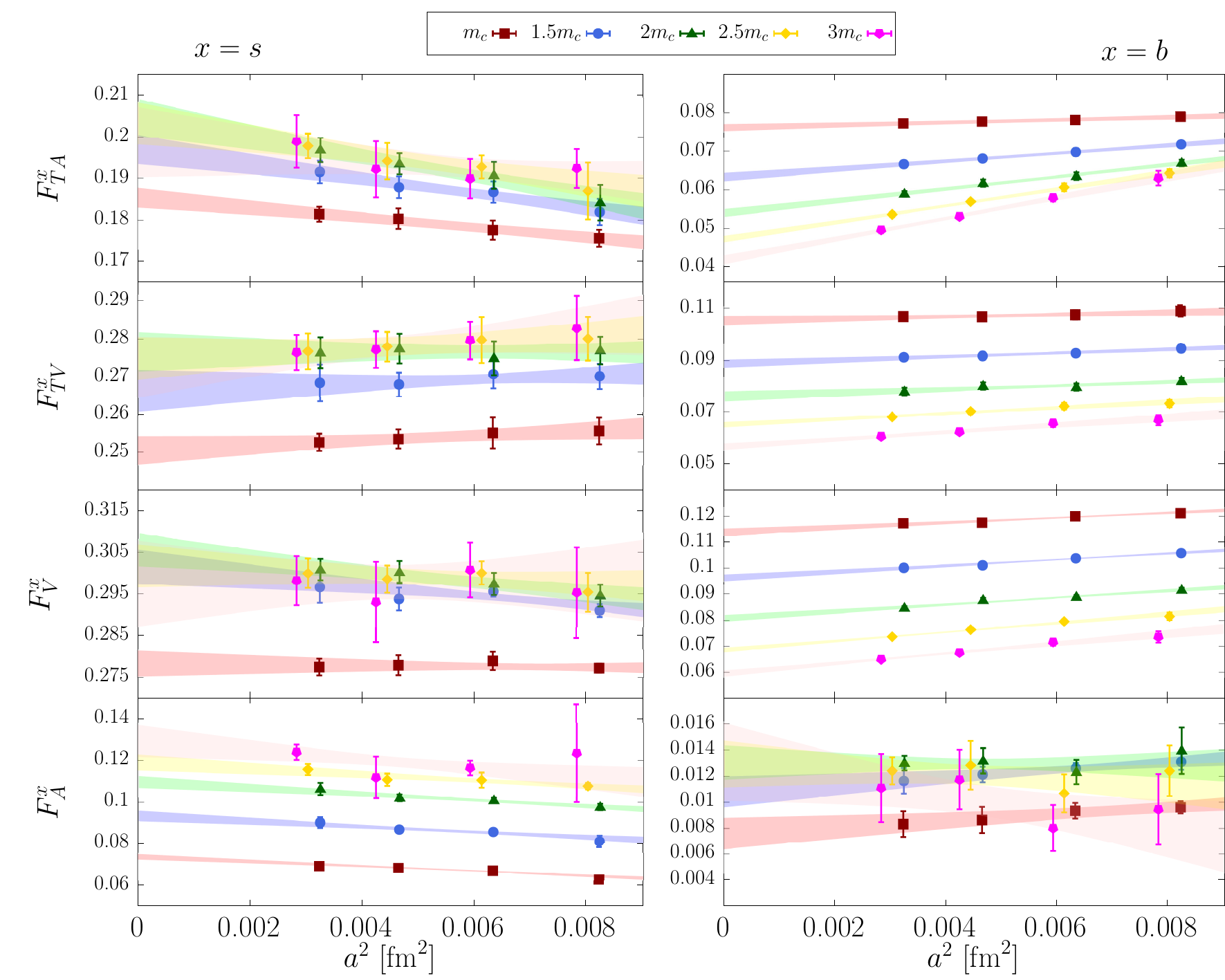}
    \caption{\small\it Same as in Figure~\ref{fig:cont_extr_ixg_1} for $x_{\gamma}=0.4$.}
    \label{fig:cont_extr_ixg_4}
\end{figure}

\begin{figure}
    \centering
    \includegraphics[scale=0.35]{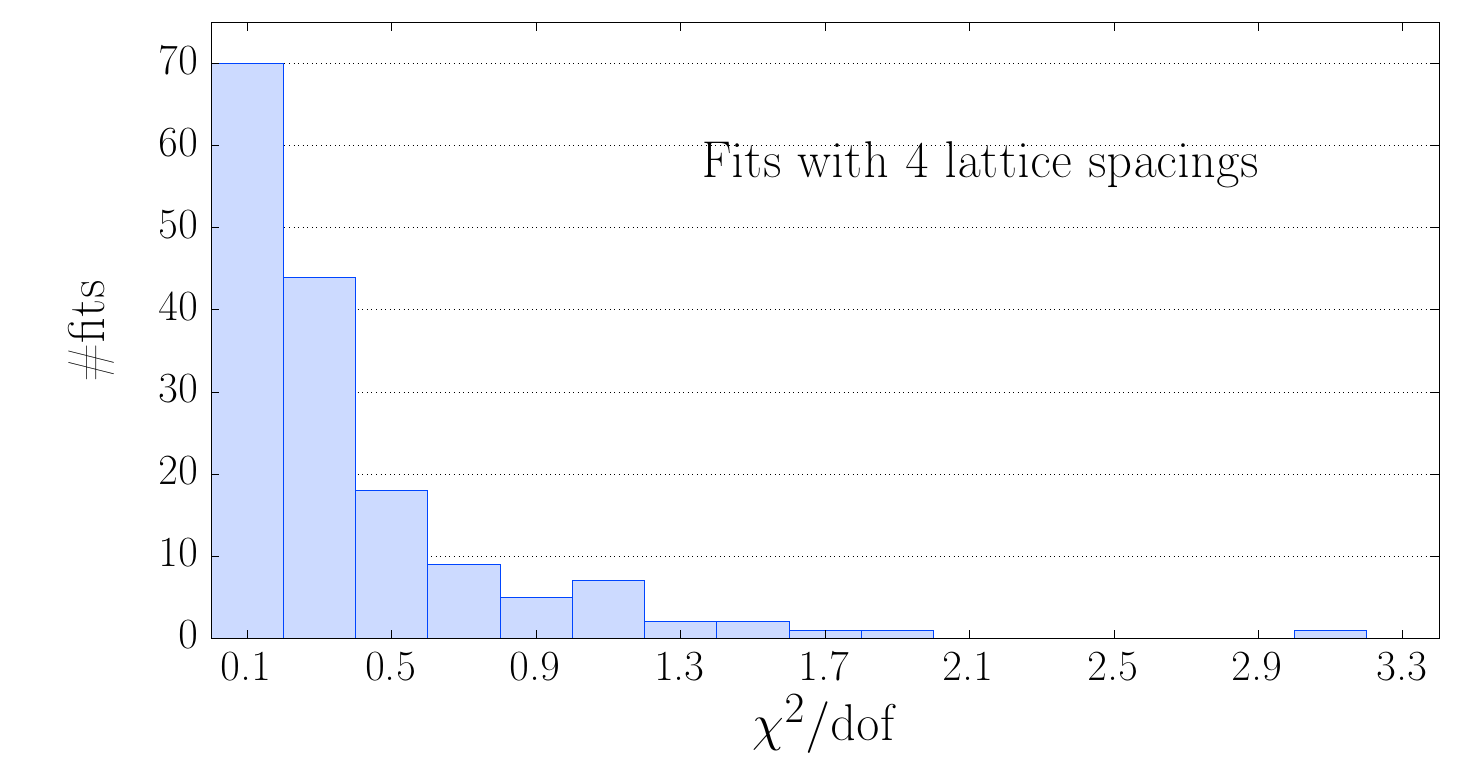}
    \includegraphics[scale=0.35]{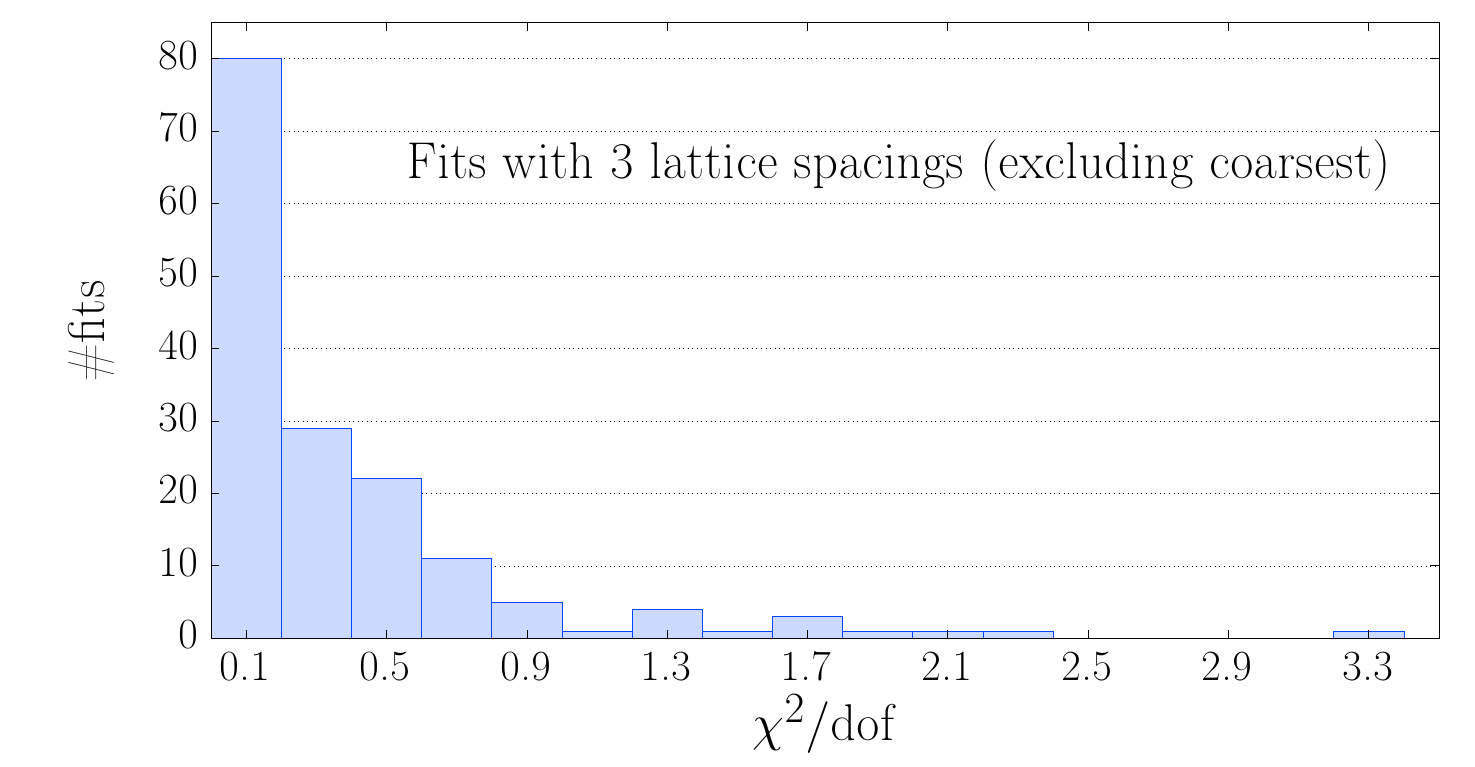}
    \caption{\small\it Histograms of the $\chi^{2}/{\rm dof}$ distribution corresponding to the 160 continuum limit extrapolations we have performed using the full dataset (left) and excluding the coarsest lattice spacing ensemble, the A48 (right).  \label{fig:histo_ch2}}
   
\end{figure}

In Figures\,\ref{fig:cont_extr_ixg_1} and~\ref{fig:cont_extr_ixg_4} we show the results of our continuum fits, for the smallest ($x_{\gamma}=0.1$) and largest ($x_{\gamma}=0.4$) simulated values of $x_{\gamma}$.
The fits shown in the figures are those for which the full dataset has been used. Clearly for large quark masses $m_{h}$, as a consequence of the Parisi-Lepage theorem\,\cite{Parisi:1983ae,Lepage:1989hd}, the statistical noise of the data rapidly increases. The quality of the fits is very good, and in Figure~\ref{fig:histo_ch2} we show the histogram of the reduced $\chi^{2}$ distribution corresponding to the $160$ continuum extrapolations we have performed. 

\subsection{Extrapolating the results for the local form factors $F_{V},F_{A}, F_{TV}, F_{TA}$ to the physical $B_{s}$ meson}
\label{sec:FF_extr}
In this section we discuss the asymptotic formulae used to extrapolate the form factors, computed for $m_{H_{s}} \in [m_{D_{s}}, 2m_{D_{s}}]$, to the physical point $m_{H_{s}}= m_{B_{s}} \simeq 5.367\,{\rm GeV}$. For heavy quark masses $m_{h}$ and energetic photons, there are elegant and simple relations relating the four form factors. 
In Ref.\,\cite{Beneke:2011nf} (see also~\cite{Korchemsky:1999qb, Descotes-Genon:2002crx, Lunghi:2002ju, Bosch:2003fc}), the authors studied in detail the behaviour of the axial and vector form factors contributing to the radiative $B\to \gamma \ell \nu$ decay amplitude in the framework of the HQET and large-photon-energy expansions. 
The relations derived in Ref.\,\cite{Beneke:2011nf} imply that up to (and including) order $\mathcal{O}(1/m_{H_{s}}, 1/E_{\gamma})$ terms in the heavy-quark and large-photon-energy expansion, the axial and vector form factors $F_{A}$ and $F_{V}$ are given by
\begin{align}
\label{eq:FV_s_ET}
\frac{F_{V}(x_{\gamma}, m_{H_{s}})}{f_{H_{s}}} &= \frac{|q_{s}|}{x_{\gamma}}\left( \frac{R(E_{\gamma},\mu)}{\lambda_{B}(\mu)} + \xi(x_{\gamma},m_{H_{s}}) + \frac{1}{m_{H_{s}}x_{\gamma}} +\frac{|q_{b}|}{|q_{s}|}\frac{1}{m_{h}}    \right)~, \\[8pt]
\label{eq:FA_s_ET}
\frac{F_{A}(x_{\gamma}, m_{H_{s}})}{f_{H_{s}}} &= \frac{|q_{s}|}{x_{\gamma}}\left( \frac{R(E_{\gamma},\mu)}{\lambda_{B}(\mu)} + \xi(x_{\gamma},m_{H_{s}}) - \frac{1}{m_{H_{s}}x_{\gamma}} - \frac{|q_{b}|}{|q_{s}|}\frac{1}{m_{h}}    \right)~,
\end{align}
where $f_{\bar{H}_{s}}$ is the decay constant of the $\bar{H}_{s}= \bar{s}h$ pseudoscalar meson of mass $m_{\bar{H}_{s}}$, $\lambda_{B}(\mu)$ is the first inverse moment of the $B_{s}$-meson light-cone distribution amplitude (LCDA), and $R(E_{\gamma},\mu) = 1 + \mathcal{O}(\alpha_{s})$ is a radiative correction factor that is the same for $F_{V}$ and $F_{A}$. Finally, $\xi(x_{\gamma}, m_{H_{s}})$ is a power-suppressed term, common to both form factors, that can be written as~\cite{Beneke:2018wjp}
\begin{align}
\xi(x_{\gamma}, m_{H_{s}}) = \frac{A}{m_{H_s}} + \frac{B}{m_{H_s}x_{\gamma}}~.
\end{align}
In Eqs.\,(\ref{eq:FV_s_ET})-(\ref{eq:FA_s_ET}), perturbative radiative corrections to the subleading terms and $\mathcal{O}(m_{s}/(m_{H_{s}}^{2}))$ terms have been neglected. 
The leading contribution to the form factors comes from the emission of the photon from the strange quark. Radiation from the heavy quark is suppressed by a factor proportional to $1/m_{H_{s}}$ and the corresponding subleading terms are proportional to $|q_{b}|$ in 
Eqs.\,(\ref{eq:FV_s_ET})-(\ref{eq:FA_s_ET}).

The large mass/photon-energy behaviour of the tensor form factors $F_{TA}$ and $F_{TV}$ including order $\mathcal{O}(1/m_{H_{s}}, 1/E_{\gamma})$ corrections has been investigated in Ref.\,\cite{Beneke:2020fot} and is given by
\begin{align}
\label{eq:FTV_s_ET}
\frac{F_{TV}(x_{\gamma}, m_{H_{s}},\mu)}{f_{H_s}} &= \frac{|q_{s}|}{x_{\gamma}}\left(\frac{R_{T}(E_{\gamma},\mu)}{\lambda_{B}(\mu)} + \xi(x_{\gamma},m_{H_{s}})  
+\frac{1-x_{\gamma}}{m_{H_{s}}x_{\gamma}} + \frac{|q_{b}|}{|q_{s}|}\frac{1}{m_{H_{s}}} \right)~,\\[8pt]
\label{eq:FTA_s_ET}
\frac{F_{TA}(x_{\gamma}, m_{H_{s}},\mu)}{f_{H_s}} &= \frac{|q_{s}|}{x_{\gamma}}\left(\frac{R_{T}(E_{\gamma},\mu)}{\lambda_{B}(\mu)} + \xi(x_{\gamma},m_{H_{s}})  
-\frac{1-x_{\gamma}}{m_{H_{s}}x_{\gamma}} + \frac{|q_{b}|}{|q_{s}|}\frac{1}{m_{H_{s}}} \right)~, 
\end{align}
where $R_{T}(E_{\gamma},\mu) = 1+\mathcal{O}(\alpha_{s})$ is the radiative correction. Again, perturbative radiative corrections to the subleading terms and $\mathcal{O}(m_{s}/(m_{H_{s}}^{2}))$ terms have been neglected. In Eqs.~(\ref{eq:FTV_s_ET}),~(\ref{eq:FTA_s_ET}) we have explicitly inserted in the l.h.s. the dependence on the renormalization scale $\mu$, which is instead absent in $F_{V}$ and $F_{A}$ which are scale-independent quantities. The previous relations imply that, neglecting power suppressed contributions and radiative corrections, one has $F_{TV}(x_{\gamma})=F_{TA}(x_{\gamma})=F_{V}(x_{\gamma})=F_{A}(x_{\gamma})$. 

We now explain that the above asymptotic relations for the form factors, being valid in the limit of large $E_{\gamma}$, are not sufficient to describe their behaviour in the range of the simulated values of $m_{H_{s}}$ and $x_{\gamma}$ because of the presence of sizeable non-asymptotic contributions from resonances. To highlight this point, we start from the canonical decomposition of the form factors in terms of intermediate-state contributions. The hadronic tensor $H_{W}^{\mu\nu}(p,k)$ in Eq.~(\ref{hadronic_tensor_Ds})
\begin{align}
H^{\mu\nu}_{W}(p,k, m_{H_{s}}) \equiv i\int d^{4}y ~ e^{iky} ~ {\rm \hat{T}} \langle 0 | J_{W}^{\nu}(0) J_{{\rm em}}^{\mu}(y) | \bar{H}_{s}(\bs{0})\rangle~,
\end{align}
can be decomposed as
\begin{align}
H^{\mu\nu}_{W}(p,k, m_{H_{s}}) &= i\int_{-\infty}^{0} dt\, e^{iE_{\gamma}t}\langle 0 | J^{\nu}_{W}(0) J^{\mu}_{\rm em}(t, \bs{k}) | \bar{H}_{s}(\bs{0})\rangle +   i\int_{0}^{\infty} dt\,e^{iE_{\gamma}t} \, \langle 0 | J^{\mu}_{\rm em}(t, \bs{k}) J^{\nu}_{W}(0)  | \bar{H}_{s}(\bs{0})\rangle \nonumber \\[8pt]
&\equiv H^{\mu\nu}_{W,1}(p,k, m_{H_{s}}) +  H^{\mu\nu}_{W,2}(p,k, m_{H_{s}})~,
\end{align}
where
\begin{align}\label{eq:Jmutk}
J_{\rm em}^{\mu}(t, \bs{k}) \equiv \int d^{3}x\, e^{-i\bs{k}\bs{x}}\, J_{\rm em}^{\mu}(t,\bs{x}) ~.
\end{align}
We now focus on the contribution from the first time-ordering, $H^{\mu\nu}_{W,1}$, which can be written as
\begin{align}
\label{eq:canonical_expansion}
H^{\mu\nu}_{W,1}(p,k, m_{H_{s}}) &= i\int_{-\infty}^{0} dt \, e^{iE_{\gamma}t} \langle 0 | J^{\nu}_{W}(0) e^{i(\hat{H} -m_{H_{s}} -i\varepsilon)t} J^{\mu}_{\rm em}(0,\bs{k}) | \bar{H}_{s}(\bs{0})\rangle \nonumber \\[8pt]
&= \langle 0| J_{W}^{\nu}(0)\frac{1}{\hat{H} + E_{\gamma} - m_{H_{s}} -i\varepsilon}J_{\rm em}^{\mu}(0,\bs{k})| \bar{H}_{s}(\bs{0})\rangle = \sum_{n} \frac{\langle 0| J^{\nu}_{W}(0) | n \rangle \langle n| J^{\mu}_{\rm em}(0,\bs{k}) | \bar{H}_{s}(\bs{0})\rangle}{2E_{n}(-\bs{k})( E_{n}(-\bs{k}) + E_{\gamma} -m_{H_{s}} ) }  ~,
\end{align}
where $\hat{H}$ is the QCD Hamiltonian.
The contributing intermediate states $|n\rangle $ are $B=-1$, $S=1$ states with $J^{P}= 1^{-}$ for $W=\{V,TV\}$ and $J^{P}=1^{+}$ for $W=\{A, TA\}$. Their energies are given by $E_{n}(-\bs{k})= \sqrt{ m_{n}^{2} + E_{\gamma}^{2}}$. In the following, in order to model the $x_{\gamma}$ and mass behaviour of the form factors, we only consider the contributions coming from the resonances that we treat as stable particles. Using the following relations ($\eta_{n}$ is the polarization of the vector meson $|n\rangle$, $k_{n} = ( E_{n}, -\bs{k})$, $p_{\gamma} = k_{n}-p$):
\begin{align}
\langle 0 |J_{V}^{\nu} | n \rangle &= \eta_{n}^{\nu}~ m_{n}~f^{V}_{n} ~, \qquad \langle 0 |J_{TV}^{\nu} | n \rangle =   f_{n}^{TV}\left[ \frac{\eta_{n}^{\nu}}{m_{H_{s}}}(k_{n}\cdot q ) -\frac{k^{\nu}_{n}}{m_{H_{s}}}(\eta_{n}\cdot q )  \right]   \nonumber \\[8pt]
\langle 0 |J_{A}^{\nu} | n \rangle &= i\eta_{n}^{\nu}~ m_{n}~f^{A}_{n} ~, \quad\,\,\,\, \langle 0 |J_{TA}^{\nu} | n \rangle = -i  f_{n}^{TA}\left[ \frac{\eta_{n}^{\nu}}{m_{H_{s}}}(k_{n}\cdot q ) -\frac{k^{\nu}_{n}}{m_{H_{s}}}(\eta_{n}\cdot q )  \right]   \nonumber \\[8pt]
\langle n | J_{\rm em}^{\mu} |\bar{H}_{s}(\bs{0}) \rangle &=  g_{n} \,\varepsilon^{\mu\nu \gamma \beta}(\eta_{n}^{\ast})_{\nu}~ k_{n,\gamma}~p_{\beta}~\qquad\qquad\qquad\qquad\qquad\,\,\,\,\, W=\{ V, TV \}~, \nonumber \\[8pt]
\langle n | J_{\rm em}^{\mu} |\bar{H}_{s}(\bs{0}) \rangle &= g'_{n}\,\big[ (\eta_{n}^{\ast})^{\mu}(p\cdot p_{\gamma}) -(\eta_{n}^{\ast}\cdot p_{\gamma})p^{\mu}\big] + \mathcal{O}(p_{\gamma}^{2}) ~,\qquad\quad W=\{ A, TA \}~, 
\end{align}
and given that at leading-order in $m_{H_{s}}$ one has 
 $f^{W}_{n} \propto f_{H_{s}}$, and assuming that at leading-order  the form factors $g_{n}$ and $g'_{n}$ are constant, one obtains that each of the intermediate states in Eq.~(\ref{eq:canonical_expansion}) gives a contribution $F_{W,n}(x_{\gamma})$ to the form factor $F_{W}(x_{\gamma})$ which scales as ($r_{n} \equiv m_{n}/m_{H_{s}}$)
\begin{align}
\label{eq:scaling_intermediate}
F_{W, n}(x_{\gamma}) \propto \frac{ f_{H_{s}}}{\sqrt{ r_{n}^{2}+ \frac{x_{\gamma}^{2}}{4}} + \frac{x_{\gamma}}{2} -1}~.
\end{align}
In the static limit, since $r_{n}$ approaches one, the scaling relations in Eqs.\,(\ref{eq:FV_s_ET})\,-\,(\ref{eq:FTA_s_ET}) are recovered. 
However, it is important to notice that for $x_{\gamma}=0$ the denominator in Eq.~(\ref{eq:scaling_intermediate}) develops a pole for $r_{n}\to 1$, signalling the fact that the scaling laws are different at $x_{\gamma}=0$ and $x_{\gamma} \ne 0$ (in this last case the denominator approaches a non-zero value $x_{\gamma}/2 + \mathcal{O}(x_{\gamma}^{2})$ in the $m_{h}\to\infty$ limit). 
For small enough values of $x_{\gamma}$, the presence of a quasi-pole may generate large corrections to the  scaling relations in Eqs.\,(\ref{eq:FV_s_ET})\,-\,(\ref{eq:FTA_s_ET}), which we now discuss. We start by recalling the following HQET relations for the  masses $m_{\bar{H}^{*}_{s}}$ and $m_{H_{s1}}$ of the lowest-lying vector mesons $\bar{H}^{*}_{s}$ and $\bar{H}_{s1}$ in the $J^{P} = 1^{-}$ and $J^{P}= 1^{+}$ channel~\cite{Neubert:1996wg}
\begin{align}
\label{eq:HQET_1minus}
m^{2}_{\bar{H}^{*}_{s}} - m^{2}_{\bar{H}_{s}} &= 2\lambda_{2}+ \mathcal{O}\left(\frac{1}{m_{h}}\right) ~, \qquad \lambda_{2} \simeq 0.24~{\rm GeV^{2}}~, \\[8pt]
\label{eq:HQET_1plus}
m_{\bar{H}_{s1}} - m_{\bar{H}_{s}} &= \Lambda_{1} + \mathcal{O}\left(\frac{1}{m_{h}}\right)~, \qquad\,\,\, \Lambda_{1} \simeq 0.5~{\rm GeV}~.
\end{align}
As is well known, the first relation comes from the fact that the ground-state pseudoscalar ($J^{P}=0^{-}$) and vector ($J^{P}=1^{-}$) mesons, are members of the same HQET spin-doublet, and so they become degenerate in the infinite heavy-quark mass limit. The mass-splitting between the ground-state pseudoscalar and axial-vector meson is instead of order $\mathcal{O}(\Lambda_{\rm QCD})$. This implies that for the lowest-lying intermediate state ($n=1$) contributing to $F_{W}$, for $W=\{ V, TV\}$, one has  
\begin{align}
\label{eq:r1_V}
r_{1} = \frac{ m_{H^{*}_{s}}}{m_{H_{s}}} \simeq 1+\frac{\lambda_{2}}{m_{H_{s}}^{2}} \qquad \implies \qquad   \sqrt{ r_{1}^{2} + \frac{x_{\gamma}^{2}}{4}} + \frac{x_{\gamma}}{2} -1 \simeq  \frac{\lambda_{2}}{m_{H_s}^{2}} +\frac{x_{\gamma}
}{2}  + \ldots~,
\end{align}
where the ellipses indicate subleading corrections at large $m_{h}$ and small $x_{\gamma}$. For $W=\{A,TA\}$, one has instead
\begin{align}
\label{eq:r1_A}
r_{1} = \frac{ m_{H_{s1}}}{m_{H_{s}}} \simeq 1+ \frac{\Lambda_{1}}{m_{H_{s}}} \qquad \implies \qquad   \sqrt{ r_{1}^{2} + \frac{x_{\gamma}^{2}}{4}} + \frac{x_{\gamma}}{2} -1 \simeq  \frac{\Lambda_{1}}{m_{H_s}} +\frac{x_{\gamma}
}{2}  + \ldots~.
\end{align}
The previous equations show that for small values of $x_{\gamma}$  the quasi-pole produces an \textit{enhancement} of the form factor of order $\mathcal{O}(m_{h}^{2})$ and $\mathcal{O}(m_{h})$, respectively in the vector-like and axial-like channels. In the following section we will combine the leading-order relations Eqs.~(\ref{eq:FV_s_ET})-~(\ref{eq:FTA_s_ET}), with the quasi-pole behaviour described by Eqs.~(\ref{eq:r1_V}),~(\ref{eq:r1_A}) in order to extrapolate the form factors to the physical $B_{s}$ meson mass.
 
\subsection{Numerical results at the physical $B_{s}$ mass}
\label{sec:num_results_I}
Guided by the analysis in the previous section, we introduce some model-dependent interpolating formulae  for the form factors  which   describe  their $q^2$ dependence in 
the resonance region and have the correct asymptotic behaviour in the limit of large $m_h$.
We have extrapolated our results for $F_{W}$, $W=\{V,A, TV, TA\}$, obtained at the five different simulated values of the heavy quark mass $m_{h}$ in Eq.~(\ref{eq:mass_list}), employing the following fit Ansatz ($z=1/m_{H_{s}}$)
\begin{align}
\label{eq:ansatz_V}
 \frac{F_{V}(x_{\gamma}, z )}{f_{H_{s}}} &= \frac{|q_{s}|}{x_{\gamma}} ~ \frac{ 1 }{ 1 + C_{V}\frac{2z^{2}}{x_{\gamma}} }~\left( K +  (1+\delta_{z})\frac{z}{x_{\gamma}} + \frac{1}{z^{-1} - \Lambda_{H}} + A_{m}z + A_{x_{\gamma}}\frac{z}{x_{\gamma}}  + B^{V}_{m}z^{2} + B^{V}_{x_{\gamma}}\frac{z^{2}}{x_{\gamma}}\right)  ~, \\[8pt]
 \label{eq:ansatz_A}
 \frac{F_{A}(x_{\gamma},z)}{f_{H_{s}}} &= \frac{|q_{s}|}{x_{\gamma}} ~\frac{ 1 }{ 1+ C_{A}\frac{2z}{x_{\gamma}} } ~ \left( K -  (1+\delta_{z})\frac{z}{x_{\gamma}} - \frac{1}{z^{-1} - \Lambda_{H}} + A_{m}z + (A_{x_{\gamma}}+2KC_{A})\frac{z}{x_{\gamma}}  + B^{A}_{m}z^{2} + B^{A}_{x_{\gamma}}\frac{z^{2}}{x_{\gamma}}\right) ~,  \\[8pt]  
 \label{eq:ansatz_TV}
 \frac{F_{TV}(x_{\gamma},z)}{f_{H_{s}}} &= \frac{|q_{s}|}{x_{\gamma}}\frac{1+2C_{V}z^{2} }{ 1 + C_{V}\frac{2z^{2}}{x_{\gamma}} }\left( K_{T} + (A^{T}_{m}+1)z + A_{x_{\gamma}}^{T}\frac{z}{x_{\gamma}} + (1 + \delta'_{z})z\frac{1-x_{\gamma}}{x_{\gamma}} + B^{T}_{m}z^{2} + B^{TV}_{x_{\gamma}}(1-x_{\gamma})\frac{z^{2}}{x_{\gamma}}    \right) ~, \\[8pt]
  \label{eq:ansatz_TA}
 \frac{F_{TA}(x_{\gamma},z)}{f_{H_{s}}} &= \frac{|q_{s}|}{x_{\gamma}} \frac{1+ 2C_{A}^{T}z }{ 1 + C_{A}^{T}\frac{2z}{x_{\gamma}}}\left( K_{T} + (A^{T}_{m}+1)z + A_{x_{\gamma}}^{T}\frac{z}{x_{\gamma}} -(1 + \delta'_{z}-2K_{T}C_{A}^{T})z\frac{1-x_{\gamma}}{x_{\gamma}}+ B^{T}_{m}z^{2} + B^{TA}_{x_{\gamma}}(1-x_{\gamma})\frac{z^{2}}{x_{\gamma}}    \right) ~,
\end{align}
where $C_{A}^{T} = C_{A}+\delta C_{A}^{T}$, $K_{T}=K+\delta K_{T}$, and $K, \delta K_{T}, \delta_{z}, \Lambda_{H}, C_{V}, C_{A}, \delta C_{A}^{T}, A_{m/x_{\gamma}}, 
 A_{m}^{T}, A_{x_{\gamma}}^{T}$, $\delta^{'}_{z}$ and $B_{m/x_{\gamma}}^{W}$ ($W=\{ V, A, TV, TA\}$) are free fit parameters. Our strategy to extrapolate the form factors to the physical $B_{s}$-meson consists in a simultaneous global fit of the mass and $x_{\gamma}$ dependence of all four form factors. 
 The phenomenological fit Ansatz described by Eqs.\,(\ref{eq:ansatz_V})-(\ref{eq:ansatz_TA}) takes into account the constraints discussed in the previous section, and contains the quasi-pole corrections to the asymptotic scaling described by Eqs.\,(\ref{eq:FV_s_ET}),~(\ref{eq:FA_s_ET}),~(\ref{eq:FTV_s_ET}) and (\ref{eq:FTA_s_ET}). We however relaxed the constraint 
 $A_{m}^{T}=A_{m}$ and $A_{x_{\gamma}}=A_{x_{\gamma}}^{T}$ due to the presence of the same function $\xi(x_{\gamma}, m_{H_{s}})$ in the expression for the tensor-like and vector-like form factors in Eqs\,(\ref{eq:FV_s_ET})-(\ref{eq:FTA_s_ET}). The position of the pole is taken to be the same in the vector and tensor-vector channel, while we allow for the possibility of having a different pole in $F_{A}$ and $F_{TA}$ (i.e. $\delta C_{A}^{T} \ne 0$). This is due to the fact that while in the vector channel vector-meson-dominance is expected to work well since the vector ($H_{s}^{*}$) and pseudoscalar ($H_{s}$) ground-state mesons become degenerate in the static limit, this is not the case in the axial channel where many resonances with masses of order $m_{H_{s}}+ \mathcal{O}(\Lambda_{\rm{QCD}})$ are present. 
 The axial pole should be considered to be an \textit{effective pole}, and its position can therefore be slightly different in the axial and tensor-axial channel due to the different couplings to the excited states. Moreover, in order to account for the fact that the tree-level equality between tensor-like and vector-like form factors is spoiled by the radiative corrections, we also allow for the possibility that $\delta K_{T} \ne 0$. Notice that in the tensor form factors, the numerator in the pole term is inserted to ensure the validity of the kinematical constraint $F_{TA}(1)= F_{TV}(1)$. Finally, we have included a parameter $\Lambda_{H}$ to account for the fact that the hadron mass $m_{H_{s}}$ differs from the heavy quark mass $m_{h}$ by an amount of order $\mathcal{O}(\Lambda_{\rm{QCD}})$ (see the last terms in Eqs.\,(\ref{eq:FV_s_ET}) and (\ref{eq:FA_s_ET})), and two parameter $\delta_{z}, \delta'_{z}$ to account for violations of the relations in Eqs.\,(\ref{eq:FV_s_ET})-(\ref{eq:FTA_s_ET}) which are only exactly valid in the limit of a massless strange quark and neglecting radiative corrections to the power suppressed terms. Our determination of the decay constant $f_{H_{s}}$, on the same configurations used for the computation of the form factors,  is discussed in detail in Appendix\,\ref{sec:fBs}. In the same appendix we also discuss our determination of $f_{B_{s}}$, for which we get the value
 \begin{align}
f_{B_{s}} = 224.5(5.0)\,{\rm MeV}\,.
 \end{align}
 Our determination of $f_{B_{s}}$ agrees with the $N_{f}=2+1+1$ FLAG average $f_{B_{s}}^{\rm FLAG} = 230.3(1.3)\,{\rm MeV}$ at the level of $1.1\sigma$, although our uncertainty is larger. 
 Using the Ans\"atze in Eqs.\,(\ref{eq:ansatz_V})-(\ref{eq:ansatz_TA}) we have performed a total of  $N \simeq \mathcal{O}(500)$ fits which differ on whether the fit parameters $\delta K_{T}$, $\delta C_{A}^{T}$, $\delta_{z}$, $\delta^{'}_{z}$ are set to zero or not, and on whether we include or not the fit parameters describing the $\mathcal{O}(1/m_{H_{s}}^{2})$ corrections. The total number of measurements is $80$ and the maximum number of fit parameters used is $14$. To stabilize the fits, large Gaussian priors are imposed on the fit parameters $\delta K_{T}, \delta C_{A}^{T}, \delta_{z}, \delta^{'}_{z}, \Lambda_{H}$ and $B^{W}_{m/x_{\gamma}}$. These are
 \begin{align}
\delta K_{T}= 0\pm 0.4~{\rm GeV^{-1}}~,\quad \delta^{(')}_{z}= 0 \pm 1~,\quad \delta C_{A}^{T} = 0 \pm 0.4~{\rm GeV}~,\quad \Lambda_{H} = 0.75 \pm 0.5~{\rm GeV}~, \quad B^{W}_{m/x_{\gamma}} = 0 \pm 2.5~{\rm GeV}~.
 \end{align}
 We minimize a correlated $\chi^{2}$ function which takes fully into account the correlations between the values of a given form factor at the different simulated values of $x_{\gamma}$ and $m_{H_{s}}$. However, in order to avoid having an ill-conditioned covariance matrix, we assume, in the construction of the $\chi^{2}$, that the different form factors are instead uncorrelated. The error on the fit parameters are always properly estimated, since they are obtained from the dispersion of the results obtained repeating the fits for each jackknife sample. 

 Many of the fit parameters entering Eqs.~(\ref{eq:ansatz_V})-(\ref{eq:ansatz_TV}) are not needed in order to obtain a good $\chi^2/\rm{dof}$, and a good description of the data is already obtained by setting $\delta K_{T}=\delta C_{A}^{T}=\delta_{z}=\delta_{z}^{'}=0$ and neglecting the $\mathcal{O}(1/m_{H_{s}}^{2})$ corrections. However, in order to estimate correctly the systematic errors due to the mass extrapolation, it is important to span over a sufficiently large number of fit Ans\"atze. 
 
 We combine the results of the $N$ different fits using two different criteria. The first one is based on the AIC discussed in Section\,\ref{subsec:formfactorsnumerical}, i.e. we assign to each of the $N$ fits a weight $w_{i}$ given by
 \begin{align}
 \label{eq:weight_AIC}
w_{i} \propto \exp( -\left( \chi^{2}_{(i)} + 2N^{(i)}_{\rm pars} - N_{\rm data}^{(i)}\right)/2 )~,\qquad \sum_{i=1}^{N} w_{i} = 1~,
\end{align}
 where $\chi^{2}_{(i)}$ is the total $\chi^{2}$ of the $i$-th fit, and $N_{\rm pars}^{(i)}$ and $N_{\rm meas}^{(i)}$ are the corresponding number of fit parameters and measurements. The second criterion consists in selecting only those fits leading to a good 
 $\chi^{2}/\mathrm{dof}$ and assigning them an uniform weight, i.e. using 
 \begin{align}
 \label{eq:weight_ch2_cut}
w_{i} \propto \theta\left( c - \chi^{2}_{(i)}/N_{\rm dof}^{(i)} \right)~, \qquad \sum_{i=1}^{N} w_{i} = 1~,
 \end{align}
 and we set $c=1.4$ which corresponds approximately to $1+2\sqrt{2/N^{(i)}_{\rm dof}}$, where $\sqrt{2/N^{(i)}_{\rm dof}}$ is the standard deviation of the reduced $\chi^{2}$ distribution with $N_{\rm dof}^{(i)}$ degrees of freedom. Then, with a given choice for the weights $w_{i}$, the final central value $\bar{x}$ is obtained from a weighted average:
 \begin{align}
  \label{eq:BMA_average}
  \bar{x} = \sum_{i=1}^{N} w_{i} x_{i}~, 
 \end{align}
 where $x_{i}$ is the result obtained from the $i$-th fit. The sum in Eq.\,(\ref{eq:BMA_average}) is evaluated in a correlated way, so that the statistical errors of the $x_{i}$ are correctly propagated to $\bar{x}$. The systematic error, which is added in quadrature to the statistical error of $\bar{x}$, is then given by
 \begin{align}
 \label{eq:BMA_syst}
 \sigma_{x, {\rm syst}}^{2} = \sum_{i=1}^{N} w_{i}~( x_{i}- \bar{x})^{2}~.
 \end{align}
We have found that the results obtained using the weights in Eq.\,(\ref{eq:weight_AIC}) and Eq.~(\ref{eq:weight_ch2_cut}) are consistent well within the uncertainties. 
However, at small $x_{\gamma}$ the errors obtained using the AIC are typically smaller than those obtained using Eq.~(\ref{eq:weight_ch2_cut}). 
In order to be conservative, we take the results obtained using the weights in Eq.\,(\ref{eq:weight_ch2_cut}) to obtain our final results for the form factors. 

\begin{figure}
\includegraphics[scale=0.46]{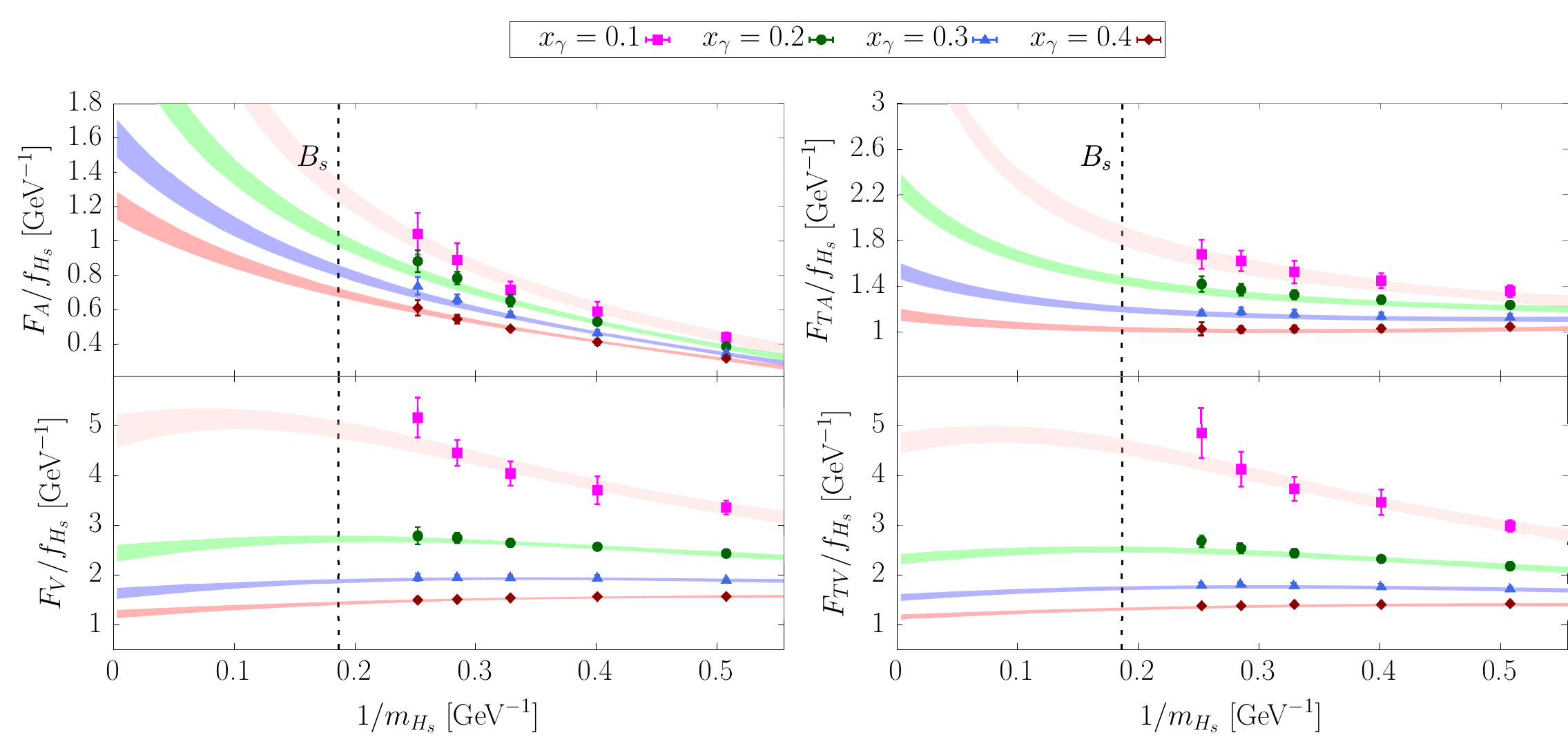}
\caption{\small\it Extrapolation to the physical $B_{s}$ meson of the four form factors $F_{A}$ (top left) , $F_{TA}$ (top right), $F_{V}$ (bottom left) and $F_{TV}$ (bottom right). The form factors are divided by the decay constant $f_{H_{s}}$ of the heavy-strange pseudoscalar meson. The different colors correspond to the different simulated values of the Lorentz invariant $x_{\gamma}$. Finally, the continuum bands correspond to the best-fit function obtained after applying the model-averaged procedure described by Eqs.\,(\ref{eq:BMA_average})-(\ref{eq:BMA_syst}) using the weights in Eq.\,(\ref{eq:weight_ch2_cut}). \label{fig:extr_ff}}
\end{figure}

The results of the extrapolation are collected in the plots of Figure\,\ref{fig:extr_ff}. 
The continuum bands in the figure correspond to the best-fit function obtained after applying the above procedure with the weights in Eq.\,(\ref{eq:weight_ch2_cut}).  We obtain for the pole coefficients $C_{V}$, $C_{A}$ and $C_{A}+\delta C_{A}^{T}$ the values
\begin{align}
C_{V} = ( 0.57(3)~{\rm GeV} )^{2} ~,\qquad C_{A} = 0.70(7)~{\rm GeV} ~, \qquad C_{A}+\delta C_{A}^{T} = 0.77(4)~{\rm GeV}~. 
\end{align}
The result for $C_{V}$ can be compared with the value expected from the HQET relation in Eq.~(\ref{eq:HQET_1minus}), namely $C_{V} \simeq \lambda_{2} \simeq (0.5~{\rm GeV})^{2}$. Although slightly larger (recall that the Ansatz we use is a phenomenological description of the full form factors where excited-states contributions are always present), our determination is in line with expectations, and provides nice evidence that the reason behind the steep rise of the vector form factors at small $x_{\gamma}$ is due to the presence of the quasi-pole. Concerning the position of the axial pole, the value we obtained for $C_{A}$ and $\delta C_{A}^{T}$ is also qualitatively in line with the expectations $C_{A}, C_{A}+\delta C_{A}^{T} \simeq \mathcal{O}(\Lambda_{1})$. We did not find clear evidences of non-zero values of $\delta_{z}$, $\delta^{'}_{z}$ and $\delta K_{T}$. We obtain
\begin{align}
1+\delta_{z} = 1.02(9) \,, \qquad 1+\delta^{'}_{z}= 1.06(8)\,, \qquad K = 1.46(10)~{\rm GeV^{-1}} \, , \qquad   K + \delta K_{T} = 1.39(6)~{\rm GeV^{-1}}\,,
\end{align}
and for the fit parameter $\Lambda_{H}$ we obtain the value $\Lambda_{H} = 0.70(17)\,{\rm GeV}$. Finally, for the parameters $A_{m}, A_{m}^{T}, A_{x_{\gamma}}$ and $A_{x_{\gamma}}^{T}$ we obtain
\begin{align}
A_{m}= 0.8(5) \,, \qquad A_{m}^{T}= 1.4(3)  \,, \qquad  A_{x_{\gamma}} = -1.0(1) \,,  \qquad A_{x_{\gamma}}^{T} = -1.0(1)~.
\end{align}
The relation $A_{x_{\gamma}}= A_{x_{\gamma}}^{T}$ which holds in the HQET and large-photon-energy expansion neglecting perturbative radiative corrections and non-zero strange-quark mass effects, appears to be well reproduced by our data. As for the relation $A_{m}=A_{m}^{T}$, we find that the fitted values of $A_{m}$ and $A_{m}^{T}$ are slightly different, which can be attributed to radiative corrections and/or $\mathcal{O}(m_{s}/m_{H_{s}})$ effects as well as to statistical fluctuations.  
In Table~\ref{tab:FF_extr} we provide our results for the four form factors, extrapolated to the physical mass $m_{B_{s}}$ and for the four simulated values of $x_{\gamma}$. The fit parameters, including their correlations, are available upon request from the authors.

\begin{table}
\setlength{\tabcolsep}{7pt}
\renewcommand{\arraystretch}{1.5}
\begin{tabular}{|c|c|c|c|c|}
\hline
$x_{\gamma}$ & $F_{V}$ & $F_{A}$ & $F_{TV}$ & $F_{TA}$ \\ \hline
$0.1$ & $1.103(38)$ & $0.290(13)$ & $1.026(35)$ & $0.413(17)$ \\ \hline
$0.2$ & $0.610(13)$ & $0.226(8)$ & $0.564(15)$ & $0.326(8)$ \\ \hline
$0.3$ & $0.422(8)$ & $0.186(6)$ & $0.389(10)$ & $0.270(6)$ \\ \hline
$0.4$ & $0.322(6)$ & $0.157(5)$ & $0.297(8)$ & $0.230(5)$ \\ \hline
\end{tabular}
\caption{\small\it Our results for the form factors $F_{V}$, $F_{A}$, $F_{TV}$ and $F_{TA}$ extrapolated to the physical mass $m_{B_{s}}$, for the four simulated values of $x_{\gamma}=0.1,0.2,0.3,0.4$.  \label{tab:FF_extr}}
\end{table}

\begin{figure}
\includegraphics[scale=0.47]{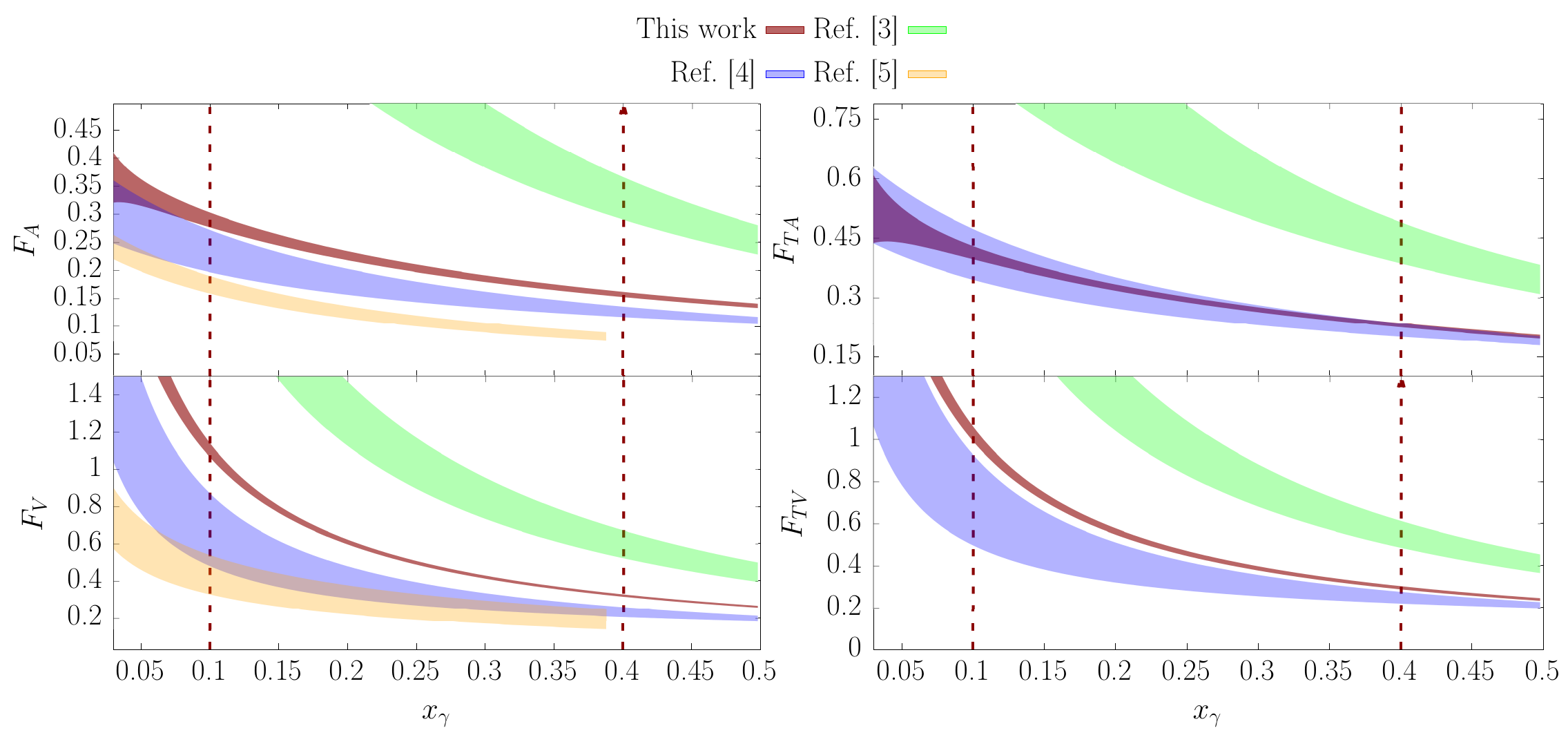}
\caption{\small\it Comparison between our results for the form factors $F_{W}$ with $W=\{V,A,TV,TA\}$ (shown in the figure by the red bands), and existing model-dependent results~\cite{Kozachuk:2017mdk, Janowski:2021yvz, Guadagnoli:2023zym}. 
The region between the vertical red dashed lines corresponds to the region of simulated $x_{\gamma}$, and therefore within this region our results are obtained through an \textit{interpolation} of our lattice data. \label{fig:comp_FF}}
\end{figure}

Our results for the form factors can be compared with available phenomenological and model estimates. 
The form factors $F_{W}$, $W=\{V,A,TV,TA\}$, have been previously obtained using relativistic dispersion relations\,\cite{Kozachuk:2017mdk}\,, light-cone sum rules\,\cite{Janowski:2021yvz} and recently a hybrid approach\,\cite{Guadagnoli:2023zym} in which the existing lattice results for the form factors $F_{V}$ and $F_{A}$ in $D_{s}\to \ell\nu\gamma$ decays are used to obtain the form factors $F_{V}$ and $F_{A}$ entering $B_{s}\to \mu^{+}\mu^{-}\gamma$ decays using a VMD-inspired ansatz. The comparison between our determination of $F_{W}$, $W=\{V,A,TV,TA\}$ and the existing model-dependent results is shown in Figure~\ref{fig:comp_FF}. 
Our results are given by the red curves, and outside the region of measured $x_{\gamma}$ are obtained by using the best-fit function obtained in the global fits discussed above. 
Our results for $F_{TV}$ and $F_{TA}$ turn out to be in rather good agreement with the estimate of Ref.\,\cite{Kozachuk:2017mdk}, taking into account that the results of the relativistic dispersion approach contain a systematic uncertainty which is difficult to quantify. However we find significant differences with respect to the results of Ref.\,\cite{Janowski:2021yvz} for $F_{TV}$ and $F_{TA}$. For the axial form factor $F_{A}$, the differences between our results and those of Ref.~\cite{Kozachuk:2017mdk} are of similar size as the one present for $F_{TV}$, while more significant deviations are observed for the vector form factor $F_{V}$. Moreover, we disagree with both the estimates given in Ref.\,\cite{Janowski:2021yvz} and Ref.\,\cite{Guadagnoli:2023zym} for $F_{V}$ and $F_{A}$. The disagreement with the light-cone sum rule calculation was somehow expected, given that large differences with respect to lattice QCD calculations have been already observed in the radiative leptonic decays of the $D_{s}$ meson\,\cite{Frezzotti:2023ygt}. The smaller value of $F_{V}$ obtained in Ref.\,\cite{Guadagnoli:2023zym} could be, at least partially, traced back to the fact that their estimate of the strange-quark contribution to the form factor $g_{D^{*}_{s}D_{s}\gamma}$, an essential input parameter of their VMD-inspired approach, turns out to be substantially smaller than the one obtained by the HPQCD Collaboration in Ref.~\cite{Donald:2013sra} and in our recent paper~\cite{Frezzotti:2023ygt} (which are instead in very nice agreement with each other). Before discussing the implications of our results for the branching fraction $\mathcal{B}(B_{s}\to\mu^{+}\mu^{-}\gamma)$, we present now our results for the form factor $\bar{F}_{T}$.

\section{The local form factor {$\bf\bar{F}_T$}}\label{sec:FbarT}
The form factor $\bar{F}_T$ can be computed from the knowledge of the hadronic tensor
\begin{align}
\label{eq:hadr_tens_virt}
H^{\mu\nu}_{\bar{T}}(p,k) =  i\int d^{4}x ~ e^{i(p-k)x} ~ {\rm \hat{T}} \langle 0 | J_{\bar{T}}^{\nu}(0) J_{{\rm em}}^{\mu}(x) | \bar{B}_{s}(\bs{0})\rangle 
=-\varepsilon^{\mu\nu\rho\sigma}k_{\rho}p_{\sigma} \frac{\bar{F}_{T}}{m_{B_{s}}}\,,
\end{align}
where 
\begin{equation}
J_{\bar{T}}^{\nu} = -iZ_{T}(\mu) \bar{s}\,\sigma^{\nu\rho}\,b \frac{k_{\rho}}{m_{B_{s}}}~.
\end{equation}
As in the case of the currents $J_{TV}^{\nu}$ and $J_{TA}^{\nu}$, we renormalize the tensor current $J_{\bar{T}}$ using the non-perturbative determination of $Z_{T}(\mu)$ in the $\msbar$ scheme at the scale $\mu=5~{\rm GeV}$ given in Table~\ref{tab:renormalization}.
Note that $H^{\mu\nu}_{\bar{T}}(p,k) = H^{\mu\nu}_{TV}(p, p-k)$, and recall that $\bar{F}_{T}(1)= F_{TV}(1) = F_{TA}(1)$ (see Eq.\,(\ref{eq:F1})). A significant complication is that the hadronic tensor $H^{\mu\nu}_{\bar{T}}(p,k)$ suffers 
from problems of analytic continuation to Euclidean spacetime. To demonstrate this, we start by writing explicitly the contributions to the hadronic tensor $H^{\mu\nu}_{\bar{T}}$ from the two time orderings, namely 
\begin{align}
\label{eq:barT_TO}
H^{\mu\nu}_{\bar{T}}(p,k) &= i\int_{-\infty}^{0} dt \, e^{i(m_{B_s}-E_{\gamma})t}\,\langle 0 |  J^{\nu}_{\bar{T}}(0)\, J^{\mu}_{\rm em}(t, -\bs{k}) | \bar{B}_{s}(\bs{0})\rangle+
i\int_{0}^{\infty} dt \, e^{i(m_{B_s}-E_{\gamma})t}\,  \langle 0 | J^{\mu}_{\rm em}(t, -\bs{k}) J^{\nu}_{\bar{T}}(0) \, | \bar{B}_{s}(\bs{0})\rangle
~.
\end{align}
Making use of 
\begin{align}
J^{\mu}_{\rm em}(t,\bs{k}) = e^{i(\hat{H}-i\varepsilon)t} \, J^{\mu}_{\rm em}(0,\bs{k}) \, e^{-i(\hat{H}-i\varepsilon)t}~,
\end{align}
where $\hat{H}$ is the QCD Hamiltonian, one has $H^{\mu\nu}_{\bar{T}}(p,k) = H^{\mu\nu}_{\bar{T},1}(p,k) + H^{\mu\nu}_{\bar{T},2}(p,k)$ with
\begin{align}
\label{eq:barT_can_deco}
H^{\mu\nu}_{\bar{T},1}(p,k) &=    \langle 0| J_{\bar{T}}^{\nu}(0)\frac{1}{\hat{H} - E_{\gamma} -i\varepsilon}J_{\rm em}^{\mu}(0,-\bs{k})| \bar{B}_{s}(0)\rangle~,\nonumber \\[8pt]
H^{\mu\nu}_{\bar{T},2}(p,k) &=  \langle 0| J_{\rm em}^{\mu}(0,-\bs{k}) \frac{1}{ \hat{H} + E_{\gamma} - m_{B_{s}} - i\varepsilon} J_{\bar{T}}^{\nu}(0)| \bar{B}_{s}(0)\rangle~.
\end{align}
 The two integrals in Eq.\,(\ref{eq:barT_TO}) can only be Wick-rotated from Minkowskian time $t$ to Euclidean time $\tau=it$ if the following positivity conditions are met
\begin{align}
\label{eq:positivity_cond}
\langle n | \hat{H} - E_{\gamma} | n \rangle > 0 , \qquad\qquad \langle m | \hat{H} + E_{\gamma} - m_{B_{s}} | m \rangle > 0~,
\end{align}
where $|n\rangle$ and $|m \rangle$ are the intermediate states contributing respectively to the first and second time ordering in Eq.\,(\ref{eq:barT_can_deco}). In the rest frame of the $\bar{B}_{s}$ meson in which we work, all intermediate states contributing to the hadronic tensor have three momentum $|\bs{k}|= E_{\gamma}$, therefore the condition $\langle n | \hat{H} - E_{\gamma} | n \rangle > 0$ is always satisfied and one can safely set $\varepsilon = 0$ in the first contribution on the r.h.s. of Eq.~(\ref{eq:barT_can_deco}). This is not the case for the second condition in Eq.~(\ref{eq:positivity_cond}) due to the presence of light unflavoured $J^{P}= 1^{-}$ intermediate states. Indeed, defining $m_{V_0}$ to be the mass of the lightest hadronic state contributing to the second time ordering, the analytic continuation is obstructed if the photon energy $E_{\gamma}$ satisfies
\begin{align}
\label{eq:forbidden_analytic_continuation}
\sqrt{ m_{V_0}^{2} + E_{\gamma}^{2}} + E_{\gamma} < m_{B_{s}} \implies x_{\gamma} < 1 - \left(\frac{m_{V_0}}{m_{B_{s}}}\right)^{2} ~.
\end{align}
As in the case of the local form factors $F_{W}$, $W=\{ V, A, TV, TA\}$, we can distinguish the two contributions $\bar{F}_{T}^{b}$ and $\bar{F}_{T}^{s}$ to the form factor $\bar{F}_{T}$, corresponding respectively to the emission of the virtual photon $\gamma^{*}$  from the bottom (Figure\,\ref{fig:ph_penguin2} left) and strange (Figure\,\ref{fig:ph_penguin2} right) quark line\,\footnote{ The two contributions are obtained by the replacements $J_{\rm em}^{\mu} \to J_{b}^{\mu}$ and $J_{\rm em}^{\mu}\to J_{s}^{\mu}$ 
in Eqs.\,(\ref{eq:barT_TO})-(\ref{eq:barT_can_deco})
(see Eq.~(\ref{eq:1PS_em})). }. 
The lightest hadronic intermediate state in the second time ordering are given,  respectively for the bottom- and strange-quark contributions, by the $\Upsilon(1S)$ resonance and by $K^{+}K^{-}$ states in a
P-wave\footnote{In the electroquenched approximation in which we work the Zweig-suppressed contributions from $u\bar{u}$, $d\bar{d}$ and $c\bar{c}$ resonances are absent.}. Given that $m_{\Upsilon} \simeq 9460~{\rm MeV} > m_{B_{s}}$, the bottom quark contribution is not affected by the problem of analytic continuation, which is only present in the strange quark contribution. Indeed, with $2m_{K}\simeq  1~{\rm GeV}$, one finds that analytic continuation is obstructed for
\begin{align}
x_{\gamma} < x_{\gamma}^{\rm th} \equiv 1 - \left( \frac{ 2m_{K}}{m_{B_{s}}}\right)^{2} \simeq 0.96~,
\end{align}
i.e. for all the values of $x_{\gamma}$ that we are considering.

Recently, some of us have proposed a novel strategy~\cite{Frezzotti:2023nun} to circumvent the problem of analytic continuation of electroweak amplitudes of the type present in Eq.~(\ref{eq:barT_TO}), i.e. involving an hadron-to-vacuum QCD matrix element of the product of two currents. In order to briefly summarise the strategy, we focus on the strange-quark contribution to $H^{\mu\nu}_{T,2}(p,k)$. To keep the notation simple, we set $E\equiv m_{B_{s}} - E_{\gamma}$ and define
\begin{align}
\label{eq:CH2defs}
C^{\mu\nu}_{q,2}(t, \bs{k}) \equiv \langle 0 | J^{\mu}_{q}(t, -\bs{k}) J^{\nu}_{\bar{T}}(0) \, | \bar{B}_{s}(\bs{0})\rangle\,, \qquad H^{\mu\nu}_{\bar{T}_q,2}(E,\bs{k}) \equiv i\int_{0}^{\infty} dt\,e^{iEt}\,C^{\mu\nu}_{q, 2}(t,\bs{k})~, \qquad q=s,b~, 
\end{align}
so that $H^{\mu\nu}_{\bar{T},2}(p,k) = H^{\mu\nu}_{\bar{T}_b,2}(E,\bs{k}) + H^{\mu\nu}_{\bar{T}_s,2}(E,\bs{k})$.
The main idea for circumventing the problem of the analytic continuation of $H^{\mu\nu}_{\bar{T}_s,2}(E,\bs{k})$, is to consider the spectral-density representation of the time-dependent correlation function $C^{\mu\nu}_{s,2}(t,\bs{k})$,
\begin{align}
\label{eq:Cs2}
C^{\mu\nu}_{s,2}(t,\bs{k}) = \int_{E^{*}}^{\infty} \frac{dE'}{2\pi} e^{-iE't} \rho^{\mu\nu}(E',\bs{k})~,
\end{align}
where $[E^{*},\infty)$ is the support of the spectral density $\rho^{\mu\nu}(E',\bs{k})$, and in our case $E^{*}= \sqrt{ m_{V_{0}}^{2} + E_{\gamma}^{2}}$ with $m_{V_{0}}= 2m_{K}$. Combining Eqs.\,(\ref{eq:CH2defs})
and (\ref{eq:Cs2}) it follows that (see Ref.\,\cite{Frezzotti:2023nun} for details)
\begin{align}
\label{eq:amplitude_spectral}
H^{\mu\nu}_{\bar{T}_s,2}(E,\bs{k}) = \lim_{\varepsilon\to 0^{+}}\int_{E^{*}}^{\infty} \frac{dE'}{2\pi} \frac{\rho^{\mu\nu}(E',\bs{k})}{E'-E-i\varepsilon} = {\rm PV} \int_{E^{*}}^{\infty} \frac{dE'}{2\pi} \frac{\rho^{\mu\nu}(E',\bs{k})}{E'-E}  + \frac{i}{2}\rho^{\mu\nu}(E,\bs{k}) ~,
\end{align}
where ${\rm PV}$ denotes the principal value of the integral.
The Minkowski correlator $C^{\mu\nu}_{q,2}(t,\bs{k})$ can always be analytically continued to Euclidean spacetime. The Euclidean correlator $\bar{C}^{\mu\nu}_{s,2}(t,\bs{k}) \equiv C^{\mu\nu}_{s,2}(-it,\bs{k})$ is then related
to the spectral density $\rho^{\mu\nu}(E',\bs{k})$ via
\begin{align}
\label{eq:laplace_transfo}
\bar{C}^{\mu\nu}_{s,2}(t,\bs{k}) = \int_{E^{*}}^{\infty} \frac{dE'}{2\pi} e^{-E' t} \rho^{\mu\nu}(E',\bs{k})~.
\end{align}
Since $\bar{C}^{\mu\nu}_{s,2}(t,\bs{k})$ can be computed using Monte Carlo simulations, we have formally solved the problem of analytic continuation; by inverting the relation in Eq.\,(\ref{eq:laplace_transfo}) to determine $\rho^{\mu\nu}(E')$ we can then obtain $H^{\mu\nu}_{\bar{T}_s,2}(E,\bs{k})$ using Eq.\,(\ref{eq:amplitude_spectral}). 
However, in order to determine $\rho^{\mu\nu}(E',\bs{k})$ using Eq.\,(\ref{eq:laplace_transfo}), an inverse Laplace transform of the Euclidean correlator $\bar{C}^{\mu\nu}_{s,2}(t,\bs{k})$ is required. This is a well-known ill-posed numerical problem when $\bar{C}^{\mu\nu}_{s,2}(t,\bs{k})$ is only known on a finite set of points in time and is affected by uncertainties, which is the typical situation encountered in a lattice calculation. 
In Ref.\,\cite{Frezzotti:2023nun} it has been proposed to use the $-i\varepsilon$ term appearing in the denominator of Eq.\,(\ref{eq:amplitude_spectral}) as a \textit{regulator} of the problem by introducing the smeared amplitude $H^{\mu\nu}_{s,2}(E,\bs{k};\varepsilon)$ 
\begin{align}
\label{eq:smeared_amplitude}
H^{\mu\nu}_{\bar{T}_s,2}(E,\bs{k} ;\varepsilon) = \int_{E^{*}}^{\infty} \frac{dE'}{2\pi} \frac{\rho^{\mu\nu}(E',\bs{k})}{E'-E-i\varepsilon} =  \int_{E^{*}}^{\infty} \frac{dE'}{2\pi}\,K(E'-E;\varepsilon)\,\rho^{\mu\nu}(E',\bs{k})~,
\end{align}
where
\begin{align}
\label{eq:Cauchy_kernel}
K(x;\varepsilon) \equiv \frac{1}{x-i\varepsilon} = \frac{x}{x^{2}+\varepsilon^{2}} + i\frac{\varepsilon}{x^{2}+\varepsilon^{2}} ~.
\end{align}
The key point is that for non-zero values of $\varepsilon$, the kernel function $K(x;\varepsilon)$ is smooth, and its convolution integral with the spectral density $\rho^{\mu\nu}(E',\bs{k})$ can be evaluated, from the knowledge of $\bar{C}^{\mu\nu}_{s,2}(t,\bs{k})$ only, using the Hansen-Lupo-Tantalo (HLT) method introduced in Ref.~\cite{Hansen:2019idp} (see also Refs.~\cite{ExtendedTwistedMassCollaborationETMC:2022sta, Evangelista:2023fmt, Bonanno:2023thi} for recent applications of the method). The idea is to numerically evaluate the smeared amplitude $H^{\mu\nu}_{\bar{T}_s,2}(E, \bs{k};\varepsilon)$ for finite values of the smearing parameter $\varepsilon$ using the HLT method (to be discussed in the next section), and then to extrapolate to $\varepsilon=0$, exploiting the fact that (see Ref.~\cite{Frezzotti:2023nun} for a proof)
\begin{align}
H^{\mu\nu}_{\bar{T}_s,2}(E,\bs{k};\varepsilon) = H^{\mu\nu}_{\bar{T}_s,2}(E,\bs{k}) + A(E,\bs{k})\varepsilon + \mathcal{O}(\varepsilon^{2})\,.    
\end{align}
We stress that the problem of evaluating $H^{\mu\nu}_{\bar{T}_s}(E,\bs{k})
$ is ill-posed only for $E > E^{*}$, i.e. if the inequality in Eq.~(\ref{eq:forbidden_analytic_continuation}) is satisfied. Instead, for $E < E^{*}$, one can directly set $\varepsilon=0$ in Eq.~(\ref{eq:amplitude_spectral}) (in this case the integrand is non-singular), and by using
\begin{align}
\frac{1}{E'-E} = \int_{0}^{\infty}\, dt\, e^{-(E'-E)t} ~\qquad ({\textrm{valid for }} E < E')~,
\end{align}
one arrives at (see Ref.~\cite{Frezzotti:2023nun})
\begin{align}
H^{\mu\nu}_{\bar{T}_s,2}(E,\bs{k}) = \int_{0}^{\infty} dt\,e^{Et}\,\bar{C}^{\mu\nu}_{s,2}(t,\bs{k})< \infty~,
\end{align}
which is the standard formula used to evaluate the form factors in absence of problems of analytic continuation (see e.g. Eq.~(\ref{eq:Cmunudef})) and the one we apply here to determine $H^{\mu\nu}_{T_{b},2}(E,\bs{k})$.

To summarize, we evaluate the hadronic tensor $H^{\mu\nu}_{\bar{T}}(p,k)$ in Eq.~(\ref{eq:hadr_tens_virt}) as the sum of the following terms
\begin{align}
H^{\mu\nu}_{\bar{T}}(p,k) \equiv H^{\mu\nu}_{\bar{T}_{b}}(E,\bs{k}) + H^{\mu\nu}_{\bar{T}_{s}}(E,\bs{k})~,
\end{align}
where
\begin{align}
\label{eq:hadr_tens_bar}
H^{\mu\nu}_{\bar{T}_{b}}(E,\bs{k}) &\equiv H^{\mu\nu}_{\bar{T}_b,1}(E,\bs{k}) + H^{\mu\nu}_{\bar{T}_b,2}(E,\bs{k})~, \\[8pt]
H^{\mu\nu}_{\bar{T}_{s}}(E,\bs{k}) &\equiv \lim_{\varepsilon\to 0^{+}} H^{\mu\nu}_{\bar{T}_s,1}(E,\bs{k}) + H^{\mu\nu}_{\bar{T}_s,2}(E,\bs{k};\varepsilon)~,
\label{eq:hadr_tens_bar}
\end{align}
and we have defined the first time ordering contribution as ($q=b,s$)
\begin{align}
H^{\mu\nu}_{\bar{T}_{q},1}(E,\bs{k}) \equiv \int_{-\infty}^{0} dt\,e^{Et}\, \bar{C}^{\mu\nu}_{q,1}(t,\bs{k})\, , \qquad \bar{C}^{\mu\nu}_{q,1}(t,\bs{k}) \equiv C^{\mu\nu}_{q,1}(-it,\bs{k})\, , \qquad  C^{\mu\nu}_{q,1}(t,\bs{k}) \equiv \langle 0 |  J^{\nu}_{\bar{T}}(0) J^{\mu}_{q}(t, -\bs{k}) \, | \bar{B}_{s}(\bs{0})\rangle\, ,
\end{align}
where the Euclidean correlator
$\bar{C}^{\mu\nu}_{q,1}(t,\bs{k})$ is the lattice input. 
On the lattice, because of the discretization of spacetime, the relations above get slightly modified, as we will discuss in the next section. 
\subsection{Numerical results for $\bar{F}_{T}$}
\label{sec:F_barnum}
In order to evaluate the form factor $\bar{F}_{T}$ we have performed simulations on a subset of the ensembles in Table\,\ref{tab:simudetails}. These are the B64 and D96 ensembles. 
The computations have been performed at all four values of $x_{\gamma}$ in Eq.\,(\ref{eq:xg_val}) but only at the following three values of the heavy quark mass, 
\begin{align}
\frac{m_{h}}{m_{c}} \simeq 1,\,1.5,\,2.5\,.
\end{align}
The reasons for reducing the number of ensembles and values of $m_h$ which we use are two-fold. 
Firstly, the use of the spectral representation technique to overcome the difficulty in the continuation to Euclidean space is computationally expensive and secondly the contribution to the differential rates from $\bar{F}_{T}$ is small and so this form factor is not required with the same precision as those studied in Section\,\ref{sec:FF_num}.

Our strategy to compute $\bar{F}_{T}$ consists in evaluating on the lattice the following three-point Euclidean correlation function
\begin{align}
M^{\mu\nu}_{\bar{T}_{q}}(t,t_{\rm sep},\bs{k}) \equiv a^{3}\sum_{\bs{x}}\langle 0 | J_{q}^{\mu}(t+t_{\rm sep},-\bs{k}) ~J_{\bar{T}}^{\nu}(t_{\rm sep}) ~\phi^{\dag}_{B_{s}}(0, \bs{x}) | 0 \rangle~\qquad q = s,b~,
\label{eq:M_munu}
\end{align}
where $\phi^{\dag}_{B_{s}}(0,\bs{x})$ is the same interpolating operator as was used in Eq.\,(\ref{eq:Cmunudef}),  while $t_{\rm sep}$ is the fixed time where the tensor FCNC $J_{\bar{T}}^{\nu}$ is inserted, which must be chosen  large enough to ensure the dominance of the ground state. In the limit of large $t_{\rm sep}$ one has
\begin{align}
\label{eq:M_muni_asympt}
M^{\mu\nu}_{\bar{T}_{q}}(t,t_{\rm sep},\bs{k}) = \frac{\langle \bar{B}_{s}(0) |   \phi^{\dag}_{B_{s}}(0) | 0 \rangle}{2m_{B_{s}}}e^{-m_{B_{s}}t_{\rm sep}}\, \left( \bar{C}^{\mu\nu}_{q}(t)  + \ldots \right)~, \qquad q=s,b~,
\end{align}
where the dots represent terms that are exponentially suppressed at large $t_{\rm sep}$, 
and 
\begin{align}
\bar{C}^{\mu\nu}_{q}(t,\bs{k}) = \theta(-t)\,\bar{C}^{\mu\nu}_{q,1}(t,\bs{k}) + \theta(t)\,\bar{C}^{\mu\nu}_{q,2}(t,\bs{k})~, \qquad q=s,b\,,
\end{align}
where the correlators $\bar{C}^{\mu\nu}_{q,1}(t,\bs{k})$ and $\bar{C}^{\mu\nu}_{q,2}(t,\bs{k})$ were introduced in the previous section.

Notice that the time $t$ in the previous equations corresponds to the time separation between the electromagnetic and tensor currents and is different from the time $t$ introduced in Eq.\,(\ref{eq:Cmunudef}).
The choice of $t_{\rm sep}$ has been adapted depending on the contribution being considered. For $q=b$ (and both $t >0$ and $t<0$) and for $q=s,\,t < 0$ we have chosen a large $t_{\rm sep} \simeq 2\,{\rm fm}$, while for $q=s$ and $t>0$, which is the only contribution requiring the spectral density reconstruction method of Eq.\,(\ref{eq:smeared_amplitude}) and for which statistical accuracy is of the upmost importance\,
\footnote{The statistical accuracy of the computed $M^{\mu\nu}_{\bar{T},q}(t, t_{\rm sep}, \bf{k})$ decreases as $t_{\rm sep}$ increases.}, we have chosen $t_{\rm sep} \simeq 1\,{\rm fm}$, after checking ground-state-dominance using the larger value $t_{\rm sep} \simeq 1.7\,{\rm fm}$. For the same reason, the inversions of the Dirac operator for $q=s$ and $t>0$ have been performed using a number of stochastic sources which is eight times larger than that used for $q=b$ and $q=s,\, t<0$.

\begin{figure}
    \centering
    \includegraphics[scale=0.4]{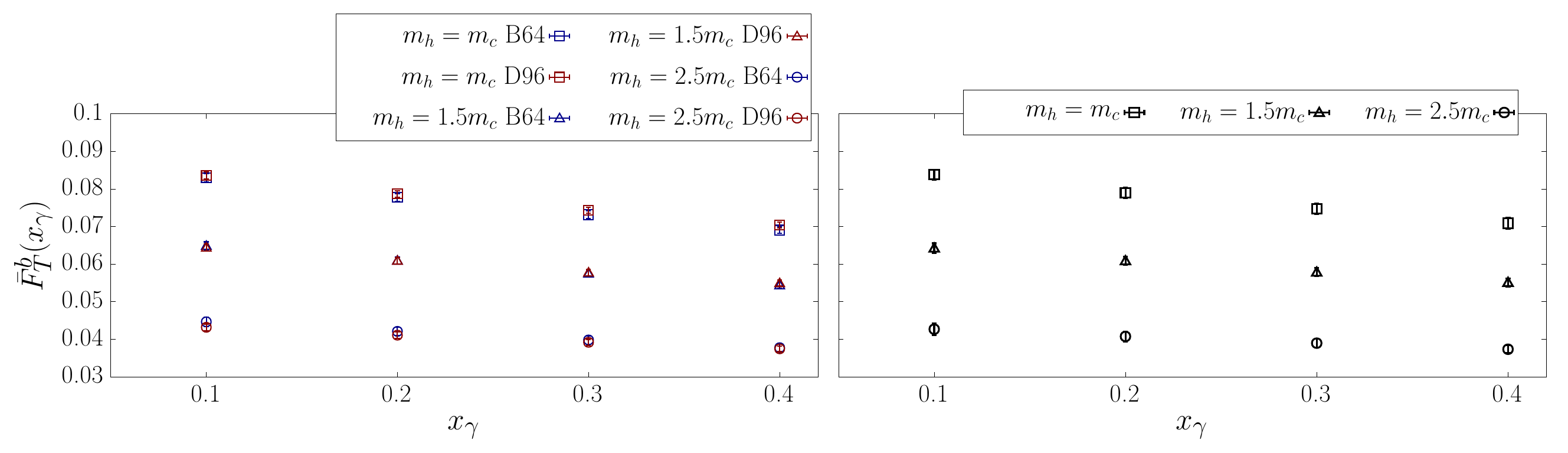}
    \caption{\small\it Left: values of $\bar{F}_{T}^{b}$ obtained on the B64 (blue data points) and D96 (red data points) ensembles as a function of $x_{\gamma}$. Right:  results for $\bar{F}_{T}^{b}$ after extrapolation to the continuum as a function of $x_{\gamma}$, obtained following the procedure described in the text. The different symbols correspond to the different values of the simulated heavy quark mass $m_{h}$.   \label{fig:FT_b}}
\end{figure}

We now discuss our determination of $\bar{F}_{T}$ starting from the b-quark contribution $\bar{F}_{T}^{b}$. In this case, since there is no problem of analytic continuation, we proceed as in Eq.~(\ref{eq:CH2defs}), and evaluate the hadronic tensor $H^{\mu\nu}_{\bar{T}_{b}}$ using\,\footnote{With respect to Eq.\,(\ref{eq:Cmunudef}), we have dropped the $e^{-ET}$ term, which is numerically negligible.}
\begin{align}
\label{eq:H_Tb_lat}
H^{\mu\nu}_{\bar{T_{b}}}(E,\bs{k}) = a\!\!\!\!\sum_{t=-t_{\rm sep}}^{T/2-t_{\rm sep}} \, e^{Et} \, \bar{C}_{b}^{\mu\nu}(t,\bs{k})~, \qquad E= m_{B_{s}} - E_{\gamma}~. 
\end{align}
For any simulated heavy-strange meson mass $m_{H_{s}}$ the corresponding energy $E$ in the previous equation is understood to be
\begin{align}
\label{eq:energy_E_scaling}
E= m_{H_{s}} - E_{\gamma} = m_{H_{s}}\left( 1 -\frac{x_{\gamma}}{2}\right)~.
\end{align}
From the knowledge of $H^{\mu\nu}_{\bar{T}_{b}}(E)$ we use Eq.\,(\ref{eq:hadr_tens_virt}) to determine the $b$-quark contribution $\bar{F}_{T}^{b}$ to the form factor $\bar{F}_{T}$. In the rest frame of the decaying meson, and with our choice of the photon momentum ($\bs{k}=(0,0,k_{z})$), the form factor can be obtained using
\begin{align}
\bar{F}_{T}^{b}(x_{\gamma}) = -\frac{1}{2k_{z}}\left( H^{12}_{\bar{T_{b}}}(E,\bs{k}) - H^{21}_{\bar{T_{b}}}(E,\bs{k})     \right)~, \qquad |k_{z}| = E_{\gamma} = m_{H_{s}}\frac{x_{\gamma}}{2}~.
\end{align}
Our determination of $\bar{F}_{T}^{b}$ for the four different simulated values of $x_{\gamma}$ and for the three different heavy-strange meson masses $m_{H_{s}}$, is shown in the left panel of Figure~\ref{fig:FT_b}. 
The blue and red colors in the left panel correspond to our results on the B64 and D96 ensemble, respectively. 
As the figure shows, we find that cut-off effects are very small, the $x_{\gamma}$ behaviour is almost linear and the form factor decreases as the heavy-quark mass $m_{H_{s}}$ increases. However, we postpone the discussion of the extrapolation to the physical mass $m_{B_{s}}$ to Section\,\ref{sec:extr_barF_T}, and concentrate here only on the issue of the continuum extrapolation. Having only two lattice spacings available, and given the smallness of the observed UV cut-off effects, we opt for extrapolating to the continuum limit at fixed $m_{H_{s}}$ and $x_{\gamma}$, employing either a constant or linear Ansatz in $a^{2}$. We then combine the results of the two extrapolations using the following criterion: if the constant fit gives a $\chi^{2}$ smaller than two, we combine the results of the linear and constant fit using the weighted average already illustrated in Eqs.~(\ref{eq:weight_ave_1})-(\ref{eq:weight_ave_2}) but using same weights for the linear and constant $a^{2}$ extrapolation, otherwise the final result is given by the result obtained using the linear $a^{2}$ Ansatz. The result of the continuum-limit extrapolation is illustrated in the right panel of Figure~\ref{fig:FT_b}.

We now turn into the discussion of the more involved strange-quark contribution.
In this case, as already discussed, the form factor cannot be obtained as in Eq.~(\ref{eq:H_Tb_lat}) since
\begin{align}
\lim_{T\to\infty} a\!\!\!\!\sum_{t=-t_{\rm sep}}^{T/2-t_{\rm sep}} e^{Et} \bar{C}_{s}^{\mu\nu}(t,\bs{k}) = \infty~,
\end{align}
due to the fact that for large and positive times $t$ the correlation function behaves approximately as
\begin{align}
\bar{C}_{s}^{\mu\nu}(t,\bs{k}) \simeq e^{-E^{*}t}, \qquad E^{*} < E ~. 
\end{align}
Our strategy to evaluate the contribution from the second time-ordering, which is the only one affected by the problem of the analytic continuation, is to consider the smeared (or regularized) hadronic amplitude introduced in Eq.~(\ref{eq:smeared_amplitude}), namely
\begin{align}
\label{eq:smeared_hadronic_amplitude}
H^{\mu\nu}_{\bar{T}_s,2}(E,\bs{k} ;\varepsilon) =  \int_{E^{*}}^{\infty} \frac{dE'}{2\pi}\,K(E'-E;\varepsilon)\,\rho^{\mu\nu}(E',\bs{k})~,
\end{align}
or equivalently, separating the real and imaginary part,
\begin{align}
\label{eq:re_smeared_amplitude}
{\rm Re}[H^{\mu\nu}_{\bar{T}_s,2}(E,\bs{k} ;\varepsilon)] &= \int_{E^{*}}^{\infty} \frac{dE'}{2\pi}\,K_{\rm Re}(E'-E;\varepsilon)\,\rho^{\mu\nu}(E',\bs{k})~, \qquad K_{\rm Re}(x;\varepsilon) = {\rm Re}[ K(x;\varepsilon)]~,\\[8pt]
\label{eq:im_smeared_amplitude}
{\rm Im}[H^{\mu\nu}_{\bar{T}_s,2}(E,\bs{k} ;\varepsilon)] &= \int_{E^{*}}^{\infty} \frac{dE'}{2\pi}\,K_{\rm Im}(E'-E;\varepsilon)\,\rho^{\mu\nu}(E',\bs{k})~, \qquad K_{\rm Im}(x;\varepsilon) = {\rm Im}[ K(x;\varepsilon)]~,
\end{align}
and then to perform the extrapolation to $\varepsilon=0$. 
Eqs.\,(\ref{eq:re_smeared_amplitude}) and (\ref{eq:im_smeared_amplitude}) can be evaluated, from the knowledge of $C^{\mu\nu}_{s,2}(t,\bs{k})$ only, using the HLT method and we now briefly summarize the main ingredients of the procedure. 
To simplify the notation, we concentrate directly on the Lorentz indices that are relevant for the determination of the form factor with our choice of kinematics (decaying meson at rest, and $k=(0,0,k_{z})$), and define
\begin{align}
\bar{C}(t,\bs{k}) &\equiv \frac{1}{2}\left[\bar{C}^{12}_{s}(t,\bs{k}) - \bar{C}^{21}_{s}(t,\bs{k})\right]~,\\[8pt]
H_{2}(E,\bs{k},\varepsilon) &\equiv \frac{1}{2}\left[ H_{\bar{T}_{s},2}^{12}(E,\bs{k},\varepsilon) - H_{\bar{T}_{s},2}^{21}(E,\bs{k},\varepsilon)\right]~,\\[8pt]
\rho(E',\bs{k}) &
\equiv \frac{1}{2}\left[\rho^{12}(E',\bs{k})  - \rho^{21}(E',\bs{k})\right]~.
\end{align}
The final goal is to find, for fixed $\varepsilon$, the best approximation of the kernel functions $K_{\rm Re/Im}(E'-E;\varepsilon)$, in terms of the basis function $\{ e^{-aE' n}
\}_{n=1,\ldots, n_{max}}$, namely
\begin{align}
\label{eq:kernel_expansion}
K_{\mathrm I}\left(E'-E;\varepsilon\right)  \simeq   \sum_{n=1}^{n_{max}} g_\mathrm{I}(n,E, \varepsilon)\, e^{-a E' n} \equiv \widetilde{K}_\mathrm{I}\left(E',E;\varepsilon)\right)~,
\end{align}
where ${\rm I}=\{\rm{Re}, \rm{Im}\}$. In this way, once the coefficients $g_\mathrm{I}$ are known, the smeared hadronic amplitude can be reconstructed, from the knowledge of $\bar{C}$, using
\begin{align}
\label{eq:HLT_master}
H_{2}(E,\bs{k}; \varepsilon) &\equiv   \int_{E^{*}}^{\infty} \frac{dE'}{2\pi} \rho(E',\bs{k})\, K(E'-E;\varepsilon)
\nonumber \\[8pt]
&\simeq\sum_{n=1}^{n_{\rm max}} \left( g_{\rm Re}(n,E,\varepsilon) + ig_{\rm Im}(n,E,\varepsilon) \right)\int_{E^{*}}^{\infty} \frac{dE'}{2\pi}\,  e^{-a E' n }\rho(E',\bs{k}) \nonumber \\[8pt]
&=  \sum_{n=1}^{n_{\rm max}} \left( g_{\rm Re}(n,E, \varepsilon) + ig_{\rm Im}(n,E, \varepsilon)\right)\, \bar{C}(na,\bs{k})~.
\end{align}
The problem of finding the coefficients $g_\mathrm{I}$ presents a certain number of technical difficulties. Any determination of the real and imaginary part of the smeared hadronic amplitude based on Eqs.\,(\ref{eq:kernel_expansion}) and (\ref{eq:HLT_master}) will inevitably be affected by both systematic errors (due to the inexact reconstruction of the kernels) and statistical uncertainties (due to the fluctuations of the correlator $\bar{C}(t,\bs{k})$), which need to be simultaneously kept under control. 
The HLT method finds an optimal balance between the size of the statistical and systematic errors. This is achieved by minimizing a linear combination
\begin{align}
\label{eq:func_W}
W_\mathrm{I}[\boldsymbol{g}] \equiv \frac{A_\mathrm{I}[\boldsymbol{g}]}{A_\mathrm{I}[\boldsymbol{0}]} + \lambda B[\boldsymbol{g}]\;,
\end{align}
of the norm-functional
\begin{align}
\label{eq:func_A}
A_\mathrm{I}[\bs{g}] = \int_{E_{\rm min}}^{\infty} \dd E'\,  \bigg| \sum_{n=1}^{n_{\rm max}} g(n) e^{-a E' n} - K_\mathrm{I}\left( E'-E; \varepsilon\right) \bigg |^{2}~, 
\end{align}
which quantifies the difference between the approximated and the target kernel, 
and of the error-functional
\begin{align}
\label{eq:func_B}
B[\boldsymbol{g}] =  B_{\rm{norm}}\sum_{n_1 , n_2 = 1}^{n_{\rm max}} g(n_{1}) ~ g(n_{2})~ {\rm{Cov}}(an_1 , an_2 )~, 
\end{align}
where ${\rm Cov}(an_{1}, an_{2})$ is the covariance matrix of the correlator $\bar{C}(an)$, and $B_{\rm norm}$ is a normalization factor introduced to render the error-functional dimensionless.  The algorithmic parameter $E_{\rm min}$ should only satisfy the constraint $E_{\rm min} < E^{*}$, and we choose $E_{\rm min}= 0.9\sqrt{ m_{\phi}^{2} + E_{\gamma}^{2}}$. 
For each simulated value of $x_{\gamma}$ and $m_{H_{s}}$, we choose $n_{\rm max}$ by requiring that the statistical error on the correlation function $\bar{C}(t,\bs{k})$ for all times $t \leq an_{\rm max}$ must be smaller than $30\%$. 
The parameter $\lambda$ in Eq.\,(\ref{eq:func_W}) is the so-called \textit{trade-off} parameter, and for a given value of $\lambda$, the minimization of the functional $W_\mathrm{I}^{\alpha}[\bs{g}]$ gives the coefficients $\bs{g^{\lambda}_\mathrm{I}}$.
In the presence of statistical errors, the second term in Eq.\,(\ref{eq:func_W}) disfavours coefficients $\bs{g}$ leading to too large statistical uncertainties in the reconstructed value of the smeared hadronic amplitude. The optimal balance between having small statistical errors (small $B[\bs{g}]$) and small systematic errors in the kernel reconstruction (small $A_\mathrm{I}[\bs{g}]$) can be achieved by tuning $\lambda$ appropriately. 
This is done performing the so-called \textit{stability-analysis}, which is discussed in detail in Refs.\,\cite{ExtendedTwistedMassCollaborationETMC:2022sta,Frezzotti:2023nun}. 
In brief, using the stability-analysis one monitors the evolution of the reconstructed values of the real and imaginary part of $H_{2}(E,\bs{k};\varepsilon)$ as a function of $\lambda$. 
The optimal value, $\lambda^{\filledstar}$, (which is generally different for the real and imaginary parts) is chosen to be in the so-called statistically-dominated regime, where $\lambda$ is sufficiently small that the systematic error due to the kernel reconstruction is smaller than the statistical one (in this region the results are therefore stable under variations of $\lambda$), but large enough to still have reasonable statistical uncertainties. 
Finally, having determined the optimal value $\lambda^{\filledstar}$, we repeat the calculation using a second (smaller) value of $\lambda = \lambda^{\filledstar\filledstar}$, which is determined by imposing the validity of the following condition
\begin{align}
\label{eq:lambda_syst}
\frac{B[\bs{g_\mathrm{I}^{\lambda^{\filledstar\filledstar}}}]}{A_\mathrm{I}[ \bs{g^{\lambda^{\filledstar\filledstar}}_\mathrm{I}}]} = \kappa\, \frac{B[\bs{g_\mathrm{I}^{\lambda^{\filledstar}}}]}{A_\mathrm{I}[ \bs{g_\mathrm{I}^{\lambda^{\filledstar}}}]}~,
\end{align}
with $\kappa=10$. Any statistically-significant difference between the values of the real and imaginary part of  $H_{2}(E,\bs{k})$ corresponding to the two choices $\lambda=\lambda^{\filledstar}$ and $\lambda=\lambda^{\filledstar\filledstar}$ is added as a systematic uncertainty in our final error. We refer the reader to Ref.\,\cite{ExtendedTwistedMassCollaborationETMC:2022sta} for further details on this point. 

At a finite lattice spacing, similarly to what had been done in Ref.\,\cite{Frezzotti:2023nun}, we adopt the kernel function
\begin{align}
\label{eq:sinh_kernel}
K(x;\varepsilon) = \frac{a}{\sinh\left[ a(x-i\varepsilon)\right]} = \frac{1}{x-i\varepsilon} + \mathcal{O}(a^{2}) ~,
\end{align}
which differs from the one in Eq.~(\ref{eq:Cauchy_kernel}) only by $\mathcal{O}(a^{2})$ cut-off effects.
A major difference in the analysis of $H_{2}(E,\bs{k})$ compared to the strategy followed for $H^{\mu\nu}_{\bar{T}_{b}}(E,\bs{k})$ and $H^{\mu\nu}_{\bar{T}_{s},1}(E,\bs{k})$, concerns the scaling of the energy $E$ with the heavy-strange meson mass, $m_{H_{s}}$. 
While the energy-scaling given by Eq.\,(\ref{eq:energy_E_scaling}) leads to a smooth mass dependence for the latter two contributions, this is not the case for $H_{2}(E,\bs{k})$: the main contributions to the spectral density $\rho(E',\bs{k})$ are expected to depend on the position of the $\phi,\,\phi(1680),\,\phi(2170)$ (and possibly heavier) resonances. 
By scaling the energy $E$ according to Eq.\,(\ref{eq:energy_E_scaling}), to our lightest simulated mass $m_{H_{s}} = m_{D_{s}}$ would correspond an energy $E$ smaller or very close to that of one of the main $s\bar{s}$ peaks. On the other hand, the energy $E= m_{B_{s}}(1- x_{\gamma}/2)$, corresponding to the physical 
 mass of the $B_s$ meson, is much larger than the energy of such resonances. 
 Since the behaviour of $H_{2}(E,\bs{k};\varepsilon)$ below (or close to) the main $s\bar{s}$ resonances is expected to be very different from the one  at much larger energies of order $\mathcal{O}(m_{B_{s}})$, the mass scaling of $H_{2}(E,\bs{k};\varepsilon)$ that would result from the use of Eq.~(\ref{eq:energy_E_scaling}) is very complicated and difficult to handle. 
 At the same time setting $E=m_{B_{s}}(1- x_{\gamma}/2)$ for all simulated $m_{H_{s}}$ is problematic, as it leads to large cut-off effects. For $H_{2}(E,\bs{k};\varepsilon)$  we thus chose to scale the energy $E$ with the heavy-strange meson mass $m_{H_{s}}$ according to
\begin{align}
\label{eq:scaling_y}
E(r) = \left( r\,m_{H_{s}}+ (1-r)\,m_{B_{s}}\right)\,\left(1 - \frac{x_{\gamma}}{2}\right)~.
\end{align}
Note that any fixed $r$ is allowed since
\begin{align}
\lim_{m_{H_{s}}\to m_{B_{s}}} E(r) = m_{B_{s}}\left(1-\frac{x_{\gamma}}{2}\right)~,
\end{align}
and for $0\leq r \leq 1$ one interpolates between the scaling in Eq.~(\ref{eq:energy_E_scaling}) ($r=1$) and the case of a fixed energy $E=m_{B_{s}}(1- x_{\gamma}/2)$ ($r=0$). For each $x_{\gamma}$ we tune the value of $r$ in such a way that for $m_{H_{s}}=m_{D_{s}}$, since $m_{H_s}(1-x_\gamma/2)$ is the closest to the resonance region,  the corresponding energy $E(r)$ is above that of the main $s\bar{s}$ peaks, and at the same time small enough to avoid large cut-off effects. We choose $r=0.65, 0.60, 0.57, 0.55$, respectively for $x_{\gamma}=0.1, 0.2, 0.3$ and $0.4$ for all three values of $m_{H_s}$. 
Finally, we define the \textit{smeared} form factor as
\begin{align}
\bar{F}_{T}^{s}(x_{\gamma};\varepsilon) &\equiv -\frac{H(E,E(r),\bs{k}; \varepsilon)}{k_{z}}~,\qquad \bs{k}=(0,0,k_{z})~,\\[8pt]
H(E,E(r),\bs{k};\varepsilon) &= H(0,\bs{k}) \,+\, H^{\rm{sub}}_{1}(E,\bs{k}) \,+\, H^{\rm{sub}}_{2}(E(r),\bs{k};\varepsilon)~, 
\end{align}
where
\begin{align}
H(0,\bs{k}) &= a\!\!\!\!\sum_{t=-t_{\rm sep}}^{T/2 - t_{\rm sep}}  \bar{C}(t,\bs{k})~, \qquad H^{\rm sub}_{1}(E,\bs{k}) = a\!\!\!\!\sum_{t=-t_{\rm sep}}^{0} \left( e^{Et} -1 \right) \bar{C}(t,\bs{k})~\\ 
H^{{\rm sub}}_{2}(E(r),\bs{k};\varepsilon) \,&= \, H_{2}(E(r),\bs{k};\varepsilon) - H_{2}(0,\bs{k};0)~.
\end{align}
In the combined $m_{H_{s}}\to m_{B_{s}}$ and $\varepsilon\to 0$ limits, the smeared form factor tends to $\bar{F}_{T}^{s}(x_{\gamma})$. The zero-energy subtraction allows us to define the contributions from the two time-orderings in such a way that cut-off effects start at order $\mathcal{O}(a^{2})$ for both time-orderings. This is because they both are now free of the contact term $\bar{C}(0,\bs{k})$. 
This contact term does not belong to either the first or second time ordering, and cannot be simply removed as this generates $\mathcal{O}(a)$ cut-off effects. Since $H_{2}(E(r),\bs{k};\varepsilon)$ is evaluated via the HLT reconstruction method using the kernel function in Eq.~(\ref{eq:sinh_kernel}), to avoid the presence of $\mathcal{O}(a)$ cut-off effects also $H_{2}(0,\bs{k},0)$ is evaluated via the HLT method using the same type of kernel function. Being able to define the two time-orderings separately turns out to be useful if a model for the spectral density $\rho(E',\bs{k})$ is used to perform the $\varepsilon\to 0$ extrapolation, as will be discussed below.

\begin{figure}
\includegraphics[scale=0.40]{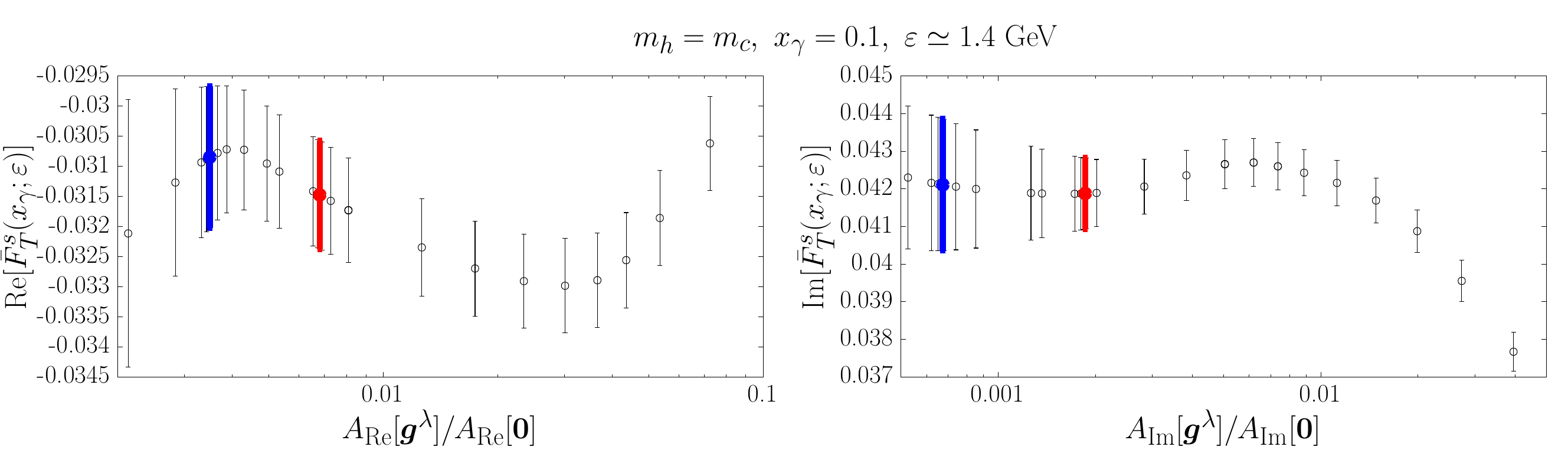}
\caption{\small\it  The real (left panel) and imaginary (right panel) part of the form factor $\bar{F}^{s}_{T}(x_{\gamma}; \varepsilon)$ on the B64 ensemble, for the lowest simulated value of $m_{h}=m_{c}$, and for $\varepsilon \simeq 1.4~{\rm GeV}$ and $x_{\gamma} \simeq 0.1$, as a function of the ratio $A_{\rm I}[\bs{g}^{\lambda}]/A_{\rm I}[\bs{0}]$ indicating the quality of the kernel reconstruction obtained employing different values of $\lambda$. The plot shows an example of our stability analysis. The red and blue data points in both panels correspond to the reconstructions obtained for $\lambda=\lambda^{\filledstar}$ and $\lambda=\lambda^{\filledstar\filledstar}$, respectively.  \label{fig:stability}}
\end{figure}

\begin{figure}
\includegraphics[scale=0.4]{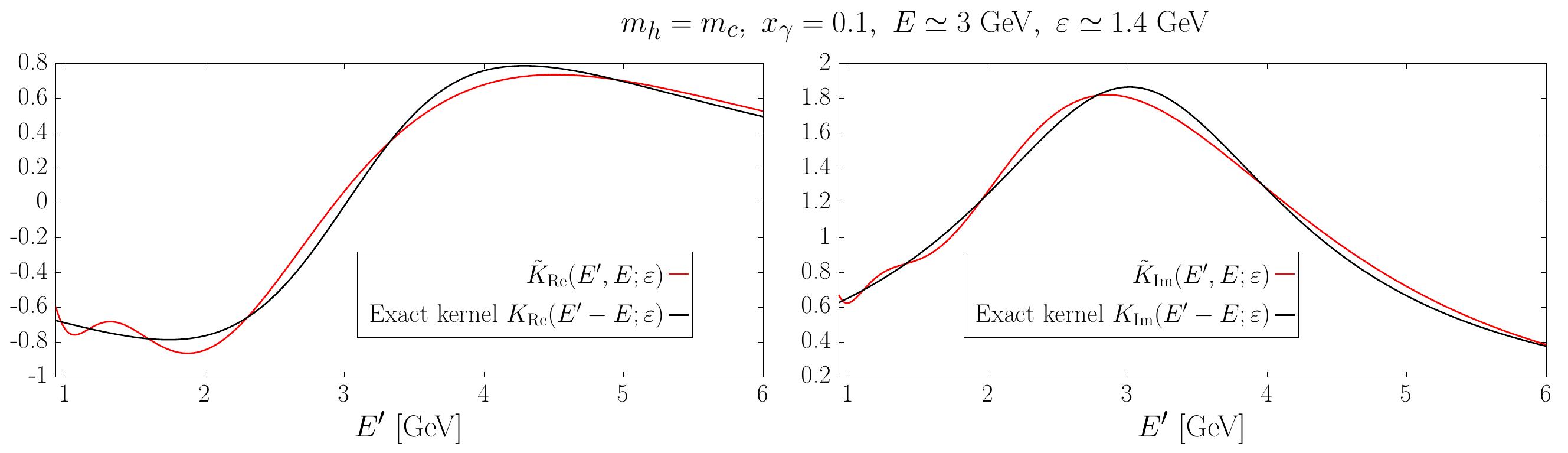}
\caption{\small\it The reconstructed smearing kernels $\widetilde{K}_\mathrm{\rm I}^{}(E', E; \varepsilon)$ obtained using the coefficients $\bs{g_\mathrm{\rm I}^{\lambda^{\filledstar}}}$ employed in the reconstruction of $\bar{F}_{T}^{s}(x_{\gamma}, \varepsilon)$ in Figure~\ref{fig:stability}. The black lines correspond to the real (right panel) and imaginary (left panel) part of the exact kernel function $K(E'-E; \varepsilon)$ of Eq.~(\ref{eq:sinh_kernel}). Th kernels are given in lattice units. \label{fig:kernel_reconstruction}}
\end{figure}

In the plot of Figure~\ref{fig:stability}, we give an example of the stability analysis in the case of the lowest simulated quark mass, and for $x_{\gamma}=0.1$ and $\varepsilon\simeq 1.4~{\rm GeV}$.  
In the figure we show the real and imaginary part of the smeared form factor $\bar{F}_{T}^{s}(x_{\gamma};\varepsilon)$, obtained employing different values of the trade-off parameter $\lambda$. The results are shown as a function of $A_{\rm I}[\bs{g}^{\lambda}]/A_{\rm I}[\bs{0}]$ which is a measure of the goodness of the reconstruction. When the systematic error due to the inexact kernel reconstruction becomes smaller than the statistical uncertainty, the reconstructed smeared form factor is stable under variation of $\lambda$. In this region we determine $\lambda^{\filledstar}$ and $\lambda^{\filledstar\filledstar}$ which are given respectively by the red and blue data points in the figure. The reconstructed kernel functions corresponding to our choice of $\lambda^{\filledstar}$ are then shown in Figure~\ref{fig:kernel_reconstruction}.

We have repeated the analysis for different values of $\varepsilon$ and for all simulated $x_{\gamma}$ and $m_{H_{s}}$. The smearing parameter $\varepsilon$ cannot be however reduced arbitrarily since the uncertainties on $F_{\bar{T}}^{s}(x_{\gamma};\varepsilon)$ generally increase as $\varepsilon$ decreases, and at the same time the reconstruction of the kernel function becomes poorer. The smallest value of $\varepsilon$ for which the errors are still under control is determined by both the statistical uncertainties on $\bar{C}(t,\bs{k})$ and by the size $n_{\rm max}$ of the exponential basis. 

\begin{figure}
    \centering
    \includegraphics[scale=0.4]{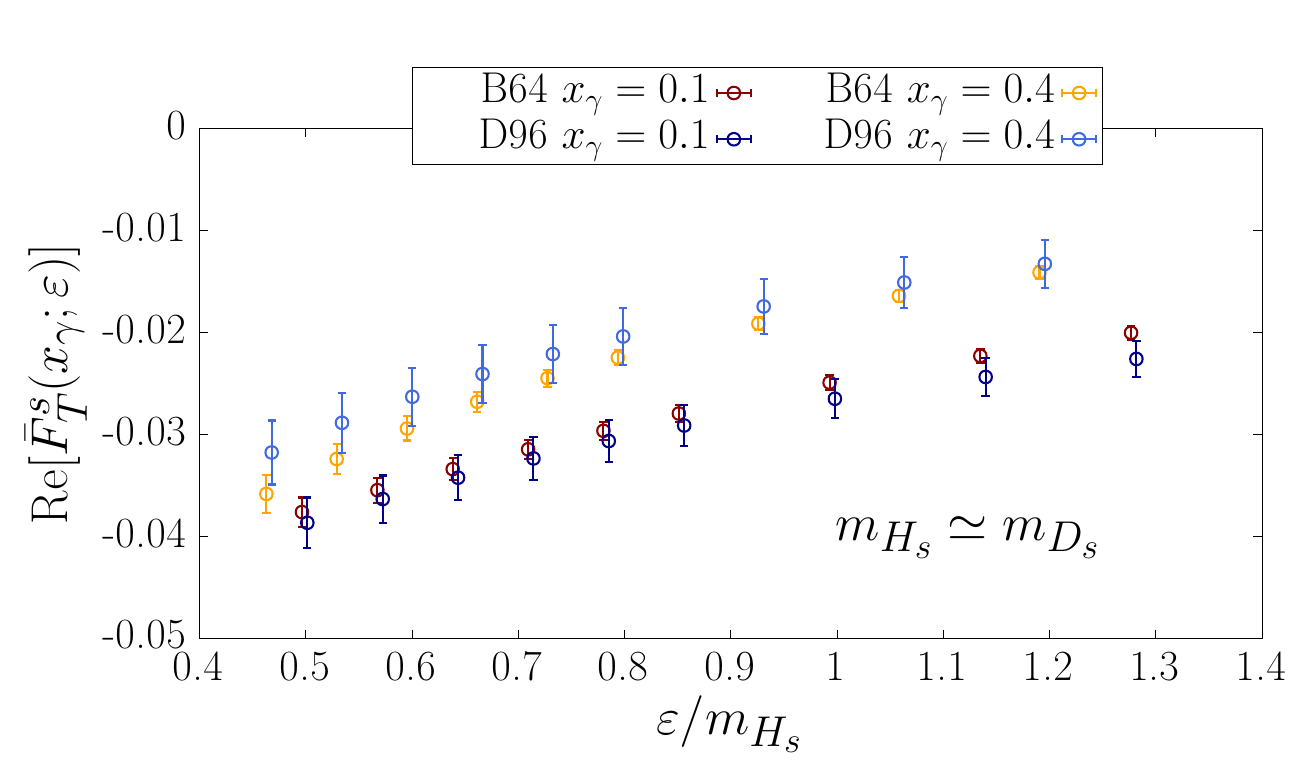}
    \includegraphics[scale=0.4]{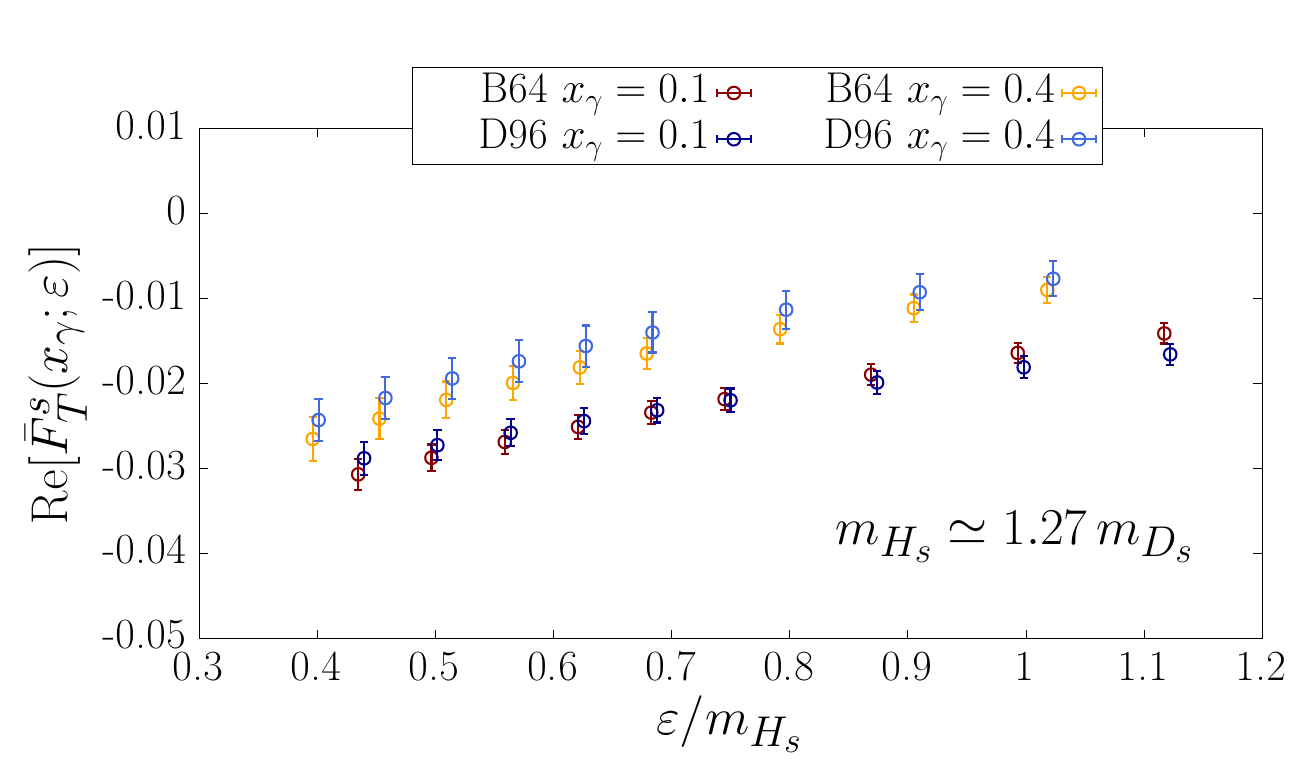} \\
    \includegraphics[scale=0.4]{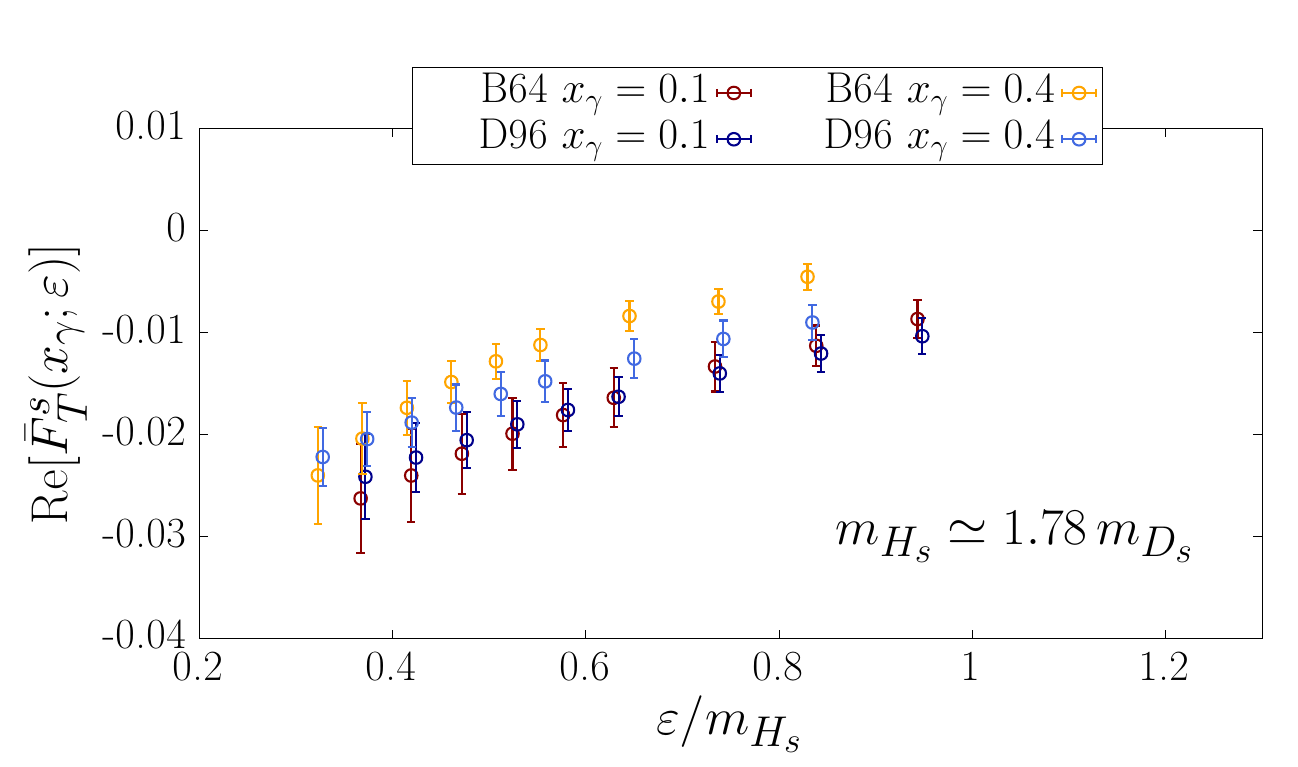}
    \caption{\small\it $\varepsilon$-dependence of the real part of the smeared form factor $\bar{F}_{T}^{s}(x_{\gamma};\varepsilon)$ for the three different masses $m_{H_{s}}\simeq m_{D_{s}}$ (top-left panel), $m_{H_{s}} \simeq 1.27\, m_{D_{s}}$ (top-right panel), and $m_{H_{s}} \simeq 1.78\,m_{D_{s}}$ (bottom panel). For each figure we show the results obtained at $x_{\gamma}=0.1$ and $x_{\gamma}=0.4$ on the B64 and D96 ensembles. \label{fig:eps_dep_RE}  }
\end{figure}

\begin{figure}
    \centering
    \includegraphics[scale=0.4]{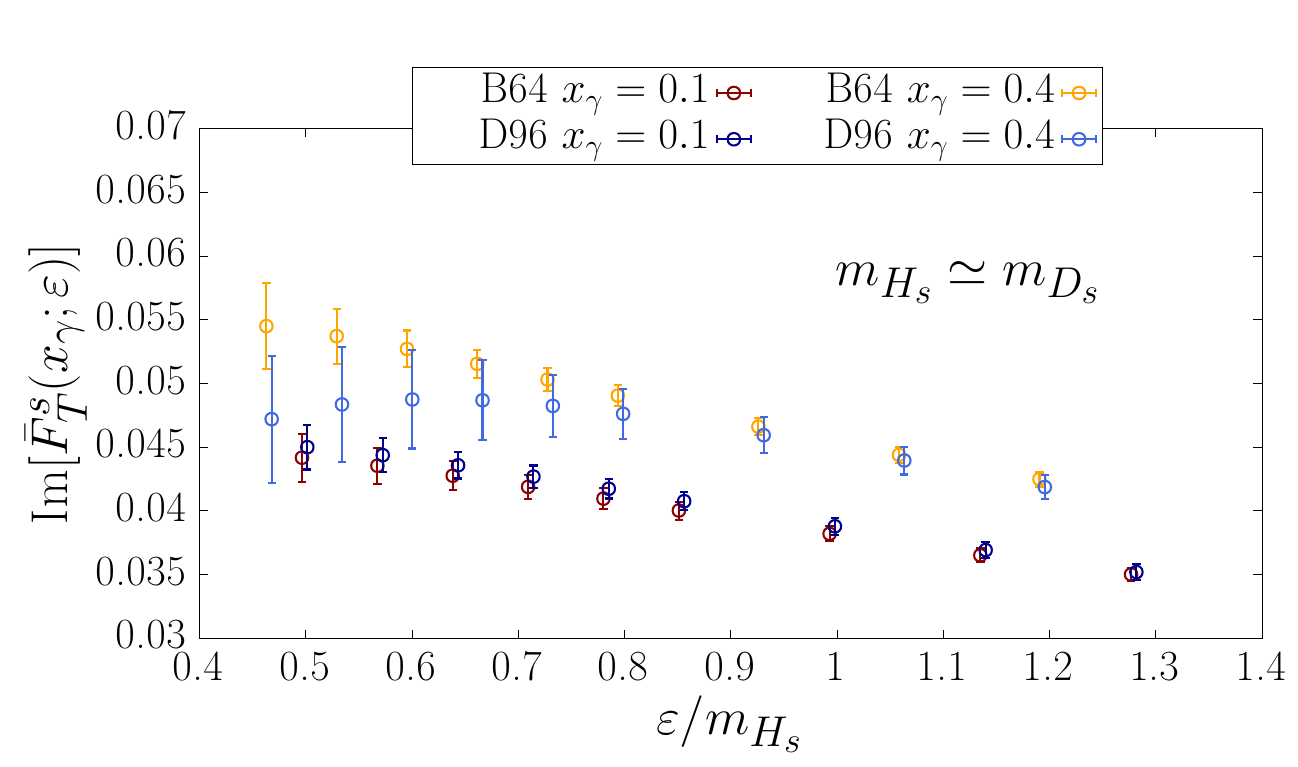}
    \includegraphics[scale=0.4]{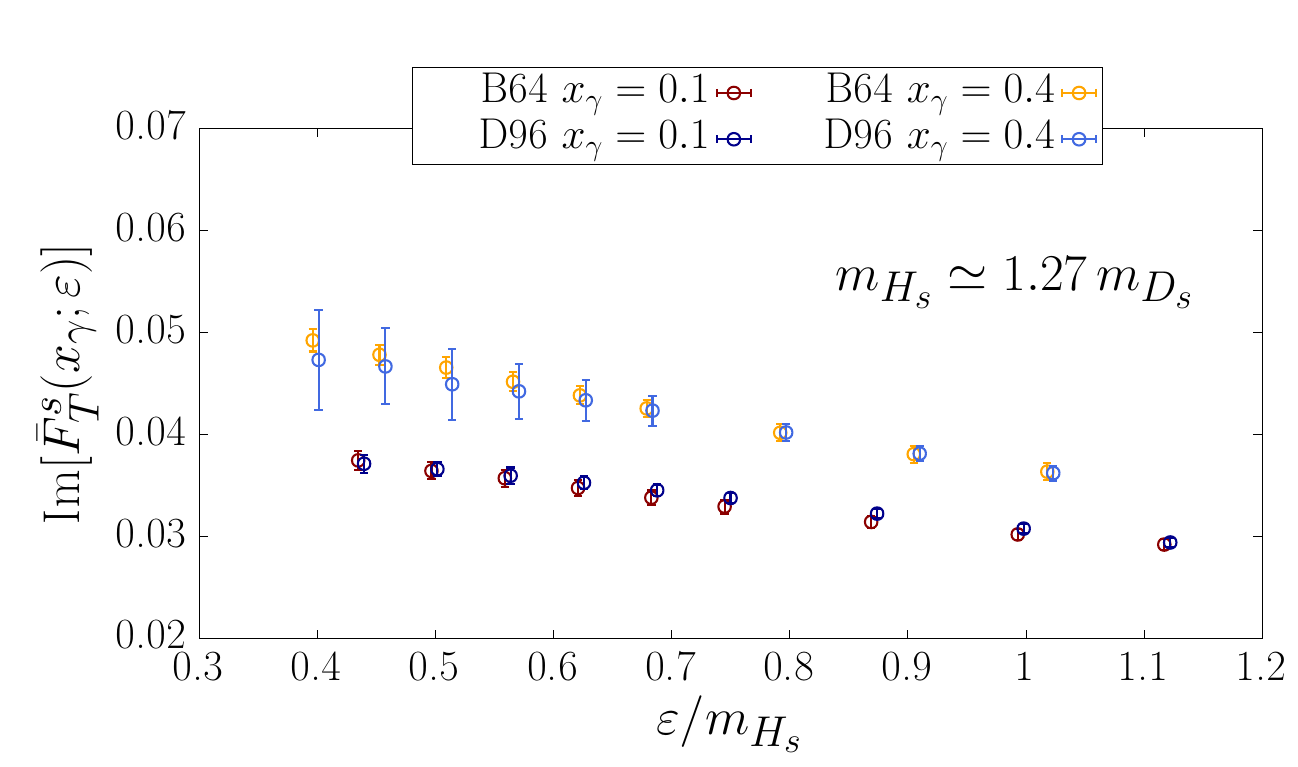} \\
    \includegraphics[scale=0.4]{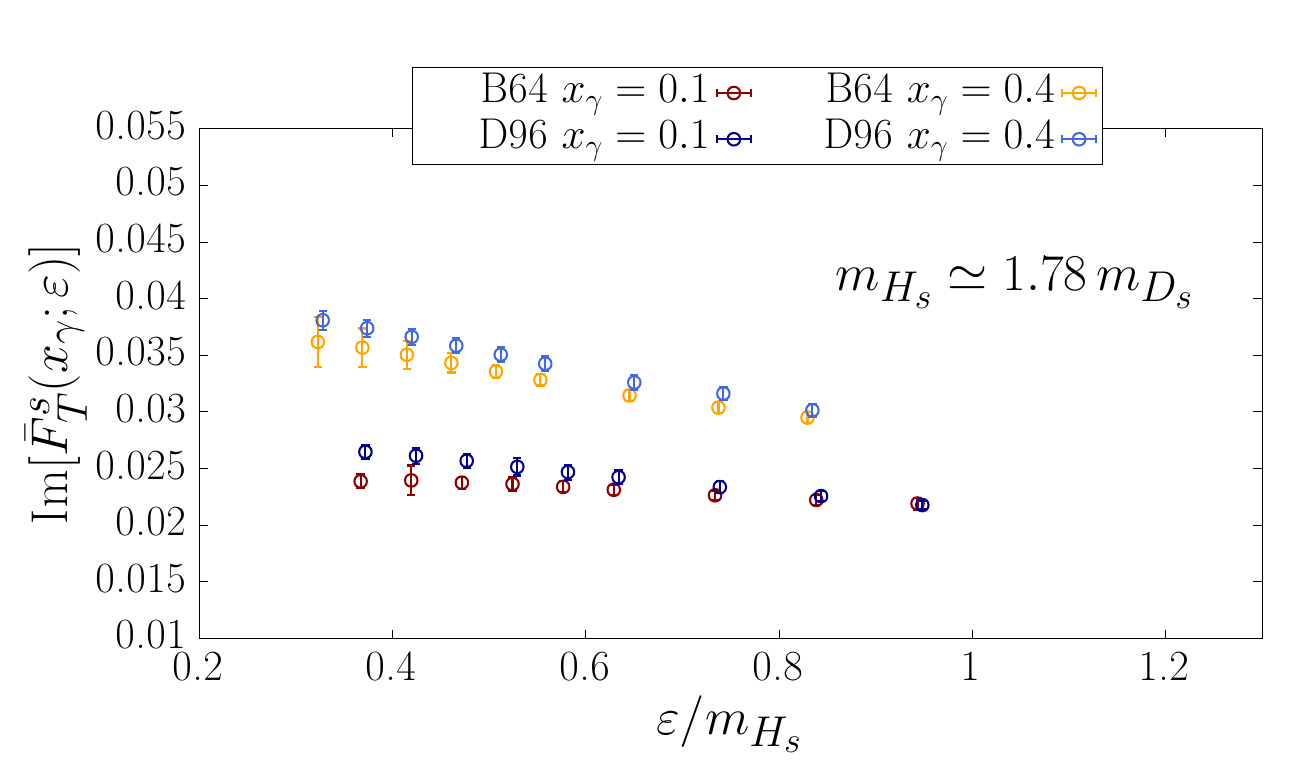}
    \caption{\small\it Same as in Figure~\ref{fig:eps_dep_RE} for the imaginary part of the smeared form factor $\bar{F}_{T}^{s}(x_{\gamma};\varepsilon)$. \label{fig:eps_dep_IM}}    
\end{figure}

In the plots of Figures\,\ref{fig:eps_dep_RE} and \ref{fig:eps_dep_IM} we show the $\varepsilon$-behaviour of the real and imaginary parts of the smeared form factors for the different simulated heavy-strange meson masses $m_{H_{s}}$, and for the smallest ($0.1$) and largest ($0.4$) simulated values of $x_{\gamma}$. In the figure we show the results obtained on both the B64 and D96 ensembles.  
Few comments are in order. First of all the observed cut-off effects are smaller or of the same size of the statistical error for all contributions, with the exception of  $\bar{F}_{T}^{s}(x_{\gamma};\varepsilon)$ for $m_{H_{s}}\simeq 1.78\, m_{D_{s}}$. Such behaviour can be expected since larger masses correspond to higher energies $E(r)$. In addition, we observe that both the real and imaginary part of $\bar{F}_{T}^{b}(x_{\gamma})$ decrease in magnitude as $m_{H_{s}}$ increases. 

We extrapolate the smeared form factor $\bar{F}_{T}^{s}(x_{\gamma};\varepsilon)$ to the continuum limit at fixed $\varepsilon$, $x_{\gamma}$ and $m_{h}$, following the same procedure used for $\bar{F}_{T}^{b}(x_{\gamma})$. Next we perform the $\varepsilon\to 0$ extrapolation at fixed $x_{\gamma}$ and $m_{h}$, which is the most delicate step of the analysis.
As already stated, using the kernel function $K(x;\varepsilon) = (x-i\varepsilon)^{-1}$, the leading corrections to the $\varepsilon=0$ limit are expected to be of the form
\begin{align}
\bar{F}_{T}^{s}(\varepsilon) = \bar{F}_{T}^{s} + A_{1}\,\varepsilon + A_{2}\,\varepsilon^{2} + \mathcal{O}(\varepsilon^{3}) ~,
\end{align}
and in the following we indicate by asymptotic regime, the regime in which the corrections to the vanishing-$\varepsilon$ limit can be described by a low-degree polynomial in $\varepsilon$.
The onset of the asymptotic regime for $H_{2}(E,\bs{k};\varepsilon)$ at a given energy $E$, as discussed in detail in Ref.\,\cite{Frezzotti:2023nun}, depends on the (unknown) typical size, $\Delta (E)$, of the interval around $E$ in which $H_{2}(E,\bs{k})$ is significantly varying. Parametrically one must then have $\varepsilon \ll \Delta(E)$ and at the same time $\varepsilon \gg 1/L$ to avoid large FSEs. Assuming that $H_{2}(E,\bs{k})$ is dominated by the contribution from a single resonance, which at fixed $x_{\gamma}$ and $m_{H_{s}}$ we approximate with a Breit-Wigner distribution centered at $M$ and of width $\Gamma$, i.e. 
\begin{align}
H_{2}(E,\bs{k}) \simeq H^{\rm BW}_{2}(E) = \frac{R}{M-E -i\frac{\Gamma}{2}} \quad\implies\quad H_{2}(E,\bs{k};\varepsilon) \simeq \frac{R}{M-E -i(\frac{\Gamma}{2} + \varepsilon)} ~,
\end{align}
then within this approximation we have $\Delta(E) = \sqrt{ (E-M)^{2} + \Gamma^{2}/4}$\,. In our case the energy $E$ is given for each $x_{\gamma}$ and $m_{H_{s}}$ by $E(r)$ in Eq.\,(\ref{eq:scaling_y}), and it ranges from $E(r)\simeq 3\,{\rm GeV}$ at the lowest mass $m_{H_{s}}=m_{D_{s}}$ to $E(r)\simeq 4\,{\rm GeV}$ at $m_{H_{s}} \simeq 1.78\,m_{D_{s}}$. 
The peaks of the main $s\bar{s}$ resonances, the $\phi$, $\phi(1680)$ and $\phi(2170)$, are at $M\simeq 1 , 1.7 , 2.2~{\rm GeV}$ respectively, with a mild dependence on the value of $|\bs{k}|$. 
In our computations we have $\varepsilon\simeq \mathcal{O}(1\,\rm{GeV})$ or higher and it is not clear whether such values of $\varepsilon$ are in the asymptotic regime, despite an approximate linear scaling in $\varepsilon$ being observed in Figure\,\ref{fig:eps_dep_RE} and~\ref{fig:eps_dep_IM}. 
To account for this source of systematic error we proceed as follows: we first carry out the extrapolation to $\varepsilon=0$ assuming that the observed behaviour is the asymptotic scaling, and perform a polynomial extrapolation in $\varepsilon$ (in practice, as explained below, we perform a quadratic extrapolation in $\varepsilon$, unless there is no signal of a $\varepsilon^2$ term, in which case we perform a linear extrapolation). In addition to the polynomial extrapolation, we follow a second approach, performing the vanishing-$\varepsilon$ extrapolation assuming the following model for the spectral density
\begin{align}
\label{eq:model_spectr}
\rho^{\rm{mod}}(E',\bs{k}) = \sum_{n=1,2,3}\theta(E_{\theta}-E')\frac{R_{n}(\bs{k})\Gamma_{n}}{\left( E_{n}(\bs{k})-E'\right)^{2} + \left(\frac{\Gamma_{n}}{2}\right)^{2}} + \theta(E'-E_{\theta}(\bs{k})) \frac{D(\bs{k})}{(E')^z}, \quad\qquad E_{n}(\bs{k}) = \sqrt{ M_{n}^{2}+ |\bs{k}|^{2}}~.
\end{align}
with $z=1/2, 1$, and where $M_{n}$ and $\Gamma_{n}$ are the mass and the decay width of the $\phi$, $\phi(1680)$ and $\phi(2170)$ resonances respectively for $n=1,2,3$, 
which we take from the PDG\cite{ParticleDataGroup:2020ssz}. The last term in Eq.\,(\ref{eq:model_spectr}) mimicks the  continuum behaviour at large energies starting at a threshold $E_{\theta}(\bs{k}) > E_{3}(\bs{k})$, and is compatible with the physical constraint $\lim_{E'\to\infty}\rho(E',\bs{k})=0$.
\footnote{The spectral density must vanish in the infinite-energy limit in order to have a finite ${\rm Re}[H_{2}(E,\bs{k};\varepsilon)]$.}. 
For any fixed values of $x_{\gamma}$ and $m_{H_{s}}$, our model for the spectral density contains four free real parameters: the three amplitudes $R_{1}(\bs{k})$, $R_{2}(\bs{k})$, $R_{3}(\bs{k})$, and the threshold energy $E_{\theta}(\bs{k})$. 
The parameter $D(\bs{k})$ is instead determined by the requirement that the spectral density is continuous at $E'=E_{\theta}(\bs{k})$. The smeared hadronic amplitude $H_{2}^{\rm mod}(E,\bs{k};\varepsilon)$ associated with $\rho^{\rm mod}(E',\bs{k})$ is then given by the convolution of $\rho^{\rm mod}(E',\bs{k})$ with the kernel function $K(E-E';\varepsilon)$. Finally, we can use $H_{2}^{\rm mod}(E',\bs{k};\varepsilon)$ to obtain the corresponding model smeared form factor,
\begin{align}
\bar{F}_{T}^{s, {\rm mod}}(x_{\gamma};\varepsilon) &\equiv -\frac{H^{\rm mod}(E,E(r),\bs{k};\varepsilon)}{k_{z}}~,\\[8pt]
H^{\rm mod}(E,E(r),\bs{k};\varepsilon) &\equiv H_{2}^{\rm mod}(E(r),\bs{k};\varepsilon) - H_{2}^{\rm mod}(0,\bs{k};0) + H_{1}^{\rm sub}(E,\bs{k}) + H(0,\bs{k})~.
\end{align}

\begin{figure}
\includegraphics[scale=0.4]{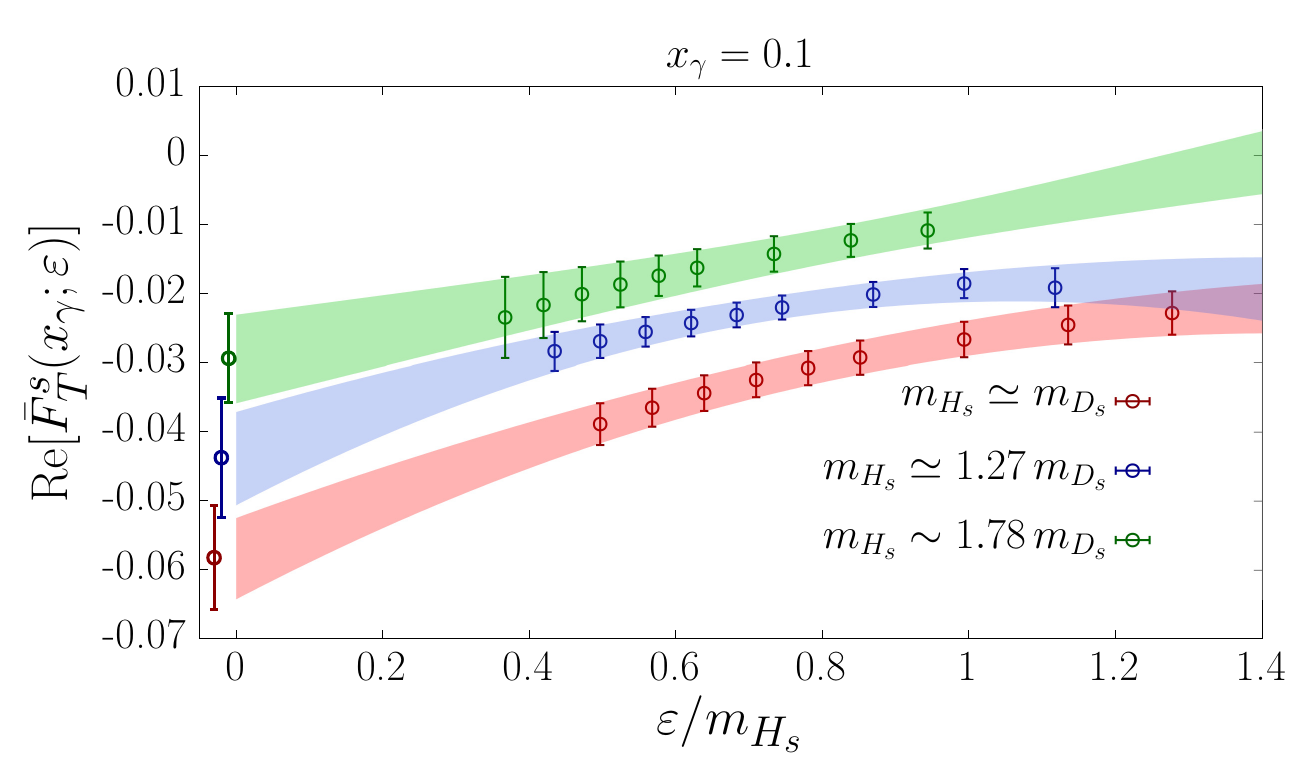}
\includegraphics[scale=0.4]{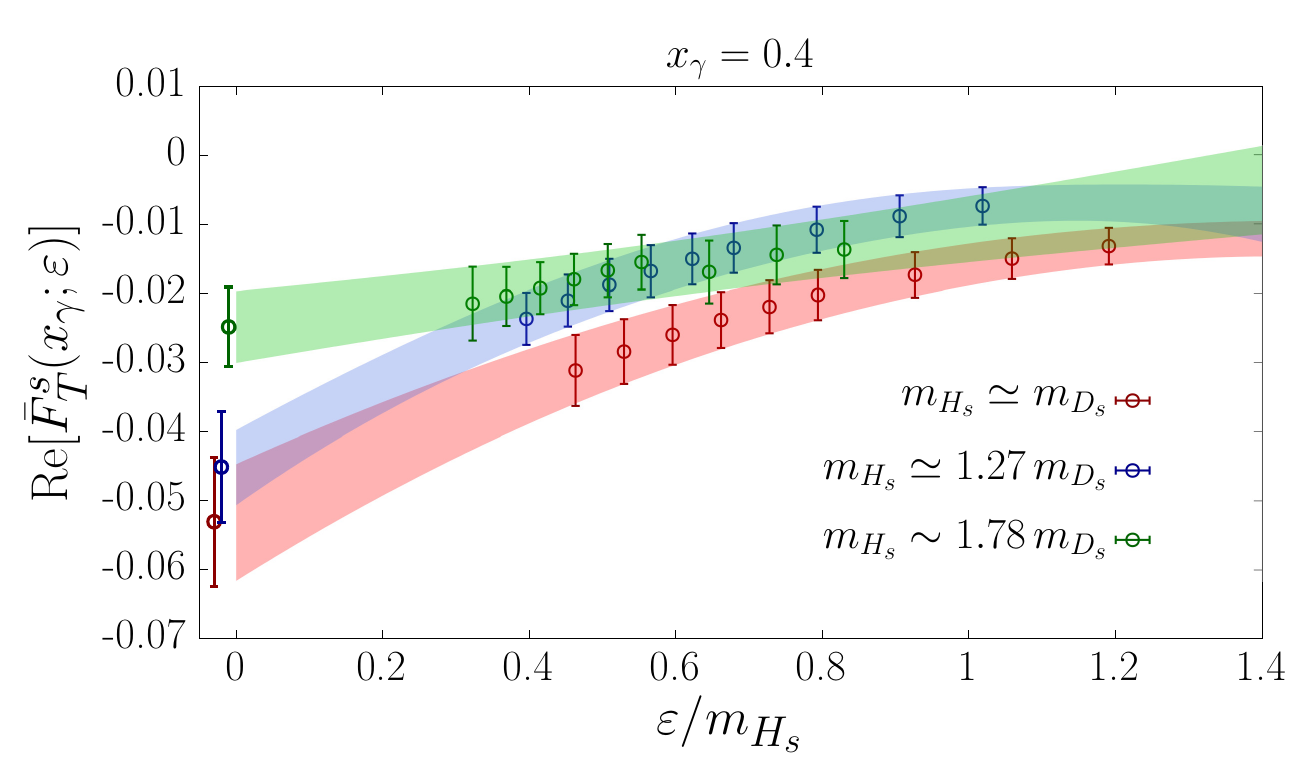}\\
\includegraphics[scale=0.4]{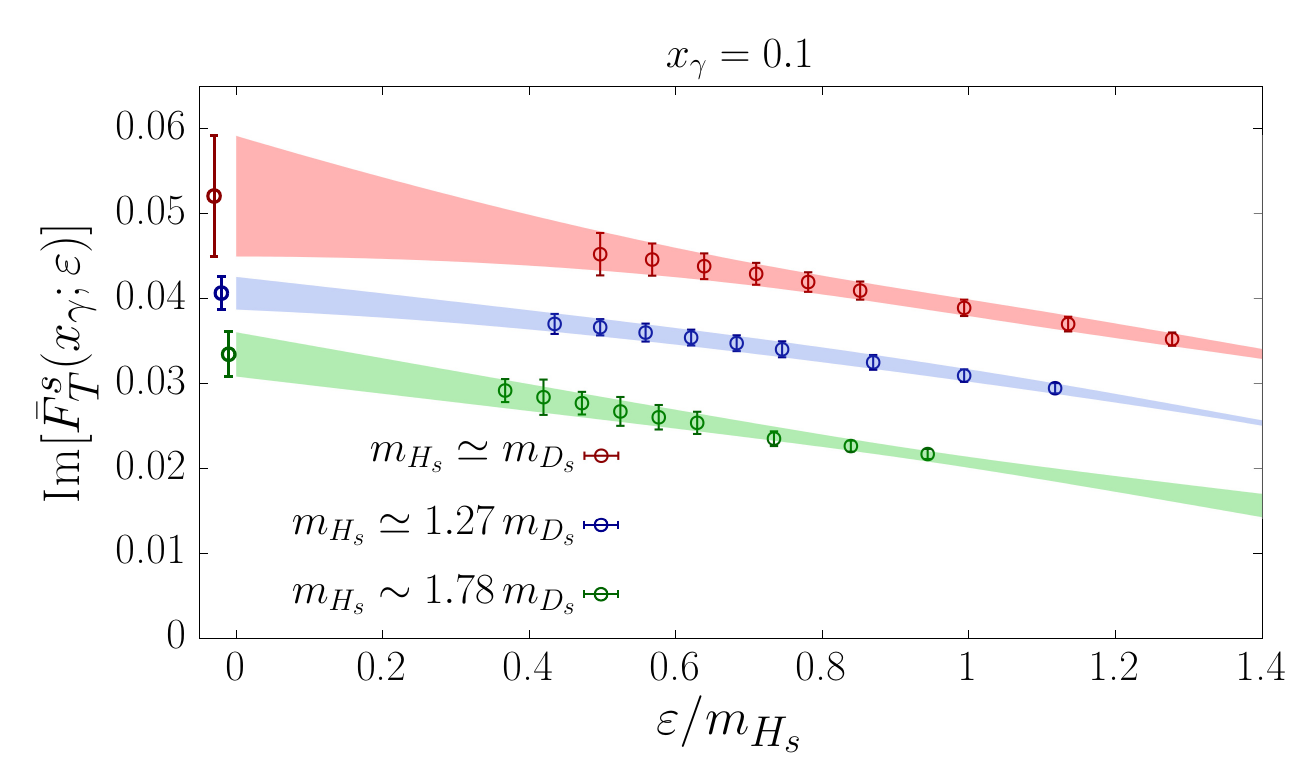}
\includegraphics[scale=0.4]{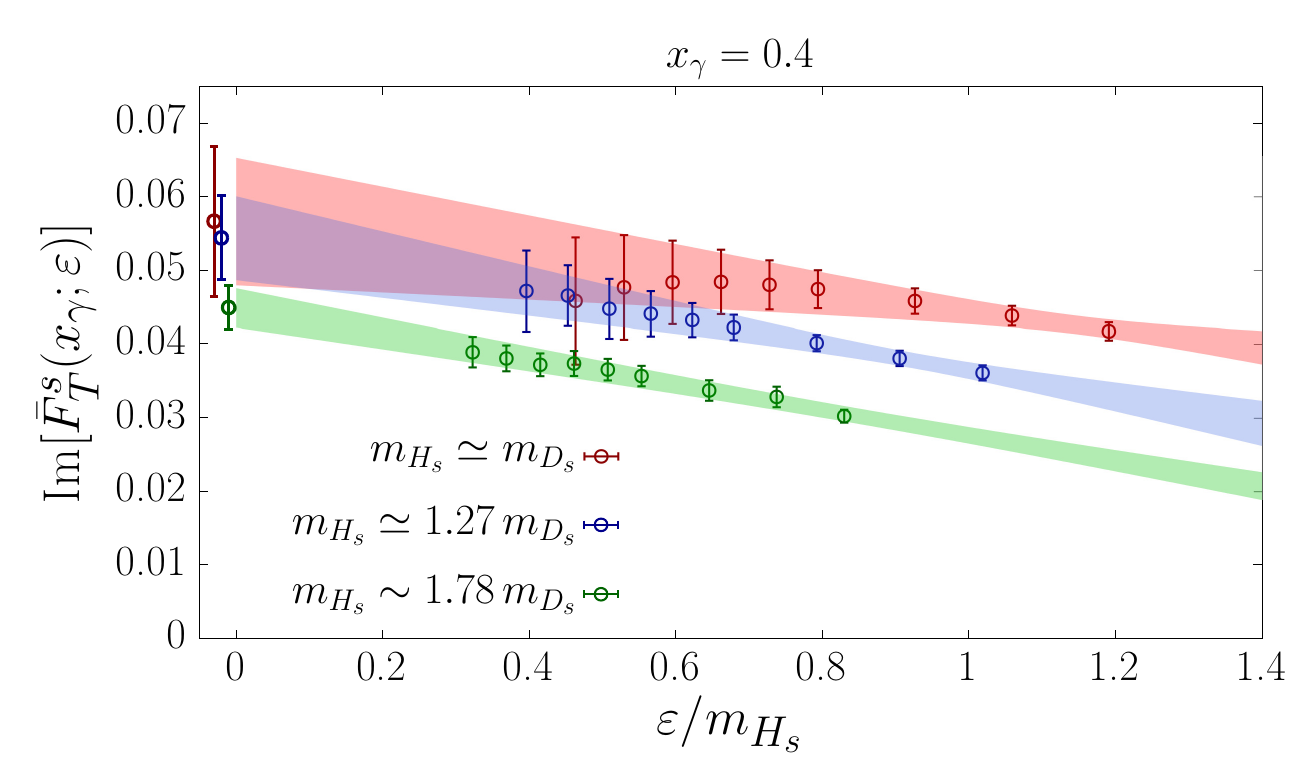}
\caption{\small\it The extrapolation in $\varepsilon$ of the real (top panels) and imaginary (bottom panels) components of the smeared form factors, using the Ansatz in Eq.\,(\ref{eq:eps_extrapolation}). For those cases when no $\varepsilon^{2}$-dependence is visible in the data, we have set $A_{2}=0$. The different colours correspond to the three different values of $m_{H_{s}}$ and the corresponding bands are the best-fit functions obtained from the fits with all values of $\varepsilon$ included. The left and right panels correspond to $x_{\gamma}=0.1$ and $x_{\gamma}=0.4$ respectively. The data points at $\varepsilon = 0$ (which in the figures are slightly shifted horizontally for better visualization) correspond to our final results after including the systematic error, determined as discussed in the text. \label{fig:eps_extr}}
\end{figure}

In Figure~\ref{fig:eps_extr} we show the results of the polynomial extrapolation to vanishing $\varepsilon$, which we perform separately for each $x_{\gamma}$ and $m_{H_{s}}$.
In the figure we show, as an illustration, the results obtained for $x_{\gamma}=0.1$ and $0.4$. The extrapolation has been carried out using the following Ansatz for the smeared form factor
\begin{align}
\label{eq:eps_extrapolation}
\bar{F}_{T}^{s}(\varepsilon) = A + A_{1}\,\varepsilon + A_{2}\,\varepsilon^{2}~,
\end{align}
where $A, A_{1}$ and $A_{2}$ are complex-valued free fit parameters, which are different for each $x_{\gamma}$ and $m_{H_{s}}$. 
In order to avoid overfitting, for those cases when there is no signal of $\varepsilon^{2}$-dependence visible in the data, we have set $A_{2}=0$. We have minimized a $\chi^{2}$-function  constructed without taking into account the correlation between the values of the smeared form factors corresponding to different $\varepsilon$, since they are too correlated, and the resulting correlation matrix is ill-conditioned. 
In this way, the reduced $\chi^{2}$ resulting from the minimization, which is always well below one, cannot be taken as a quantitative measure of the quality of the fit. 
To estimate the systematic error of the polynomial extrapolation, we have also performed for all the cases a second fit, linear in $\varepsilon$, using only the five smallest simulated values of $\varepsilon$. 
Any statistically-significant deviation from the results obtained in the fit with all simulated values of $\varepsilon$ included (i.e. those whose resulting best-fit functions are given by the coloured bands of Figure\,\ref{fig:eps_extr}) is then added as a systematic error. 
In Figure\,\ref{fig:eps_extr} the data points at $\varepsilon = 0$ correspond to our final results from the polynomial extrapolation, after including the systematic error determined following the procedure described above. 
As is clear from the figure the real and imaginary part of the form factor $\bar{F}_{T}^{s}(x_{\gamma})$ decrease in magnitude as the mass $m_{H_{s}}$ increases, and already for $m_{H_{s}} \simeq 1.78\, m_{D_{s}}$ they are both one order of magnitude smaller than the tensor form factors $F_{TV}$ and $F_{TA}$ determined in the previous section. 

As discussed above, since the simulated values of $\varepsilon$ may not be in the asymptotic regime, we have also performed  non-polynomial extrapolations in $\varepsilon$ using the model in Eq.\,(\ref{eq:model_spectr})
with $z=1/2$ and $z=1$. The fits have been performed imposing Gaussian priors on all four fit parameters. The prior corresponding to  the amplitude $R_{1}(\bs{k})$ of the $\phi$ resonance is
\begin{align}
R_{1}(\bs{k}) = \langle R_{1}(\bs{k}) \rangle \, ( 1 \pm 0.1)\,,
\end{align}
where $\langle R_{1}(\bs{k})\rangle$ has been estimated from an effective residue analysis of the correlation function $\bar{C}(t,\bs{k})$ at large times. We use $\langle R_{1}(\bs{k}) \rangle/|\bs{k}| = -0.045, -0.042, -0.040~{\rm GeV}$, respectively for $m_{H_{s}}/m_{D_{s}} \simeq 1, 1.27, 1.78$. The priors corresponding to the two amplitudes $R_{2}(\bs{k})$ and $R_{3}(\bs{k})$ are instead
\begin{align}
R_{2}(\bs{k}) = \frac{\langle R_{1}(\bs{k}) \rangle}{2} \, ( 1 \pm 1)~,\qquad R_{3}(\bs{k}) =  \frac{\langle R_{1}(\bs{k}) \rangle}{2} \, (  1 \pm 1)\,,
\end{align}
i.e. we assume that, within one standard deviation, they are
at most of the same size as the contribution from the $\phi$ resonance. Finally the prior on the threshold parameter $E_{\theta}(\bs{k})$ is
\begin{align}
E_{\theta}(\bs{k}) = E_{3}(\bs{k}) + (0.5 \pm 0.5)~{\rm GeV}~, 
\end{align}
i.e. we assume that the onset of the perturbative regime occurs at an energy which is $\mathcal{O}(\Lambda_{\rm QCD})$ larger than that of the heaviest known $s\bar{s}$ resonance.
We have found that both values of $z$ describe the data well at all masses and $x_{\gamma}$.\footnote{Additionally, we have tried to fit our data using the model in Eq.\,(\ref{eq:model_spectr}) with $z=2$, but found that it does not provide an equally good description of the smeared form factor.} 

For the real part of $\bar{F}^{s}_{T}(x_{\gamma})$, the results of the extrapolation to $\varepsilon=0$, obtained using the model with either $z=1/2$ and $1$ are in good agreement with those of the polynomial extrapolation. For the imaginary part, instead, we find that the model results (in particular for $z=1/2$) are significantly smaller than those obtained from the polynomial extrapolation.
\begin{figure}
    \centering
    \includegraphics[scale=0.4]{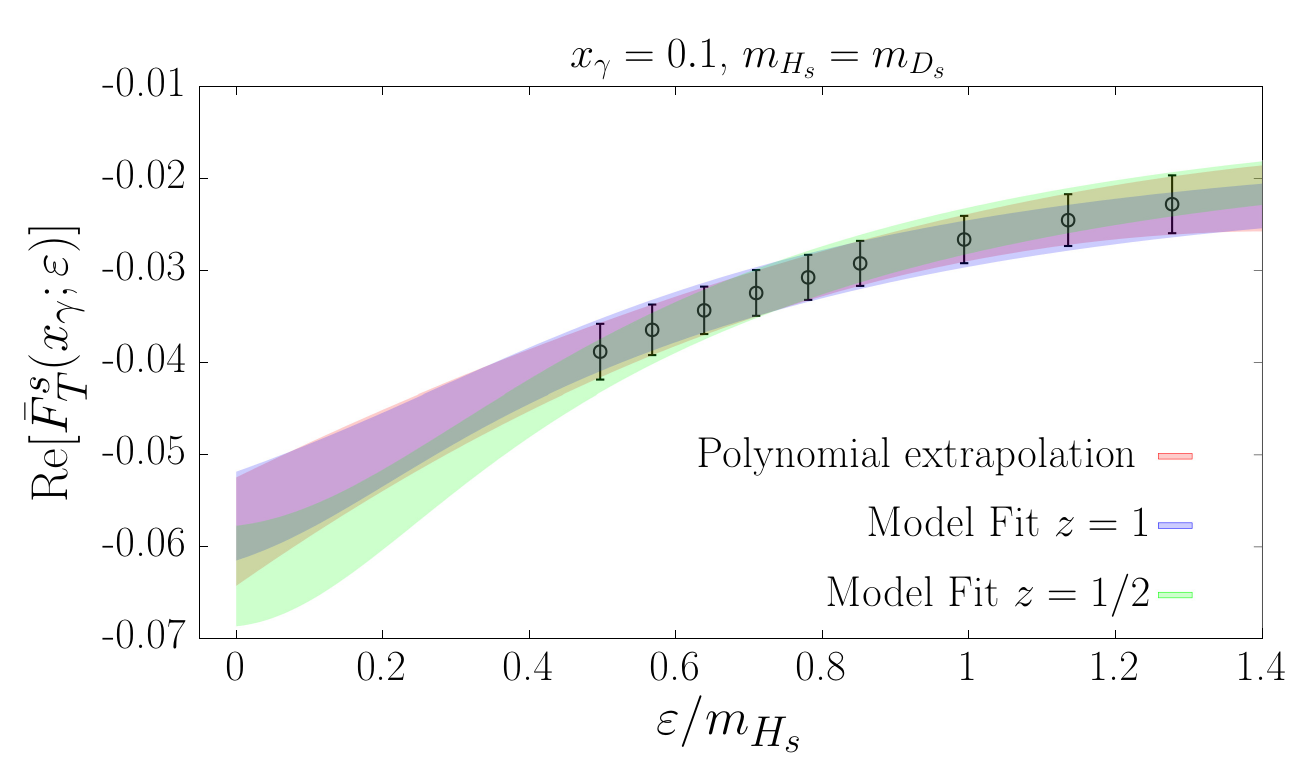}
    \includegraphics[scale=0.4]{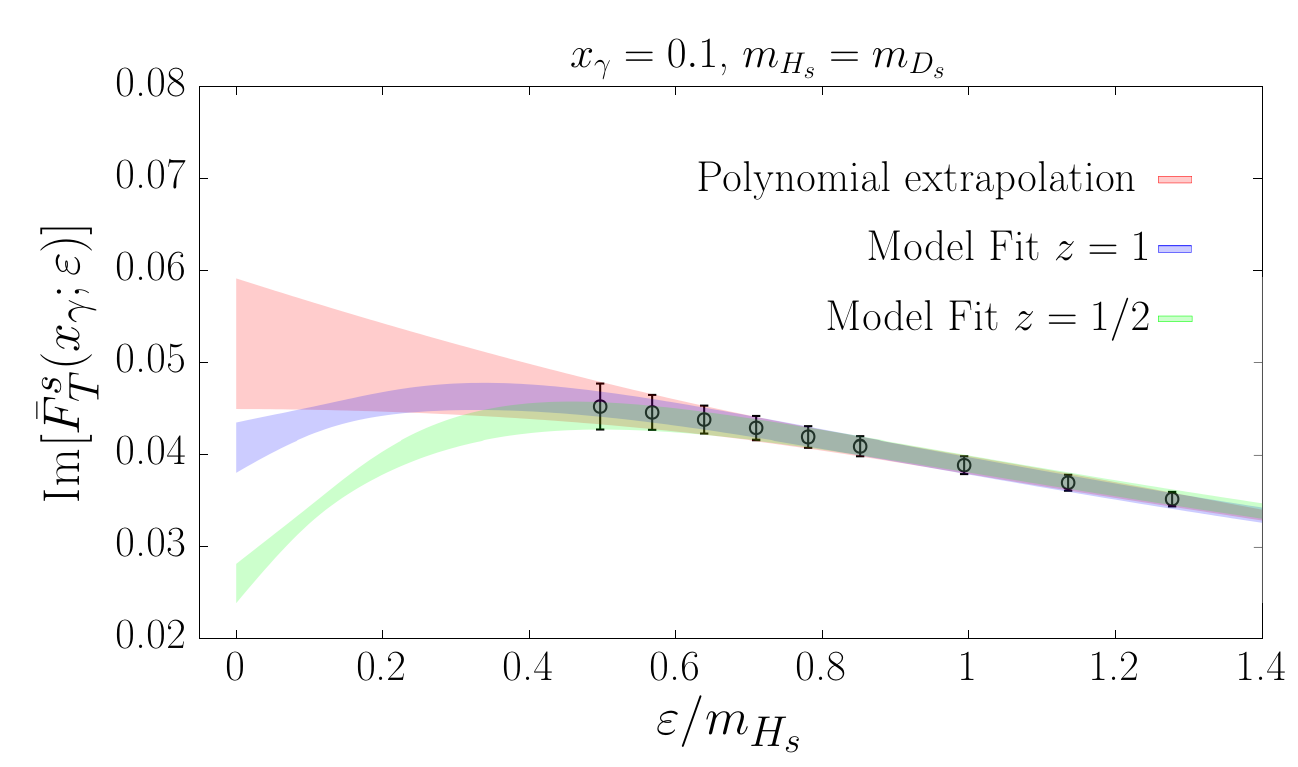} \\
    \includegraphics[scale=0.4]{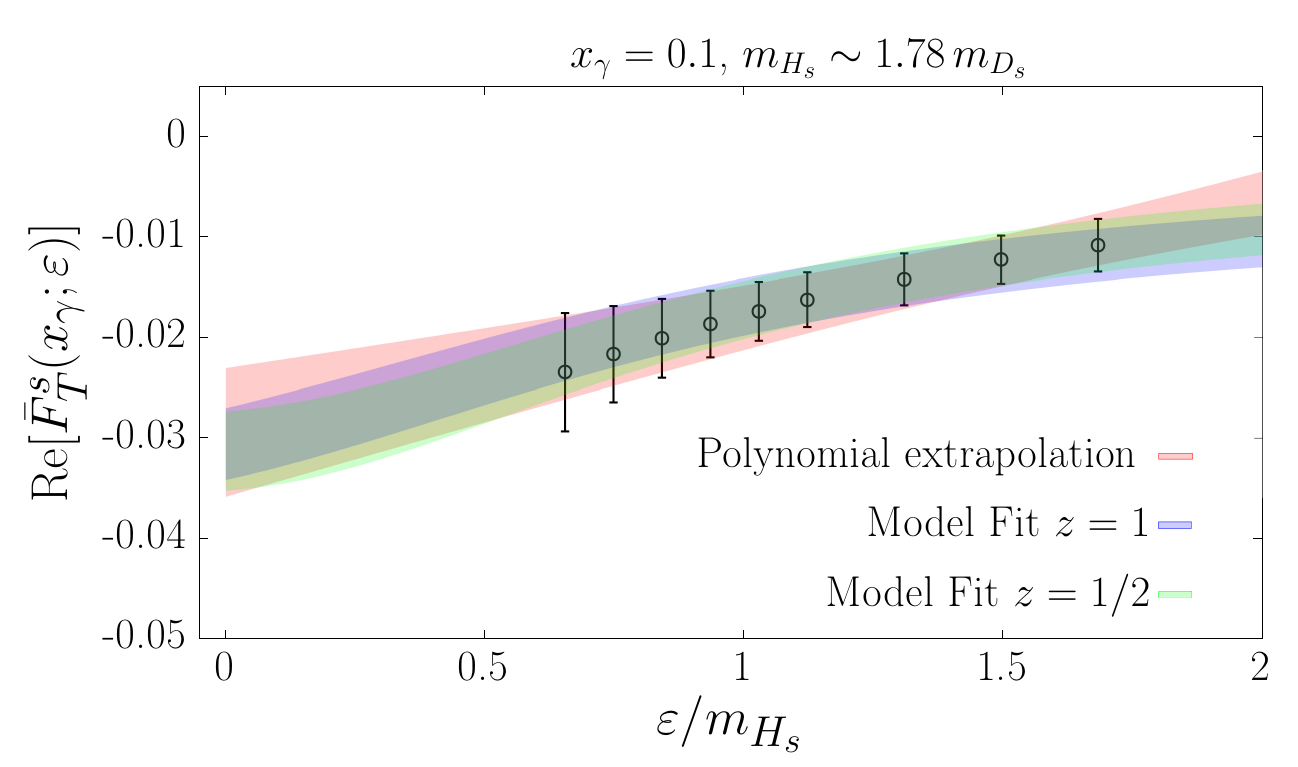}
    \includegraphics[scale=0.4]{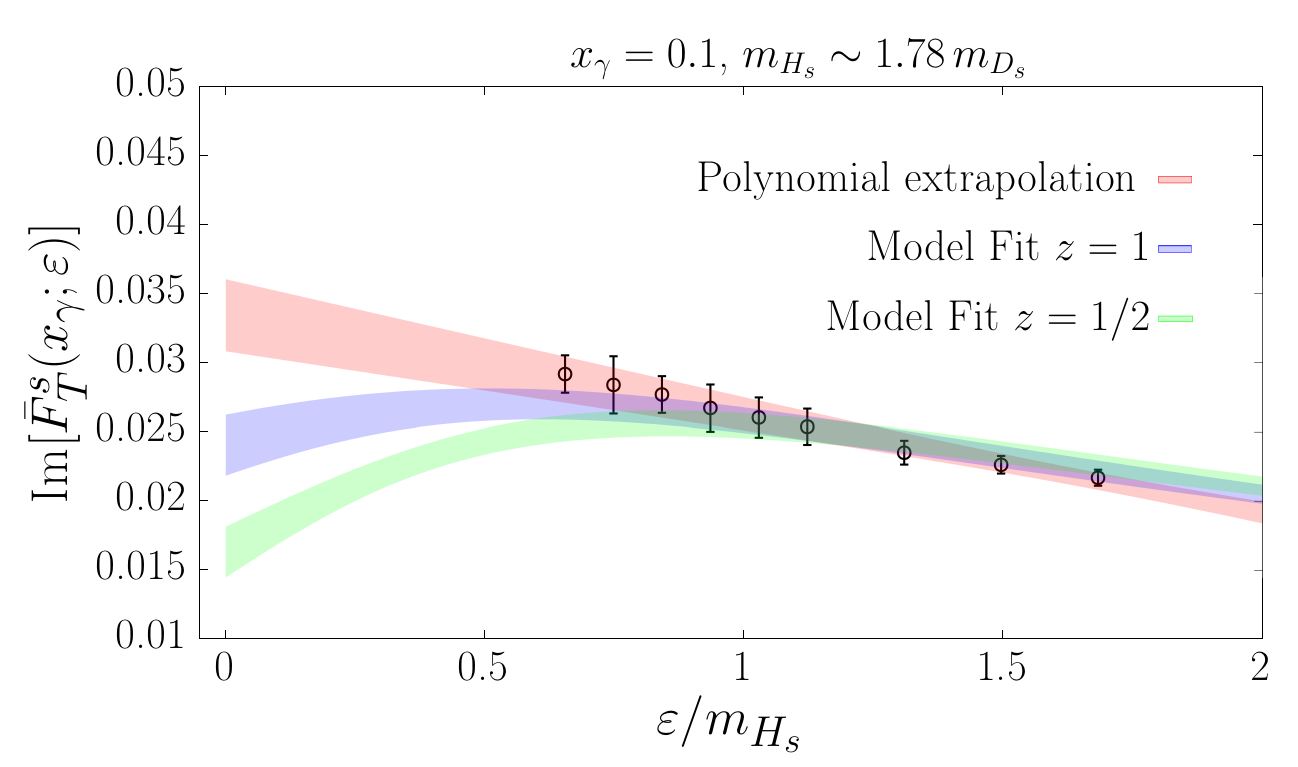}
    \caption{\small\it Comparison between the results from the polynomial extrapolation (given by the red band), and those obtained fitting the smeared form factor using the model in Eq.\,(\ref{eq:model_spectr}) with $z=1/2$ (given by the green band) and $z=1$ (given by the blue band). The top panels correspond to $x_{\gamma}=0.1$, $m_{H_{s}}=m_{D_{s}}$, while the bottom panels to $x_{\gamma}=0.1$, $m_{H_{s}}\sim 1.78 m_{D_{s}}$. \label{fig:comparison_model}}
    \end{figure}
The comparison is shown in Figure~\ref{fig:comparison_model}, for the case $x_{\gamma}=0.1$ and for both $m_{H_{s}} = m_{D_{s}}$ and $m_{H_{s}} \sim 1.78 m_{D_{s}}$. 
All the other cases are very similar. 
The lower value obtained for ${\rm Im}[\bar{F}_{T}^{s}(x_{\gamma})]$ assuming the model $\rho^{\rm mod}(E',\bs{k})$ for the spectral density, could be due to the fact that at the simulated values of $\varepsilon$, the imaginary part of the kernel function still has a sizeable overlap with the peaks of the nearby resonances (e.g. the $\phi(2170)$ resonance). 
In this case, the imaginary part of the smeared form factor, ${\rm Im}[\bar{F}_{T}^{s}(x_{\gamma}; \varepsilon)]$, is expected to decrease in value for smaller, presently unreachable, values of $\varepsilon$. This behaviour cannot be captured by the polynomial extrapolation, but is in-built in our model for the spectral density. To have a realistic estimate of the systematic uncertainty for $\bar{F}_{T}^{s}(x_{\gamma}; \varepsilon)$, we average the results of the polynomial and model-dependent extrapolation with $z=1/2$, and include a systematic error equal to half the difference between the two results.

\subsection{Extrapolating $\bar{F}_{T}$ to the mass of the physical $B_{s}$-meson}
\label{sec:extr_barF_T}
We now discuss the extrapolation of the form factor
$\bar{F}_{T}$ to the mass of the physical $B_{s}$-meson. We start from the $b$-quark contribution $\bar{F}_{T}^{b}(x_{\gamma})$, which we determined for three values of the heavy-strange meson mass $m_{H_{s}}/m_{D_{s}} \simeq 1, 1.27$ and $1.78$ as shown in the right panel of Figure\,\ref{fig:FT_b}. To perform the mass extrapolation we make use of a phenomenological VMD-inspired Ansatz to describe the combined $x_{\gamma}$ and $m_{H_{s}}$ dependence of the form factor. At the physical $B_{s}$ mass point, the form factor $\bar{F}_{T}^{b}(x_{\gamma})$ is expected to be dominated by the contributions of neutral, $J^{P} = 1^{-}$, $b\bar{b}$ resonance states (e.g. $\Upsilon(1S)$, $\Upsilon(2S)$, $\Upsilon(3S)$, $\ldots$). The contribution to the form factor $\bar{F}_{T}^{b}(x_{\gamma})$ of a given vector resonance state containing an heavy quark $(h)$ and an heavy anti-quark ($\bar{h}$), 
for a given value of $m_{H_{s}}$, and approximating the resonance as a stable state, is of the form
\begin{align}
\label{eq:pole_contr_FTb}
\bar{F}_{T, n}^{b}(x_{\gamma}) =  \frac{ q_{b} \, f_{n}\, m_{n}\, g_{n}^{+}(0)}{ E_{n}(E_{n} + E_{\gamma} - m_{H_{s}})} + \textrm{regular terms}~,
\end{align}
where $E_{n} = \sqrt{ m_{n}^{2} + E_{\gamma}^{2}}$, and $m_{n}$ and $f_{n}$ are respectively the mass and the electromagnetic decay constant of the vector resonance. The latter is defined through
\begin{align}
\langle 0 | \bar{h}\gamma^{\mu}h | n(-\bs{k},\eta) \rangle = \eta^{\mu}\, f_{n} \, m_{n}~,
\end{align}
where $| n(-\bs{k}, \eta) \rangle$ is the vector resonance state with given polarization $\eta$.
The coupling $g_{n}^{+}(0)$ is defined by~\cite{Kozachuk:2017mdk}
\begin{align}
\langle n(-\bs{k}, \eta) | \,\bar{s}\sigma^{\mu\nu} h\, | \bar{H}_{s} (\bs{0})\rangle = i\eta^{*}_{\beta}\epsilon^{\mu \nu\beta\gamma}\left[ g_{n}^{+}(p_{\gamma}^{2})(p+q_{n})_{\gamma} + g_{n}^{-}(p_{\gamma}^{2})(p-q_{n})_{\gamma}\right] +   i g_{n}^{0}(p_{\gamma}^{2})(\eta^{\ast}\cdot p)\epsilon^{\mu \nu\beta\gamma}p_{\beta}q_{n,\gamma} ~,
\end{align}
with $q_{n}=(E_{n}, -\bs{k})$, $p_{\gamma}=p-q_{n}$\,. In the heavy-quark limit, $m_{h}\to \infty$, the following scaling laws hold
\begin{align}
f_{n} \propto \frac{1}{\sqrt{ m_{h}}} +\ldots \propto \frac{1}{\sqrt{ m_{H_{s}}}} +\ldots ~, \qquad \frac{m_{n}}{m_{H_{s}}} = 2 + \frac{\Lambda_{T}^{n}}{m_{H_{s}}} + \ldots~,
\end{align}
where $\Lambda_{T}^{n} \simeq \mathcal{O}(\Lambda_{\mathrm{QCD}})$, and the ellipses represent higher-order terms in the heavy-quark expansion. In light of the previous relations $\bar{F}_{T, n}^{b}(x_{\gamma})$ can be further approximated with
\begin{align}
\bar{F}_{T,n}^{b}(x_{\gamma}) = \frac{q_{b}}{m_{H_{s}}} \frac{f_{n}\,g_{n}^{+}(0)}{  1+ \frac{x_{\gamma}}{2} + \frac{\Lambda^{n}_{T}}{m_{H_{s}}} }\left( 1 + \mathcal{O}\left( x_{\gamma}, \frac{\Lambda_{\rm QCD}}{m_{H_{s}}}\right)\right)~.
\end{align} 
Our strategy to extrapolate $\bar{F}_{T}^{b}(x_{\gamma})$ to the physical mass $m_{B_{s}}$, consists in approximating the tower of contributions of type~\ref{eq:pole_contr_FTb}, with a single effective pole. This is achieved through the use of the following fit Ansatz for the combined mass and $x_{\gamma}$ dependence of the form factor
\begin{align}
\label{eq:ansatz_FTb}
\bar{F}_{T}^{b}(x_{\gamma}, m_{H_{s}}) = \frac{1}{m_{H_{s}}} \frac{A + B\,x_{\gamma}}{1 + \frac{x_{\gamma}}{2} + \frac{\Lambda_{T}}{m_{H_{s}}}}~, 
\end{align}
where $A,B$ and $\Lambda_{T}$ are free fit parameters, and the effective-pole mass is $m_{\rm eff} = 2m_{H_{s}} + \Lambda_
{T}$. We assume that $A$ and $B$ are mass-independent, which is consistent with our data, as illustrated below\footnote{The Ansatz in Eq.\,(\ref{eq:ansatz_FTb}) assumes that
$g_{n}^{+} \propto \sqrt{ m_{H_{s}}}$ for which however we are not aware of any formal proof in the HQET.}. 

\begin{figure}
\includegraphics[scale=0.5]{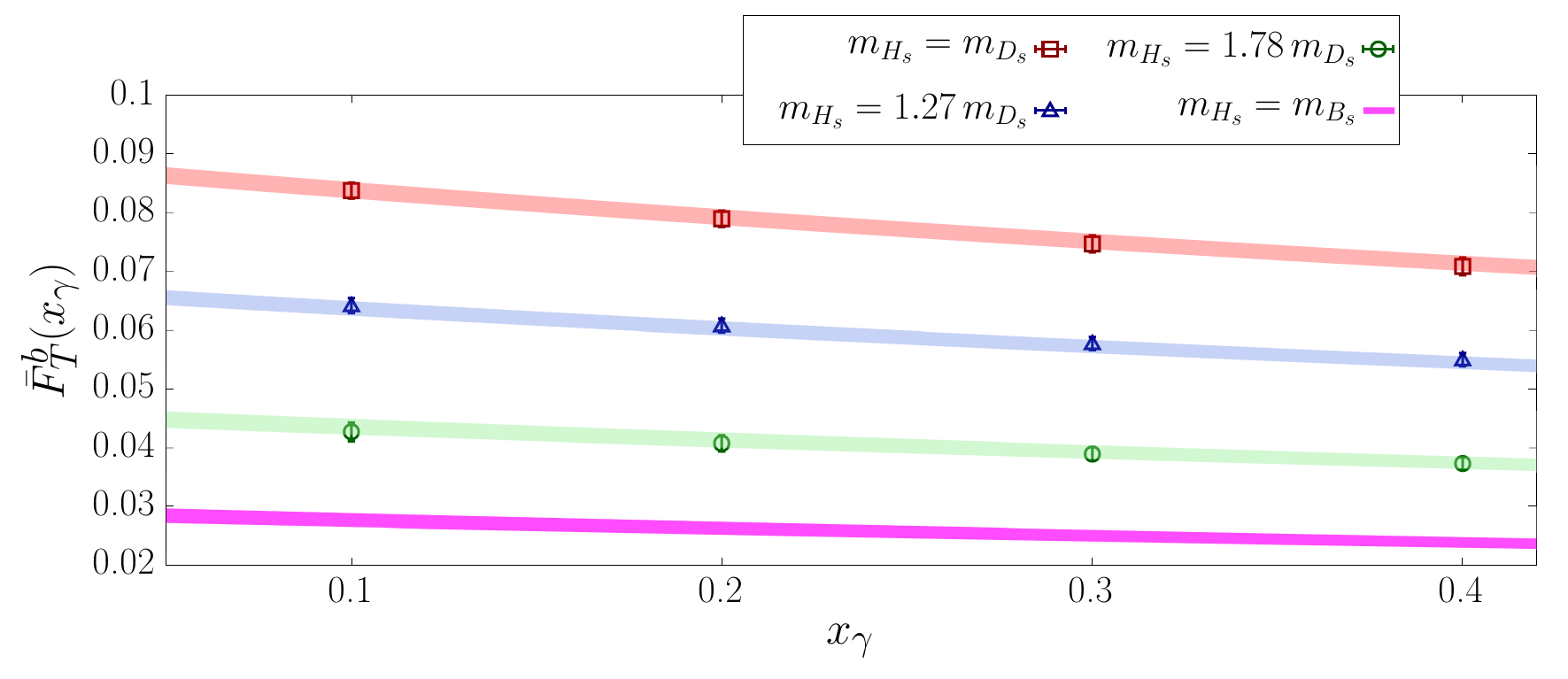}
\caption{\small\it Extrapolation to $m_{B_s}$ of the form factor $\bar{F}_{T}^{b}(x_{\gamma})$ using the Ansatz in Eq.\,(\ref{eq:ansatz_FTb}). The red, blue and green bands correspond to the best fit-function obtained for $m_{H_{s}}/m_{D_{s}} \simeq 1, 1.27, 1.78$, respectively. The magenta band correspond to our final result for $\bar{F}_{T}^{b}$ at $m_{H_{s}}=m_{B_{s}}$.                    \label{fig:FT_b_mass_extr}}
\end{figure}

Using the Ansatz in Eq.~(\ref{eq:ansatz_FTb}) we have performed a combined fit of the $x_{\gamma}$ and $m_{H_{s}}$ dependence of our data. The total number of measurement entering the $\chi^{2}$ minimization is $12$, and the number of fit parameters is $3$. The $\chi^{2}/{\rm dof}$ resulting from the minimization is very good and well below unity, although in this case we have employed an uncorrelated $\chi^{2}$ function, since we find that the covariance matrix is ill-conditioned. To illustrate the quality of the fit, we show in Figure\,\ref{fig:FT_b_mass_extr} the best-fit functions obtained from the global fit. 
As is clear from the figure, the VMD-inspired Ansatz  perfectly captures both the mass and $x_{\gamma}$ behaviour of our data. The resulting value of the parameter $\Lambda_{T}$ is
\begin{align}
\Lambda_{T} = -0.32\,(11)~{\rm GeV}~,
\end{align}
which implies that at the physical $B_{s}$ mass the effective pole is located at $m_{\rm eff} = 10.4~(1)\,{\rm GeV}$, i.e. around the mass of the $\Upsilon(2S)$ resonance. To check for possible systematic errors due to the mass extrapolation we have repeated the fit setting to zero the parameter $B$ in Eq.~(\ref{eq:ansatz_FTb}). However, we did not find significant differences within uncertainties. The magenta band in Figure~\ref{fig:FT_b_mass_extr} correspond to our final result for $\bar{F}_{T}^{b}(x_{\gamma})$ at $m_{H_{s}}=m_{B_{s}}$. This contribution turns out to be small compared to the tensor form factors $F_{TV}(x_{\gamma})$ and $F_{TA}(x_{\gamma})
$ described in Section~\ref{sec:FF_extr}, which are more than one order of magnitude larger. In Table~\ref{tab:FT_b_extr} we give our results for $\bar{F}_{T}^{b}$ extrapolated at the physical mass $m_{B_{s}}$, for the four simulated values of $x_{\gamma}$.

\begin{table}
\setlength{\tabcolsep}{7pt}
\renewcommand{\arraystretch}{1.5}
\begin{tabular}{|c|c|c|c|c|}
\hline
 &  \multicolumn{4}{c|}{$x_{\gamma}$} \\ \hline
 & $0.1$ & $0.2$ & $0.3$ & $0.4$ \\ \hline
 $\bar{F}_{T}^{b}$ & $0.028(1)$ & $0.026(1)$ & $0.025(1)$ & $0.0239(9)$     \\ \hline
\end{tabular}
\caption{\small\it Our results for the form factor $\bar{F}_{T}^{b}$ extrapolated to the physical mass $m_{B_{s}}$, for the four simulated values of $x_{\gamma}=0.1,0.2,0.3,0.4$.  \label{tab:FT_b_extr}}
\end{table}

\begin{figure}
\includegraphics[scale=0.4]{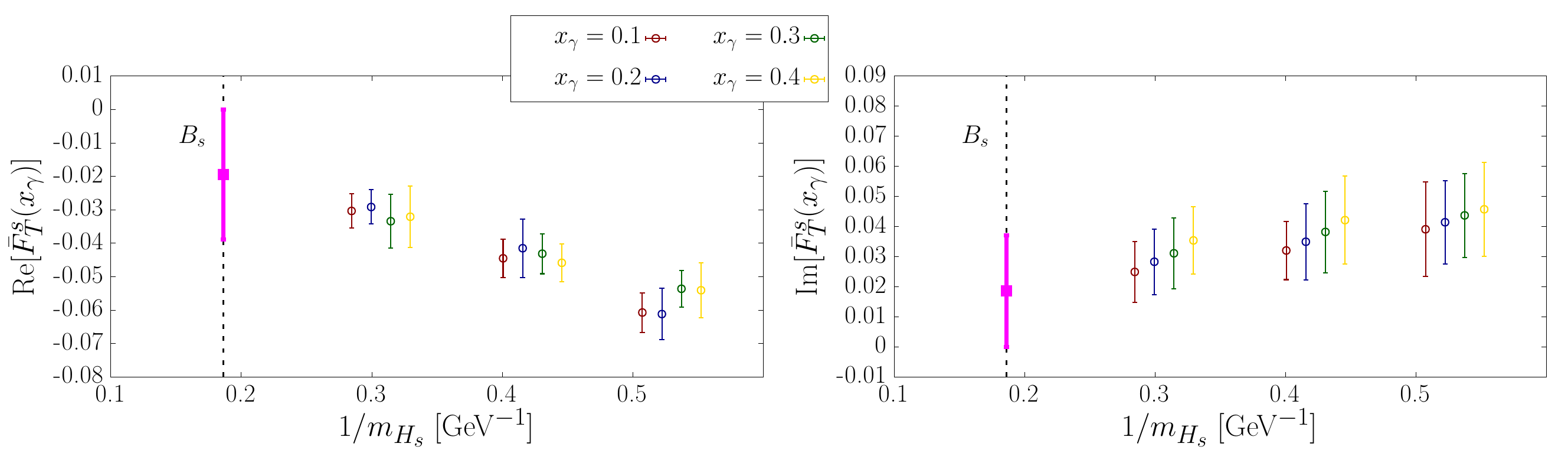}
\caption{\small\it Mass dependence of the real (left panel) and imaginary (right panel) part of $\bar{F}_{T}^{s}(x_{\gamma})$ for the four simulated values of $x_{\gamma}$. The data points corresponding to the four simulated values of $x_{\gamma}$ have been slightly shifted horizontally for better visualization. The data point in magenta corresponds to our final (conservative) estimate for $\bar{F}_{T}^{s}(x_{\gamma})$ at the physical mass $m_{H_{s}}=m_{B_{s}}$. \label{fig:FT_s_mass_extr}}
\end{figure}

We now turn into the discussion of the mass extrapolation of $\bar{F}_{T}^{s}(x_{\gamma})$. In this case the uncertainties are significantly larger than those affecting $\bar{F}_{T}^{b}(x_{\gamma})$. 
Moreover, after including the systematic errors due to the $\varepsilon\to 0$ extrapolation, only a very smooth $x_{\gamma}$-dependence is visible in the data within uncertainties. This is shown in Figure\,\ref{fig:FT_s_mass_extr}, where we plot  the real and imaginary part of $\bar{F}_{T}^{s}(x_{\gamma})$ as a function of $1/m_{H_{s}}$ for all simulated values of $x_{\gamma}$.
As is clear from the figure, both the real and imaginary part of $\bar{F}_{T}^{s}(x_{\gamma})$ decrease as $m_{H_{s}}$ increases. 
This is expected since the form factor vanishes in the static limit. In this case, to have a conservative error estimate, we take the results at the largest simulated mass $m_{H_{s}}\simeq 1.78\, m_{D_{s}}$ as a bound for the value of the form factor at the physical point, $m_{H_{s}}=m_{B_{s}}$. Since no clear $x_{\gamma}$-dependence is visible in the data, we associate the same central value and errors to all $x_{\gamma}$. Our final determination is
\begin{align}\label{eq:FbarTs}
{\rm Re}[\bar{F}_{T}^{s}(x_{\gamma})] = -0.019(19)~, \qquad {\rm Im}[\bar{F}_{T}^{s}(x_{\gamma})] = 0.018(18)\,,
\end{align}
which correspond to the data points in magenta in the panels of Figure\,\ref{fig:FT_s_mass_extr}.


\section{The $B_{s}\to \mu^{+}\mu^{-}\gamma$ decay rate}
\label{sec:rate}
The doubly-differential cross section for the $B_{s}\to\mu^{+}\mu^{-}\gamma$ decay can be written as
\begin{align}\label{eq:d2Gamma}
\frac{{\rm d}^2\Gamma}{ {\rm d} x_{\gamma}\, {\rm d}(\cos{\theta})} = \frac{{\rm d}^2\Gamma^{({\rm PT})}}{{\rm d} x_{\gamma}\, {\rm d}(\cos{\theta})} + \frac{{\rm d}^2\Gamma^{({\rm INT})}}{{\rm d} x_{\gamma}\, {\rm d}(\cos{\theta})} +\frac{{\rm d}^2\Gamma^{({\rm SD})}}{{\rm d} x_{\gamma}\, {\rm d}(\cos{\theta})}~,
\end{align}
where the superscript {\rm(PT)}
refers to the point-like contribution (which becomes negligible for large $x_\gamma$), {\rm (SD)} labels the structure-dependent contribution and {\rm (INT)} labels the
contribution from the interference between the point-like and
structure-dependent terms in the amplitude. 
In Eq.\,(\ref{eq:d2Gamma}) $\theta$ is the angle between the three-momenta of the $\mu^+$ and the photon in the rest frame of the $\mu^+\mu^-$ pair. 
Recalling that $x_\gamma=2p\cdot k/m_{B_s}$, $\cos\theta$ is written in terms of Lorentz invariant quantities in Eq.\,(\ref{eq:thetadef}) below. We now present the expressions for the three terms on the right-hand side
of Eq.\,(\ref{eq:d2Gamma}), neglecting the contributions from $\mathcal{O}_{1-6,8}$ except for the charming penguin diagram in Figure~\ref{fig:penguin} which is included in the effective Wilson coefficient $C_{9}^{\rm eff}$.

The structure-dependent contribution, which depends quadratically on the form factors,  can be written as~\cite{Kozachuk:2017mdk}
\footnote{In Ref.\,\cite{Kozachuk:2017mdk} the authors chose $\hat{s}=(p-k)^2/m_{B_s}^2=1-x_\gamma$ and $\hat{t}=(p-p_1)^2/m_{B_s}^2$, where $p_1$ is the four-momentum of the $\mu^+$, as the independent variables. We choose $x_\gamma$ and $\cos\theta$ and $J(x_\gamma)$ is the Jacobian relating the two sets of variables.}
\begin{align}
\label{eq:SD_contr}
\frac{{\rm d}^2\Gamma^{({\rm SD})}}{{\rm d} x_{\gamma}\, {\rm d}(\cos{\theta})}\, &=\, 
\frac{G^2_F\,\alpha^3_{\rm em}\, m^5_{B_s}}{2^{10}\,\pi^4}\, 
\left |V_{tb}\, V^*_{ts} \right |^2
 J(x_{\gamma})\left [ 
x_{\gamma}^2\, B_0\left (x_{\gamma}\right ) +
x_{\gamma}\,\xi\left (x_{\gamma},\hat t\right )\,\tilde B_1\left (x_{\gamma}\right )
+  
\xi^2\left (x_{\gamma},\hat t\right )\,\tilde B_2\left (x_{\gamma}\right ) 
\right ], 
\end{align}
where $\hat{t}=(p-p_1)^2/m_{B_s}^2$ and $p_1$ is the momentum of the $\mu^+$ lepton.
The function $\xi(x_\gamma,\hat{t})$ is defined as
\begin{equation}
    \xi( x_{\gamma}, \hat{t}) = x_{\gamma} + 2\hat{m}_{\mu}^{2} - 2\hat{t}\,,
\end{equation}
where $\hat{m}_\mu=m_\mu/m_{B_s}$
\footnote{$\xi( x_{\gamma}, \hat{t})$ can also be written as $\hat{u}-\hat{t}$, where $\hat{u}=(p-p_2)^2/m_{B_s}^2$ and $p_2$ is the momentum of the $\mu^-$.}, in terms of which 
the angle $\theta$ is given by 
\begin{equation}
  \label{eq:thetadef}
\cos\theta\, =\,\frac{ \xi(x_{\gamma},\hat{t}) }
          {x_{\gamma}\,\sqrt{1\, -\, 4\hat m^2_{\mu}/(1-x_{\gamma})}} \,. 
\end{equation}
The Jacobian $J(x_\gamma)$ is
\begin{equation}
   J(x_{\gamma}) = \frac{x_{\gamma}}{2}\sqrt{ 1 - \frac{4\hat{m}_{\mu}^{2}}{1-x_{\gamma}}}~.
\end{equation}
Finally the functions $B_0(x_\gamma)$,
$\tilde{B}_1(x_\gamma)$ and 
$\tilde{B}_2(x_\gamma)$ are as follows:
\begin{align}
\qquad\qquad B_0\left (x_{\gamma}\right )\, &=\,
    \left ( 1 - x_{\gamma}\, +\, 4\hat m^2_{\mu} \right )
    \left (F_1\left(x_{\gamma}\right )\, +\, F_2\left(x_{\gamma}\right )\right)\, -\, 
    8\hat m^2_{\mu}\,\left |C_{10}(\mu)\right |^2
    \left (F^2_V\left(x_{\gamma}\right )\, +\, F^2_A\left(x_{\gamma}\right )\right)~ , 
   \\[12pt]
\qquad\qquad \tilde B_1\left (x_{\gamma}\right )\, &=\,
     8\,\big[
            (1-x_{\gamma})\, F_V(x_{\gamma})\, F_A(x_{\gamma})\, 
                {\rm Re}\left (C^{{\rm eff}\, *}_{9}(\mu, x_{\gamma})\, C_{10}(\mu)\right )\, 
        \nonumber\\[12pt]
&   +\,  
                \hat m_b\, F_V(x_{\gamma})\, {\rm Re}\left (C^*_{7}(\mu)\, 
                F^{{\rm eff}\ast}_{TA}(x_{\gamma})\, C_{10}(\mu)\right )
              + \hat m_b\, F_A(x_{\gamma})\, {\rm Re}\left (C^*_{7}(\mu)\, 
                F^{{\rm eff}\ast}_{TV}(x_{\gamma})\, C_{10}(\mu)\right ) 
        \big]~, \\[12pt]
 \qquad\qquad\tilde B_2\left (x_{\gamma}\right )\, &=\,(1-x_{\gamma}) \, 
    \left (F_1\left(x_{\gamma}\right )\, +\, F_2\left(x_{\gamma}\right )\right)~,
\end{align}
 where $\hat{m}_b=m_b/m_{B_s}$,
 \begin{align}
 \qquad\qquad F_1\left (x_{\gamma}\right )\, &=\, 
   \left (\left |C^{\rm eff}_{9}(\mu, x_{\gamma}) \right |^2\, +\,
   \left |C_{10}(\mu) \right |^2  \right)F^2_V(x_{\gamma})
   \, +\,
   \left (\frac{2\hat m_b}{1-x_{\gamma}}\right )^2
   \left |C_{7}(\mu)\, F^{\rm eff}_{TV}(x_{\gamma})\right |^2\\[8pt]
&+\,\frac{4\hat m_b}{1-x_{\gamma}}\, F_V(x_{\gamma})\, 
   {\rm Re}\left (C_{7}(\mu)\, F^{\rm eff}_{TV}(x_{\gamma})\, C^{{\rm eff}\, *}_{9}(\mu, x_{\gamma}) 
     \right ),\nonumber\\[12pt]
 \qquad\qquad F_2\left (x_{\gamma}\right )\, &=\, 
   \left (\left |C^{\rm eff}_{9}(x_{\gamma}, \mu) \right |^2\, +\,
   \left |C_{10}(\mu)\right |^2  \right)F^2_A(x_{\gamma})\, +\,
   \left (\frac{2\hat m_b}{1-x_{\gamma}}\right )^2
   \left |C_{7}(\mu)\,  F^{\rm eff}_{TA}(x_{\gamma})\right |^2\\[12pt]
&+\,\frac{4\hat m_b}{1-x_{\gamma}}\, F_A(x_{\gamma})\, 
   {\rm Re}\left (C_{7}(\mu)\, F^{\rm eff}_{TA}(x_{\gamma})\, C^{{\rm eff}\, *}_{9}(\mu, x_{\gamma}) 
     \right )
\end{align}
and
\begin{align}
\label{eq:bar_FTV}
F^{\rm eff}_{TV}(x_{\gamma})&=F_{TV}(x_{\gamma})+\bar{F}_{T}(x_{\gamma})~, \\
\label{eq:bar_FTA}
{F}^{\rm eff}_{TA}(x_{\gamma})&=F_{TA}(x_{\gamma})+\bar{F}_{T}(x_{\gamma})~.
\end{align} 

The interference and point-like contributions are given by
\begin{eqnarray}
\frac{{\rm d}^2\Gamma^{({\rm INT})}}{{\rm d} x_{\gamma}\, {\rm d}(\cos{\theta})} &=&
-\frac{G^2_F\,\alpha^3_{\rm em}\, m^5_{B_{s}}}{2^{10}\,\pi^4}\, 
\left |V_{tb}\, V^*_{ts} \right |^2\,\frac{16\, f_{B_s}}{m_{B_{s}}}\, \hat m^2_{\mu} J(x_{\gamma})
\,\frac{x_{\gamma}^2}{(x_{\gamma} + \hat m^2_{\mu} - \hat{t})(\hat t\, -\,\hat m^2_{\mu})}
\nonumber\times
 \\[8pt]
&&\hspace{-1.3in}
 \Bigg[\frac{2x_\gamma\hat{m}_b}{1-x_\gamma}\,{\rm Re}
 \left(C_{10}^\ast(\mu)
 C_{7}(\mu)
 F_{TV}^{\rm eff}(x_\gamma)
 \right)
 +\, x_{\gamma}\, F_V(x_{\gamma})\, {\rm Re}\left (C^*_{10}(\mu)C^{\rm eff}_{9}(\mu, x_{\gamma})\right ) + 
  \,\xi(x_{\gamma},\hat t)\, F_A(x_{\gamma})\,\left |C_{10}(\mu) \right |^2 
  \bigg]
\end{eqnarray}
and
\begin{eqnarray}
\frac{{\rm d}^2\Gamma^{({\rm PT})}}{{\rm d} x_{\gamma}\, {\rm d}(\cos{\theta})} &=&  
\frac{G^2_F\,\alpha^3_{\rm em}\, m^5_{B_s}}{2^{10}\,\pi^4}\, 
\left |V_{tb}\, V^*_{ts} \right |^2\,
\left (\frac{8\, f_{B_s}}{m_{B_s}}\right )^2\,\hat m^2_{\mu}\,
\left |C_{10}(\mu) \right |^2 J(x_{\gamma}) 
\bigg[
   \frac{1-x_{\gamma}\, +\, x_{\gamma}^2/2}
         {(x_{\gamma} +\hat m^2_{\mu} - \hat{t})(\hat{t}\, -\,\hat m^2_{\mu})}\, 
- \nonumber \\[8pt]
&&\,\left(\frac{x_{\gamma}\,\hat m_{\mu}}
        {(x_{\gamma} + \hat m^2_{\mu} - \hat{t})\, (\hat{t}\, -\,\hat m^2_{\mu})}
   \right )^2\, 
   \bigg] ~.      
\end{eqnarray}

In the following we will use the Wilson coefficients evaluated in the $\msbar$ scheme at the scale $\mu=5\,{\rm GeV}$, which corresponds to the same scheme and scale at which we calculated the tensor form factors. In the calculation of the rate we input the value $m_{b} = m_{b}(5\,{\rm GeV}) = 4.073(11)\,{\rm GeV}$ obtained from $m_{b}(m_{b}) = 4.203(11)\,{\rm GeV}$~\cite{FlavourLatticeAveragingGroupFLAG:2021npn} using the four-loop quark-mass anomalous dimension~\cite{Chetyrkin:1997dh}.

We now discuss our strategy for estimating in a conservative way the systematic error due to the charming-penguin diagram in Figure\,\ref{fig:penguin}, corresponding to the emission of the $\mu^{+}\mu^{-}$ pair from the $c\bar{c}$ loop. As already discussed in Section~\ref{sec:eff_weak}, this contribution can be written as a process and $q^{2}= m^{2}_{B_{s}}(1- x_{\gamma})$ dependent shift of the Wilson coefficient $C_{9}\to C_{9}^{\rm eff}(q^{2}) = C_{9}+ \Delta C_{9}(q^{2})$, and we rely on the phenomenological parameterization in Eq.(\ref{eq:delta_C9V}) which we rewrite here for convenience,
\begin{align}
\label{eq:penguin_shift_C9V}
\Delta C_{9}(q^{2}) = -\frac{9 \pi}{\alpha_{\rm em}^2} \, (C_{1}+\frac{C_{2}}{3}) \,  \sum_V |k_{V}| e^{i\delta_{V}} 
 \frac{m_V \, B(V \to \mu^+ \mu^-) \, \Gamma_V}{q^2 - m_V^2 + i m_V \Gamma_V}~.
\end{align}
The values of the parameters $m_{V}, \Gamma_{V}, B(V\to \mu^{+}\mu^{-})$  are known experimentally for the lowest-lying resonances, and are collected in Table\,\ref{tab:charmonium_resonances}.
\begin{table}
\begin{center}
    \begin{tabular}{|c||c|c|c|}
    \hline
     $V_{c\bar{c}}$ ~ & ~ $M_{V_{c\bar{c}}}~[{\rm GeV}]$ ~ & ~ $\Gamma~[{\rm MeV}]$ ~ & ~ $\mathcal{B}(V_{c\bar{c}}\to \mu^{+}\mu^{-})$ \\
     \hline
     \hline
     $J/\psi$ ~ & ~ $3.096900(6)$ ~ & ~ $0.0926(17)$ ~ & $0.05961(33)$ \\
     $\Psi(2S)$ ~ & ~ $3.68610(6)$ ~ & ~ $0.294(8)$ ~ & $8.0(6)\cdot 10^{-3}$ \\
     $\Psi(3770)$ ~ & ~ $3.7737(4)$ ~ & ~ $27.2(1.0)$ ~ & $^{\ast}9.6(7)\cdot 10^{-6}$ \\
     $\Psi(4040)$ ~ & ~ $4.039(1)$ ~ & ~ $80(10)$ ~ & $^{\ast}1.07(16)\cdot 10^{-5}$ \\
     $\Psi(4160)$ ~ & ~ $4.191(5)$ ~ & ~ $70(10)$ ~ & $^{\ast}6.9(3.3)\cdot 10^{-6}$ \\
     $\Psi(4230)$ ~ & ~ $4.2225(24)$ ~ & ~ $48(8)$ ~ & $3.2(2.9)\cdot 10^{-5}$ \\
     $\Psi(4415)$ ~ & ~ $4.421(4)$ ~ & ~ $62(20)$ ~ & $2(1)\cdot 10^{-5}$ \\
     $\Psi(4660)$ ~ & ~ $4.630(6)$ ~ & ~ $72^{+14}_{-12}$ ~ & \textrm{ not seen} \\
     \hline
   
    \end{tabular}
      \caption{ Masses, decay widths and branching fractions into  a $\mu^{+}\mu^{-}$ pair for the lowest-lying charmonium resonances $V_{c\bar{c}}$~\cite{ParticleDataGroup:2020ssz}. For some of the charmonium resonances, in absence of information on the branching into $\mu^{+}\mu^{-}$, we provide the branching into $e^{+}e^{-}$, which is expected to provide a good approximation of $\mathcal{B}(V_{c\bar{c}}\to \mu^{+}\mu^{-})$ given that $M_{V_{c\bar{c}}}\gg m_{\mu}, m_{e}$. In those cases where the branching into $e^{+}e^{-}$ is given, the numerical value of the branching is preceded by an asterisk. For the last resonance in the table, the $\Psi(4660)$ resonance, neither the branching into $e^{+}e^{-}$ or $\mu^{+}\mu^{-}$ has been measured, and we input the fiducial value $\mathcal{B}(\Psi(4660)\to \mu^{+}\mu^{-}) = 1(1) \cdot 10^{-5}$. \label{tab:charmonium_resonances} }
    \end{center}
\end{table}
Instead, the value of the coefficients $|k_{V}|$ and the phases $\delta_{V}$ are largely unknown: $\delta_{V}=|k_{V}|-1=0$ only holds in the factorization approximation. In order to estimate the systematic error induced in the parameterization of Eq.\,(\ref{eq:penguin_shift_C9V}) by the poor knowledge of some of the parameters, we follow a (conservative) procedure similar to the one adopted in Ref.\,\cite{Guadagnoli:2023zym}. We assume for each resonance the value $|k_{V}| = 1.75(0.75)$\footnote{It has been found~\cite{Lyon:2014hpa} that $|k_{V}| \simeq 2.5$ well describes the $B\to K \mu^{+}\mu^{-}$ experimental data. The $1\sigma$ interval we choose for $|k_{V}|$ thus spans the region between $|k_{V}|=2.5$ and $|k_{V}|=1$ which corresponds to the value obtained in the factorization approximation.}, and that the phases $\delta_{V}$ are completely unknown.  Furthermore we assume that the resonance parameters are completely uncorrelated.  To correctly propagate the uncertainty on $|k_{V}|$ and $\delta_{V}$, as well as the one coming from all other input parameters (e.g. from the CKM matrix elements $|V_{tb}|$ and $|V_{ts}|$), we generate a large bootstrap sample of size $N_{b}=\mathcal{O}(1000)$ (and we assume that the parameters $\delta_{V}$ are uniformly distributed in the interval $[0,2\pi)$), and repeat the calculation of the rate for each bootstrap value of the input parameters. Central values and standard errors are then obtained from the usual bootstrap average and dispersion formulae. 

\begin{figure}
    \centering
    \includegraphics[scale=0.6]{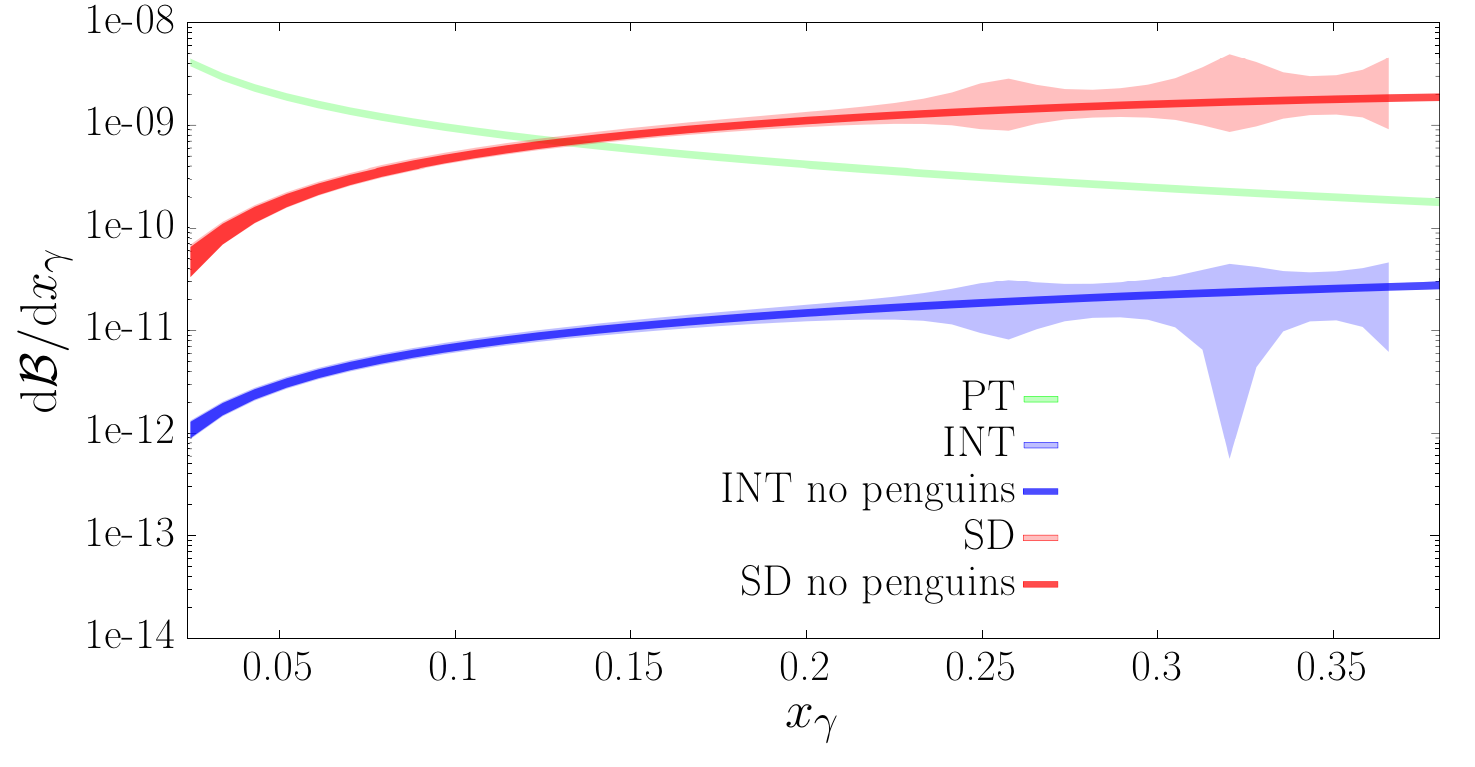}
    \caption{\small\it Our determination of the differential branching ${\rm d}\mathcal{B}/{\rm d}x_{\gamma}$ for $x_{\gamma} \, {\mathlarger{\mathlarger{\in}}} \, [0.025, 0.4]$. We give separately the point-like (light-green band), interference (light-blue band), and structure-dependent (light-red band) contributions. The blue and red bands correspond respectively to the determination of the interference and structure-dependent contribution obtained neglecting the charming-penguin diagrams.}
    \label{fig:diff_rate}
\end{figure}

For the Wilson coefficients we take the values from Ref\,\cite{Beneke:2020fot}, which in the basis of operators which we use, correspond to
\begin{align}
C_{1}(5\,{\rm GeV})= 0.147~,\quad C_{2}(5\,{\rm GeV})=-1.053~, \quad C_{7}(5\,{\rm GeV}) = 0.330~,\quad\,\, C_{9}(5\,{\rm GeV}) = -4.327~, \quad\,\, C_{10}(5\,{\rm GeV}) = 4.262~,
\end{align}
and for the remaining input parameters we take\,\cite{ParticleDataGroup:2020ssz}
\begin{align}
|V_{tb}| = 1.014(29)~,\qquad |V_{ts}| = 4.15(9) \times 10^{-2}~,\qquad \tau_{B_{s}} = 1.521(5) \times 10^{-12}~ {\rm s} ~, 
\end{align}
where $\tau_{B_{s}}$ is the average between the lifetimes of the $B_{sH}$ and $B_{sL}$ mesons, which are the mass eigenstates of the $B_{s}-\bar{B}_{s}$ system. 
The Wilson coefficients computed in Ref.\,\cite{Beneke:2020fot} include next-to-leading 
logarithm corrections. 
This has a particularly large relative effect on $C_1$ which is reduced by approximately 40\% compared to the leading logarithmic result\,\cite{Buras:1994dj,Buchalla:1995vs}, and subsequently on the magnitude of the combination $C_1+C_2/3$ entering the charming-penguin parameterization in Eq.\,\ref{eq:penguin_shift_C9V}, which is increased by more than $60\%$. 
In the plot of Figure\,\ref{fig:diff_rate} we provide our determination of the differential branching fraction
\begin{align}
\frac{ {\rm d}\mathcal{B}}{{\rm d}x_{\gamma}}  \equiv \tau_{B_{s}} \frac{{\rm d}\Gamma}{{\rm d}x_{\gamma}}~,
\end{align}
as a function of $x_{\gamma} \,\mathlarger{\mathlarger{\in}}\, [0.025,0.4]$. 
We give separately the point-like, interference, and structure-dependent contributions. As the figure shows the point-like contribution becomes subleading for $x_{\gamma} \gtrsim 0.15$, while the interference contribution turns out to be orders of magnitude smaller than the structure-dependent one on the entire range of $x_{\gamma}$ explored. At large $x_{\gamma}\gtrsim 0.2$, the uncertainty stemming from the missing charming-penguin contributions is dominant over all other sources of uncertainties, and therefore in order to improve the precision of the differential branching at large $x_{\gamma}$ a rigorous treatment of the charming penguin diagrams is necessary. 

\begin{figure}
    \centering
    \includegraphics[scale=0.6]{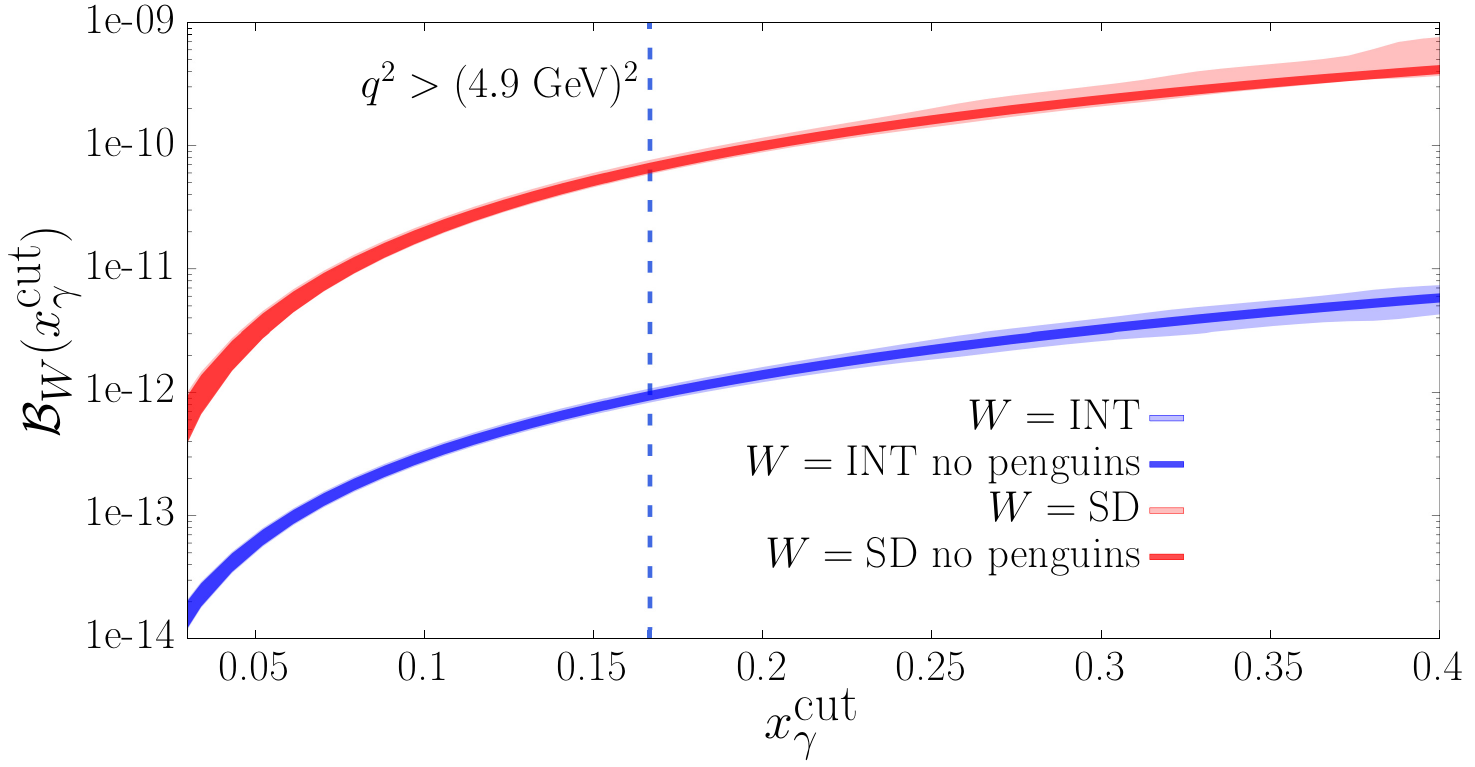}
    \caption{\small\it Our determination of the IR-finite structure-dependent ($\mathcal{B}_{\rm SD}(x_{\gamma}^{\rm cut})$, red band) and interference ($\mathcal{B}_{\rm INT}(x_{\gamma}^{\rm cut})$, blue band) contributions to the partial branching fractions $\mathcal{B}(x_{\gamma}^{\rm cut})$. The vertical blue line corresponds to the experimental cut imposed on the photon energy by the LHCb collaboration in Refs.\,\cite{LHCb:2021awg,LHCb:2021vsc}. }
    \label{fig:total_branching}
\end{figure}

We now proceed to discuss the determination of the total branching fraction
\begin{align}
\mathcal{B}(x_{\gamma}^{\rm cut}) = \int_{0}^{x_{\gamma}^{\rm cut}}\!\!\!\!\!\!\! {\rm d}x_{\gamma}\, \frac{ {\rm d}\mathcal{B}}{{\rm d}x_{\gamma}}~,
\end{align}
as a function of the upper bound $E_{\gamma}^{\rm cut} = m_{B_{s}}x_{\gamma}^{\rm cut}/2$ on the measured photon energy. 
As is well known, $\mathcal{B}(x_{\gamma}^{\rm cut})$  suffers from an infrared divergence generated by the point-like contribution to ${\rm d}\mathcal{B}/{\rm d}x_{\gamma}$ which at small $x_{\gamma}$ behaves as $1/x_{\gamma}$. The infrared divergence appearing in the decay rate with a real photon in the final state is then cancelled by the $\mathcal{O}(\alpha_{\rm em})$ virtual photon contribution
to the $B_{s}\to \mu^{+}\mu^{-}$ decay amplitude, through the usual Block-Nordsieck mechanism~\cite{PhysRev.52.54}.
The interference ($\mathcal{B}_{\rm INT}(x_{\gamma}^{\rm cut}))$ and structure-dependent ($\mathcal{B}_{\rm SD}(x_{\gamma}^{\rm cut}))$ contributions are instead IR finite. In the experimental analysis made by the LHCb Collaboration in Refs.~\cite{LHCb:2021vsc,LHCb:2021awg} the point-like contribution (called the final-state-radiation (FSR) contribution in Refs.\,\cite{LHCb:2021vsc, LHCb:2021awg}) has been included in the analysis of the $B_{s}\to \mu^{+}\mu^{-}$
invariant-mass distribution, as a radiative tail. For the IR-finite structure-dependent contribution (called the initial-state-radiation (ISR) contribution in Refs.~\cite{LHCb:2021vsc, LHCb:2021awg}) LHCb quotes the following upper-bound
\begin{align}
\label{eq:LHCb_res}
\mathcal{B}^{{\rm LHCb}}_{\rm SD}(0.166) < 2 \times 10^{-9}~. 
\end{align}
In agreement with our results, the interference contribution has been instead considered negligible in Refs.\,\cite{LHCb:2021vsc, LHCb:2021awg} on the basis of the results of  Ref.\,\cite{Kozachuk:2017mdk} obtained using the relativistic dispersion approach. In Figure\,\ref{fig:total_branching} we provide our determination of the IR-finite structure-dependent ($\mathcal{B}_{\rm SD}(x_{\gamma}^{\rm cut})$) and interference ($\mathcal{B}_{\rm INT}(x_{\gamma}^{\rm cut})$) contributions to $\mathcal{B}(x_{\gamma}^{\rm cut})$. The blue vertical line corresponds to the experimental cut $x_{\gamma}^{\rm cut} \simeq 0.166$ adopted in the experimental result of Eq.~(\ref{eq:LHCb_res}).
For $x_{\gamma}^{\rm cut} = 0.166$  we obtain
\begin{align}
\mathcal{B}_{\rm SD}(0.166) = 6.9(9) \times 10^{-11}~, 
\end{align}
while the interference contribution is completely negligible. Our result is well within the bound set by the LHCb Collaboration (Eq.~(\ref{eq:LHCb_res})). In Table~\ref{tab:branching} we collect our results for the sum $\mathcal{B}_{\rm SD+INT}(x_{\gamma}^{\rm cut})\equiv \mathcal{B}_{\rm SD}(x_{\gamma}^{\rm cut}) + \mathcal{B}_{\rm INT}(x_{\gamma}^{\rm cut})$ of the interference and structure dependent contribution to the partial branching fraction, for different values of $x_{\gamma}^{\rm cut}$.
\begin{table}
\setlength{\tabcolsep}{7pt}
\renewcommand{\arraystretch}{1.5}
\begin{tabular}{|c|c|c|c|c|c|c|}
\hline
 & \multicolumn{6}{c|}{$\sqrt{q^{2}_{\rm cut}}~ [\rm{GeV}] = m_{B_{s}}\sqrt{ 1 - x_{\gamma}^{\rm cut}}$} \\ \cline{2-7}
 &  $4.1$ & $4.2$ & $4.3$ & $4.4$ & $4.5$ & $4.6$ \\ \hline
 $\mathcal{B}_{\rm SD+INT} \times 10^{10}$  &  $ 6.1(2.1)$ & $ 5.3(1.7)$ & $ 3.99(88)$ & $ 3.31(74)$ & $ 2.57(50)$ & $ 2.02(39)$ \\ \hline
 & $4.7$ & $4.8$ & $4.9$ & $5.0$ & $5.1$ & $5.2$  \\ \hline
 $\mathcal{B}_{\rm SD+INT}\times 10^{10}$  & $ 1.47(22)$ & $ 1.04(14)$ & $ 0.685(90)$ & $ 0.399(55)$ & $ 0.188(29)$ & $ 0.057(12)$  \\ \hline
\end{tabular}
\caption{\small\it Our results for the partial branching fraction $\mathcal{B}_{\rm SD+INT}(x_{\gamma}^{\rm cut}) \equiv \mathcal{B}_{\rm SD}(x_{\gamma}^{\rm cut}) + \mathcal{B}_{\rm INT}(x_{\gamma}^{\rm cut})$ for different values of $x_{\gamma}^{\rm cut}$. The interference contribution $\mathcal{B}_{\rm INT}(x_{\gamma}^{\rm cut})$ is orders of magnitude smaller than $\mathcal{B}_{\rm SD}(x_{\gamma}^{\rm cut})$ and completely negligible within uncertainties.  \label{tab:branching}}
\end{table}

\begin{figure}
\includegraphics[scale=0.58]{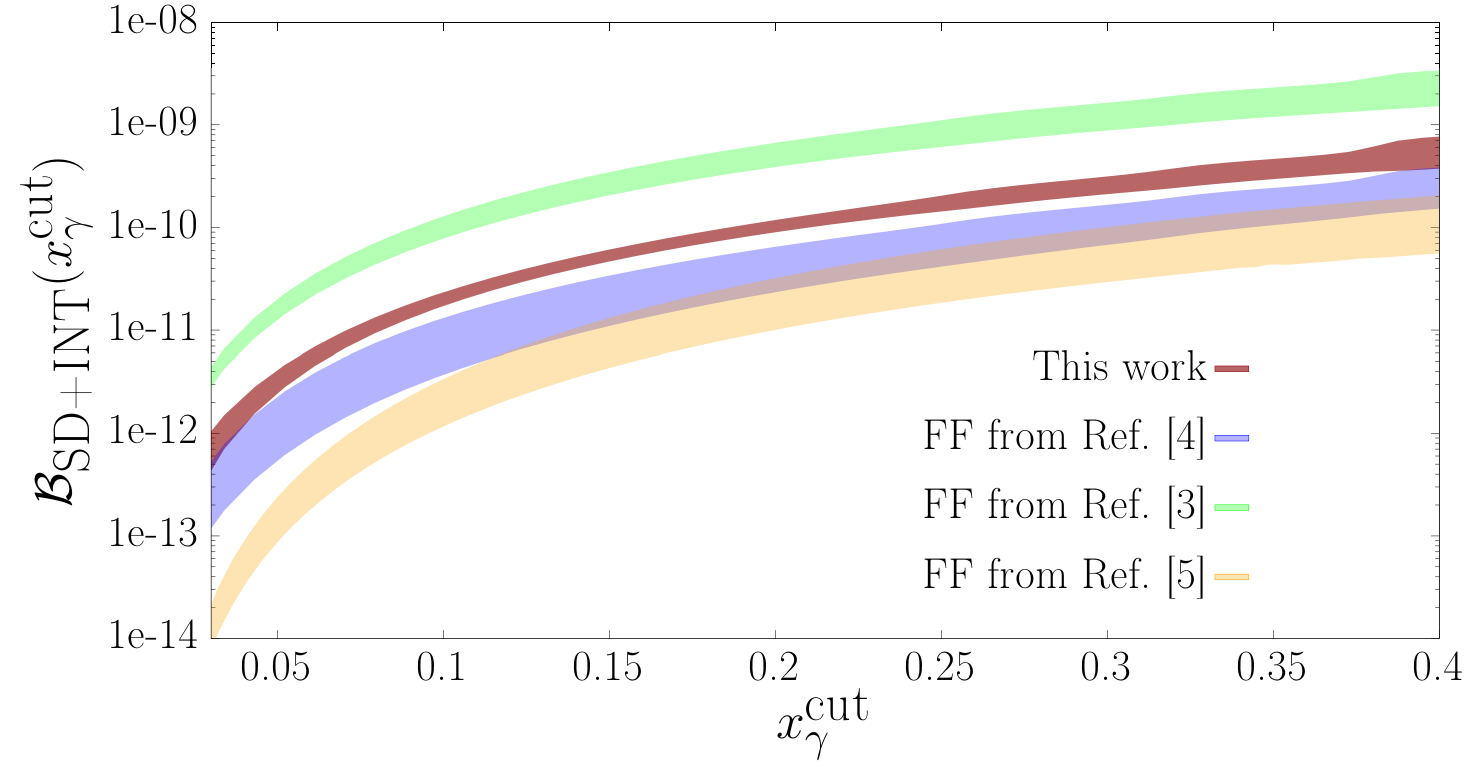}
\caption{\small\it Comparison between our results for $\mathcal{B}_{\rm SD+INT}(x_{\gamma}^{\rm cut})$ (shown in the figure by the red band), the ones obtained in Ref.\,\cite{Guadagnoli:2023zym} (shown by the orange band), and the ones we obtained using the form factors $F_{V}$, $ F_{A}$, $F_{TV}$ and $F_{TA}$ from Ref.\,\cite{Janowski:2021yvz} (shown in the figure by the green band), and Ref.~\cite{Kozachuk:2017mdk} (shown in the figure by the blue band). We used the estimate of $\bar{F}_{T}$ given in Ref.\,\cite{Kozachuk:2017mdk} to produce the results corresponding to the blue band, while the green band has been obtained setting $\bar{F}_{T}=0$. The impact of $\bar{F}_{T}$ on the branching fraction is however extremely modest, and negligible within uncertainties. \label{fig:comparison_branching}}
\end{figure}

We can further compare our results with the ones obtained using the model-dependent determination of the form factors $F_{V}, F_{A}, F_{TV}$ and $F_{TA}$ from Refs.~\cite{Janowski:2021yvz, Kozachuk:2017mdk, Guadagnoli:2023zym}. The results of the comparison are shown in Figure~\ref{fig:comparison_branching}. 
As the figure shows, our results for $\mathcal{B}_{\rm INT+SD}$ are smaller than those obtained using the form factors from Ref.~\cite{Janowski:2021yvz}, and larger than those obtained using the form factors from Refs.~\cite{Kozachuk:2017mdk, Guadagnoli:2023zym} (w.r.t. Ref.~\cite{Kozachuk:2017mdk} the difference is however less pronounced). This is not surprising given that the same trend is observed for the form factors (see Figure~\ref{fig:comp_FF}). Finally, we repeat that
in order to obtain a more accurate theoretical prediction for $\mathcal{B}_{\rm SD}(x_{\gamma}^{\rm cut})$ at large values of  $x_{\gamma}^{\rm cut}$, a first-principles calculation of the charming-penguin contributions is needed, since our model-dependent estimate presently represents the main source of uncertainty for large $x_{\gamma}^{\rm cut}$.

\section{Conclusions and future perspectives}
\label{sec:conclusions}
The rare radiative leptonic decay $B_s\to\mu^+\mu^-\gamma$ is a flavour-changing-neutral current transition which is forbidden at tree level in the Standard Model and is therefore particularly sensitive to potential New Physics contributions. 
Although there is an additional factor of $\alpha_\mathrm{em}$ in the amplitude for this process compared to that for the widely-studied $B_s\to\mu^+\mu^-$ decay, the presence of the final state photon removes the helicity suppression making the rates for the two processes comparable. On the other hand, while the leading hadronic effects in the $B_s\to\ell^+\ell^-$ ($\ell=e,\mu,\tau$) decay amplitude depend only on the $B_s$-meson decay constant $f_{B_s}$, which is known to sub-percent precision from lattice computations, the determination of the amplitude for the $B_s\to\mu^+\mu^-\gamma$ decay is much more complex. 
In this case the non-perturbative hadronic effects depend not only on local form factors, but also on resonance (including ``charming penguin") and other long-distance contributions. 
In this paper, we have presented a first-principles calculation of the local form factors $F_{V}, F_{A}, F_{TV}, F_{TA}$ and $\bar{F}_{T}$, which provide the main contributions to the amplitude for the $B_{s}\to \mu^{+}\mu^{-}\gamma$ decay at large di-muon invariant masses $\sqrt{q^{2}}>4.16$\,GeV, above the peaks of the lowest charmonium resonances. 
In order to determine the amplitude, we combine our results for the form factors with previous phenomenological estimates of the remaining contributions, in particular those from charming penguins. 
Whilst we find that the dominant contribution to the differential branching fraction is given by the well-determined form factor $F_V$, the largest contribution to the uncertainty for $q^{2} \lesssim (4.8\,{\rm GeV})^{2}$ at present comes from the charming penguins.

The fitted results for the form factors are plotted as functions of $x_\gamma$ in Figure\,\ref{fig:comp_FF}, where they are also compared to earlier estimates obtained using different techniques\,\cite{Janowski:2021yvz,Kozachuk:2017mdk,Guadagnoli:2023zym}. It can be seen that, with a few exceptions, our results for the form factors differ significantly from the earlier estimates (which also differ from each other).
In particular our results for the form factor $F_V$, which gives the largest contribution to the amplitude, are significantly smaller than that obtained in Ref.\,\cite{Janowski:2021yvz} and larger than those in Refs.\,\cite{Kozachuk:2017mdk,Guadagnoli:2023zym}.

In evaluating $\bar{F}_{T}^{s}$, the contribution to the form factor $\bar{F}_{T}$ in which  the virtual photon is emitted from the strange anti-quark, one encounters the difficulty of performing the analytic continuation to Euclidean spacetime due to the presence of intermediate vector $s\bar{s}$ states with masses below $\sqrt{q^2}$. 
As explained in detail in Sec.\,\ref{sec:FbarT}, in order to overcome this problem, we have employed the novel spectral-density reconstruction technique developed in Ref.\,\cite{Frezzotti:2023nun}.  
Since the contribution of $\bar{F}_{T}$ to the differential rate is small, and in view of the computational expense of implementing the spectral representation technique, we have evaluated it at the same four values of $x_\gamma$ as the four other form factors, but only on two ensembles and at three values of the heavy-quark mass ($m_h/m_c=1, 1.5, 2.5$). 
We do not observe any significant dependence on $x_\gamma$ in $\bar{F}_{T}^{s}(x_\gamma)$ and present our results for its real and imaginary parts in Eq.\,(\ref{eq:FbarTs}). 
There is no difficulty in the continuation to Euclidean space for $\bar{F}_{T}^{b}$, the contribution to the form factor $\bar{F}_{T}$ in which  the virtual photon is emitted from the b-quark, and we find that $\bar{F}_{T}^{b}$ is an order of magnitude smaller than $F_{TV}$ and $F_{TA}$. 
The results at the four values $x_\gamma=0.1,\,0.2,\,0.3$ and 0.4 are presented in Table\,\ref{tab:FT_b_extr}.

 We use our results for the local form factors to evaluate the $B_{s}\to\mu^{+}\mu^{-}\gamma$ amplitude for $q^{2} > (4.16\,{\rm GeV})^{2}$, taking into account the systematic uncertainties due to the contributions that we have not computed in the present work, in particular those from the charming-penguin diagrams.
 We present our results for the partial branching fractions as a function of the upper cut-off on $x_\gamma$ (or equivalently on the lower cut-off on $\sqrt{q^2}$) in Table\,\ref{tab:branching}.
 Imposing the same cut on the photon energy $(q^{2} > (4.9~{\rm GeV})^{2}$, i.e. $x_\gamma<0.166$) as adopted by the LHCb Collaboration, we obtain a value for the structure-dependent contribution to the branching fraction $\mathcal{B}_{\rm SD}(0.166)= 7.0(8)\times10^{-11}$, 
 which is well within the bound set by the LHCb collaboration\,
 $\mathcal{B}_{\rm SD}(0.166)<2.0\times10^{-9}$\,
 \cite{LHCb:2021awg, LHCb:2021vsc}. However, as illustrated in Figures\,\ref{fig:comp_FF} and\,\ref{fig:comparison_branching}, our results disagree with the LCSR and model/effective-theory determinations of the branching fractions from Refs.\,\cite{Janowski:2021yvz, Kozachuk:2017mdk, Guadagnoli:2023zym}; in particular they are smaller than the result in Ref.\,\cite{Janowski:2021yvz} and larger than those in Refs.\,\cite{Kozachuk:2017mdk, Guadagnoli:2023zym}. The difference can be traced back to the fact that our result for the form factor $F_{V}$, which is the dominant contribution to the rate, is larger (smaller) than those obtained in Refs.\,\cite{Kozachuk:2017mdk, Guadagnoli:2023zym} (\cite{Janowski:2021yvz}) by about a factor of $1.5-2$. 

At present our results for the branching fractions have uncertainties ranging from $\mathcal{O}(15\%)$ for $\sqrt{q^{2}_{\rm cut}} = 4.9~{\rm GeV}$, to $\mathcal{O}(30\%)$ for $\sqrt{q^{2}_{\rm cut}} = 4.2~{\rm GeV}$. Our uncertainties should already be at the level of precision that can be obtained in the future experimental measurements of $\mathcal{B}(B_{s}\to\mu^{+}\mu^{-}\gamma)$ at LHCb. Our analysis shows that in order to further improve the accuracy of the theoretical predictions in the low-$q^{2}$ region, it is necessary to obtain a first-principles determination of the (currently missing) charming-penguin contributions, which presently constitute the main source of uncertainty in the differential branching fraction for $q^{2} \lesssim (4.8\,{\rm GeV})^{2}$. 
This can be seen as the difference between the light-red band in Figure\,\ref{fig:diff_rate}, which is our full result, and the dark-red curve in which the charming penguin contributions have been neglected.

\section{Acknowledgements}
We thank Dmitri Melikhov for useful discussions, Martin Beneke for correspondence on the next-to-leading-order determination of the Wilson coefficients, Yasmine Sara Ambis for correspondence clarifying the status of LHCb measurements, Diego Guadagnoli, Camille Normand and Ludovico Vittorio for providing us with the form factors and the partial branching fractions of Ref.\,\cite{Guadagnoli:2023zym}. We thank all members of the ETMC for a most enjoyable collaboration and for providing us with the preliminary results of the renormalization constant $Z_{T}$\cite{ExtendedTwistedMass_RCs}. 
We acknowledge CINECA for the provision of CPU time under the specific initiative INFN-LQCD123 and IscrB\_S-EPIC. 
V.L. F.S. R.F. and N.T. are supported by the Italian Ministry of University and Research (MUR) under the grant PNRR-M4C2-I1.1-PRIN 2022-PE2 Non-perturbative aspects of fundamental interactions, in the Standard Model and beyond F53D23001480006 funded by E.U. - NextGenerationEU.
S.S. is supported by MUR under grant 2022N4W8WR. 
F.S. G.G and S.S. acknowledge MUR for partial support under grant PRIN20172LNEEZ. 
F.S. and G.G acknowledge INFN for partial support under GRANT73/CALAT. C.T.S. was partially supported by an Emeritus Fellowship from the Leverhulme Trust and by STFC (UK) grant ST/T000775/1.
F.S. is supported by ICSC – Centro Nazionale di Ricerca in High Performance Computing, Big Data and Quantum Computing, funded by European Union – NextGenerationEU.

\bibliography{biblio}
\bibliographystyle{JHEP}

\appendix

\section{Determination of the decay constant $f_{B_{s}}$}
\label{sec:fBs}
In this appendix we discuss our determination of the decay constant of the $B_{s}$ meson, which enters in the extrapolation formulae of Section~\ref{sec:num_results_I}. 
For this calculation,
in order to account for any correlations in the determination of the decay constants and form factors,
we use the same configurations and masses as in the determination of the form factors $F_{W}$, $W=\{V,A, TV, TA\}$ discussed in Section\,\ref{subsec:formfactorsnumerical}.

For each simulated value of the heavy-light meson mass $m_{H_s}$, and for each lattice spacing, we use two different estimators of the decay constant $f_{H_{s}}$. The first determination of $f_{H_{s}}$ is obtained from the pseudoscalar-pseudoscalar two-point correlation functions
\begin{align}
\label{eq:A1}
C_{\rm PS}^{\textrm{sm-loc}}(t) \equiv \sum_{\vec{x}}\langle P_{5}^{\rm loc}(t,\vec{x})P_{5}^{\dag {\rm sm}}(0)\rangle ,\qquad C_{\rm PS}^\textrm {sm-sm}(t) \equiv \sum_{\vec{x}}\langle P_{5}^{\rm sm}(t,\vec{x})P_{5}^{\dag {\rm sm}}(0)\rangle~,
\end{align}
where the labels sm and loc indicate ``smeared" and ``local" respectively,
\begin{align}
\label{eq:A2}
P_{5}^{\rm loc}(x) = \bar{s}(x)\gamma^{5}h(x)~,  \qquad\quad P_{5}^{\rm sm}(x) = \bar{s}(x)\gamma^{5}H_{k}^{N}(x,y)h(y)  ~,
\end{align}
and where 
\begin{align}
H_{k}(x,y) = \frac{1}{1+6k}\left(  1 + k\sum_{\hat{j}}U_{j}(x)\delta(x+a\hat{j},y)\right)
\end{align}
is the gauge-field-dependent Gaussian-smearing operator that has been used to construct the interpolating operator of the $\bar{H}_{s}$ meson in the Euclidean three-point correlation function $B_{W}^{\mu\nu}(t;k,p)$ in Eq.~(\ref{eq:Cmunudef}). The Wilson parameter of the $\bar{s}$ and $h$ valence quarks entering the bilinears are always chosen to be opposite. 
In the large time limit $t \gg a,$ $t \ll T$, the correlation functions $C_{\rm PS}^{\textrm{sm-loc}}(t)$ and $C_{\rm PS}^{\textrm{sm-sm}}(t)$  behave as
\begin{align}
\label{eq:A3}
C_{\rm PS}^{\textrm{\,sm-loc}}(t) &=  \frac{|Z_{\rm PS}^{\textrm{\,sm-loc}}|^{2}}{2m_{H_{s}}}\left( e^{-m_{H_s}t} + e^{-m_{H_{s}}(T-t)}   \right) + \ldots ~, \nonumber \\[8pt]
C_{\rm PS}^{\rm\,sm-sm}(t) &=  \frac{|Z_{\rm PS}^{\textrm{\,sm-sm}}|^{2}}{2m_{H_{s}}}\left( e^{-m_{H_s}t} + e^{-m_{H_{s}}(T-t)}   \right) + \ldots        ~,
\end{align}
where the ellipsis indicate terms that are subleading in the limit of large time separations. From the knowledge of $Z_{\rm PS}^{\textrm{\,sm-loc}}$ and $Z_{\rm PS}^{\text{\,sm-sm}}$ it is possible to determine the decay constant $f_{H_s}$, without the need of any renormalization constant, making use of\,\cite{EuropeanTwistedMass:2014osg}
\begin{align}
\label{eq:A4}
Z_{\rm PS} \equiv \langle \bar{H}_s | \bar{s}\gamma^{5}h | 0 \rangle =  \frac{|Z_{\rm PS}^{\textrm {\,sm-loc}}|^{2}}{|Z_{\rm PS}^{\textrm{\,sm-sm}}|}~,\qquad   f_{H_{s}} = a(m_{s}^{\rm bare}+m_{h}^{\rm bare}) \frac{|Z_{\rm PS}|}{m_{H_{s}}\sinh(am_{H_{s}})}~,
\end{align}
where $m_{s}^{\rm bare}$ and $m_{h}^{\rm bare}$ are the simulated values of the bare strange and heavy quark mass. The use of smeared interpolating operators is essential in order to improve the signal for $f_{H_{s}}$, as it allows the ground state contribution to be isolated at much smaller times (as compared to the standard local interpolating operator), where the correlation function is generally more precise.

\begin{figure}
\includegraphics[scale=0.355]{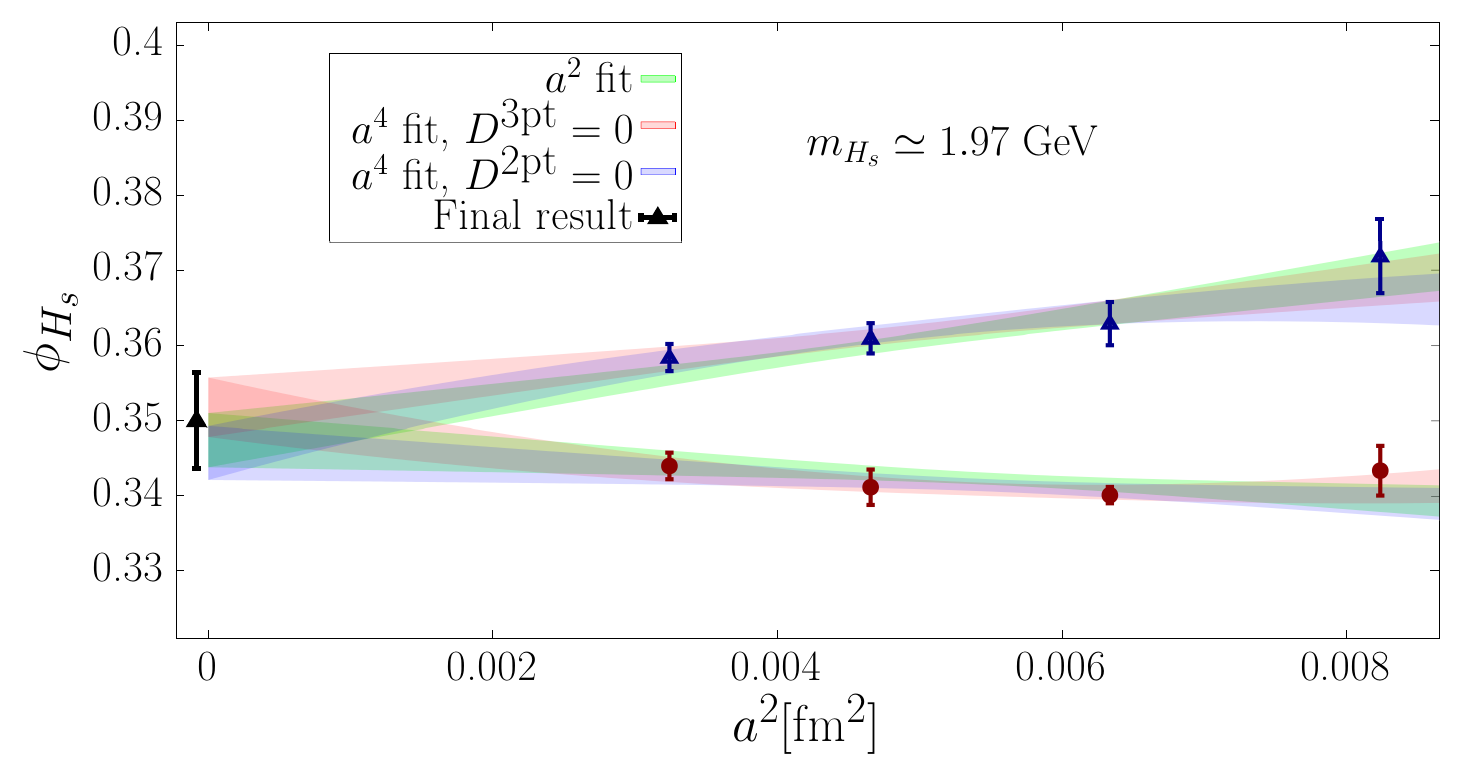}
\includegraphics[scale=0.355]{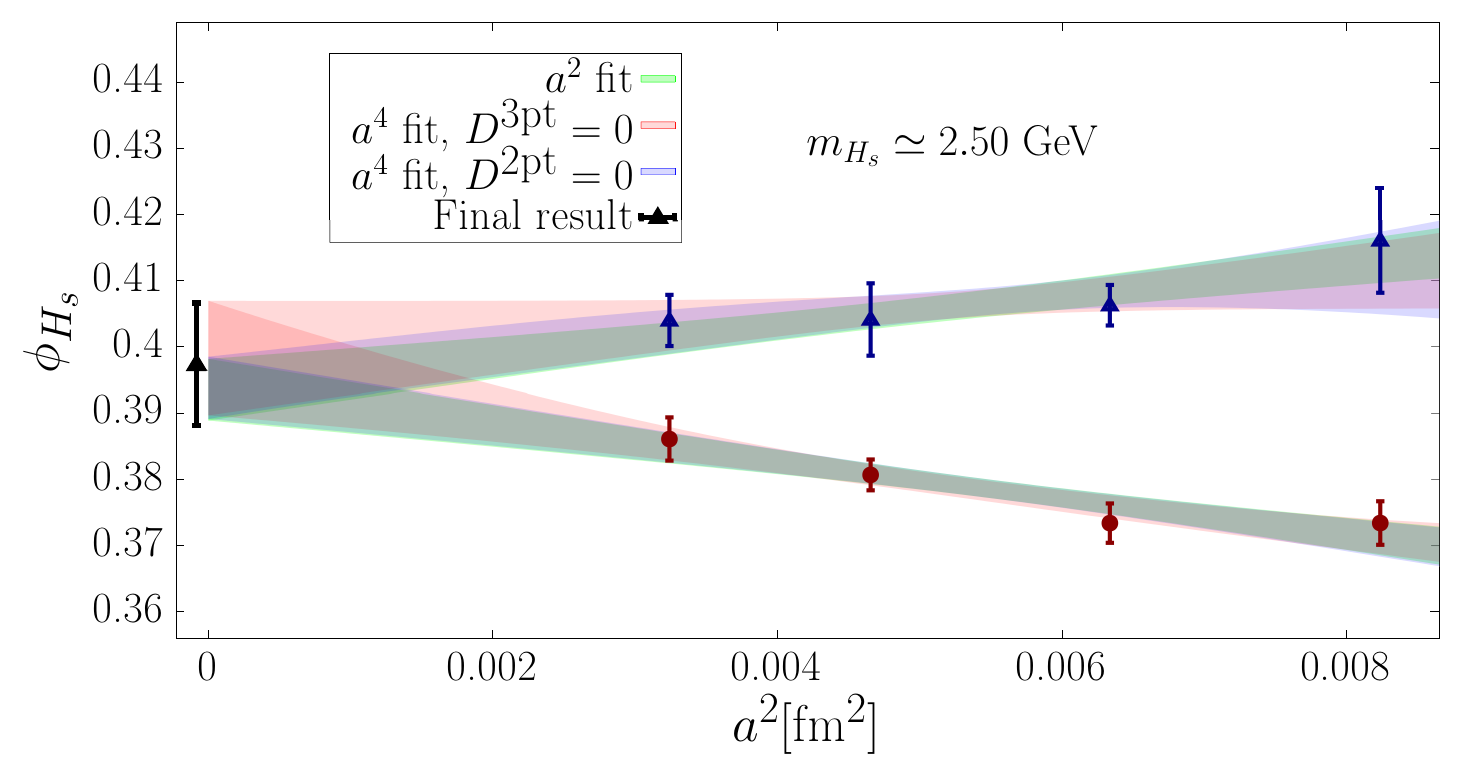} \\
\includegraphics[scale=0.355]{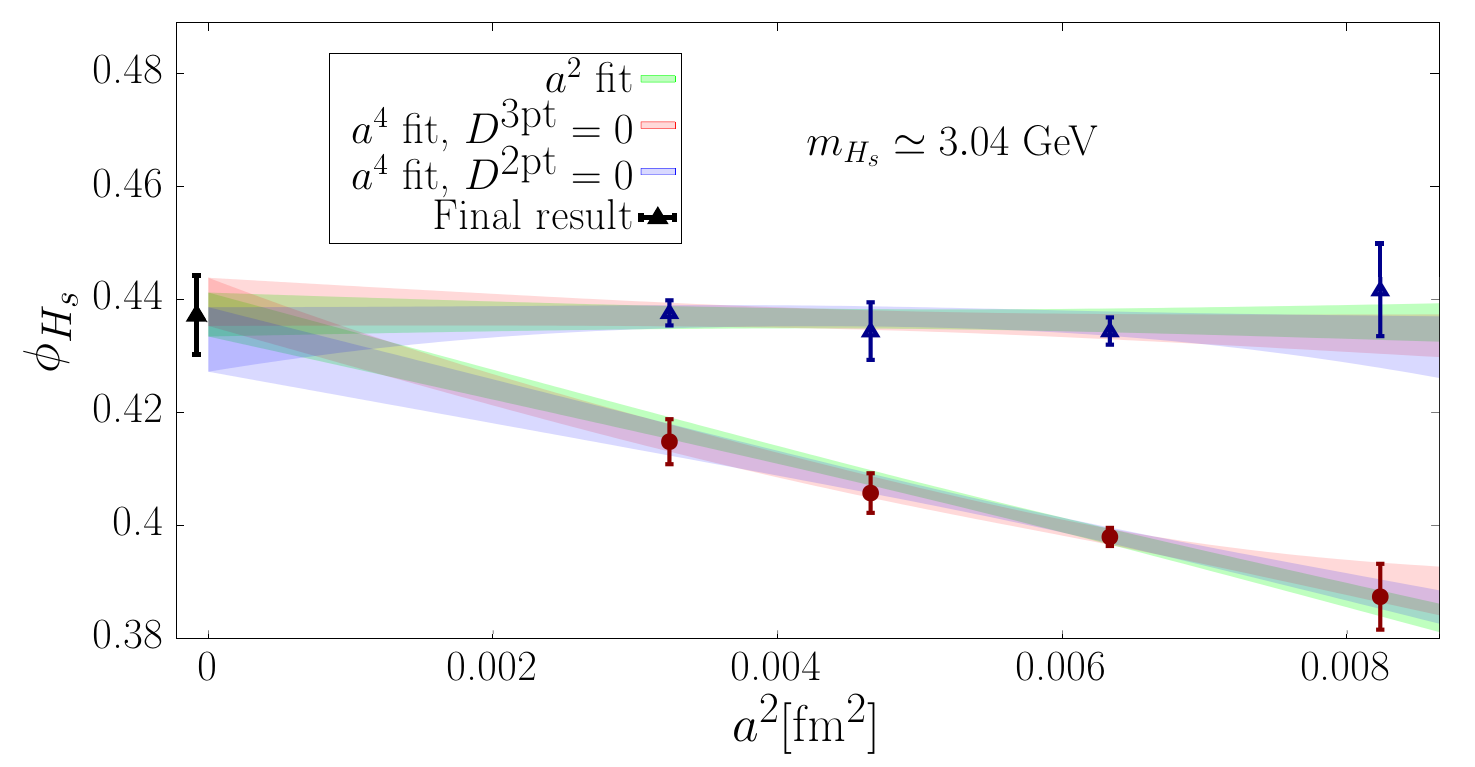}
\includegraphics[scale=0.355]{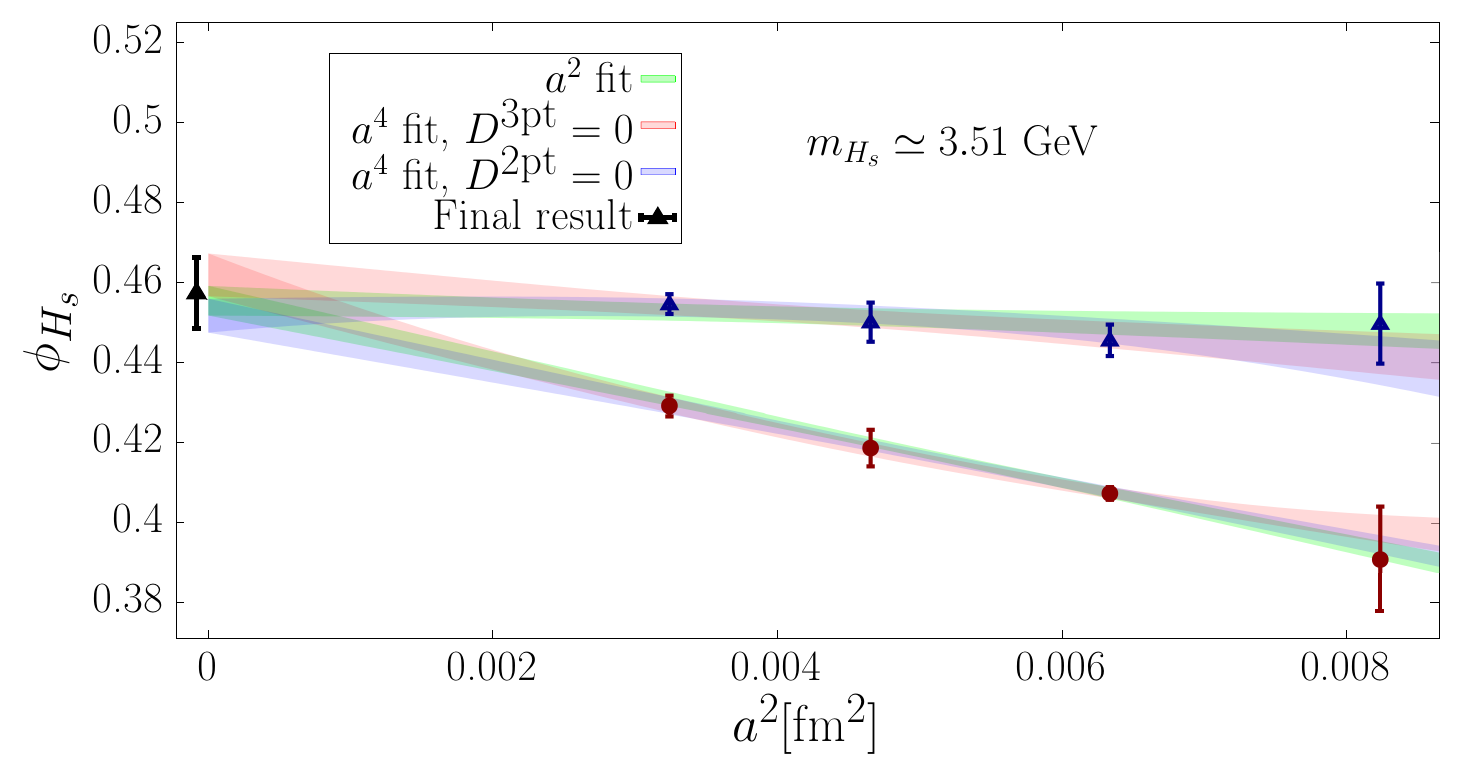} \\
\includegraphics[scale=0.355]{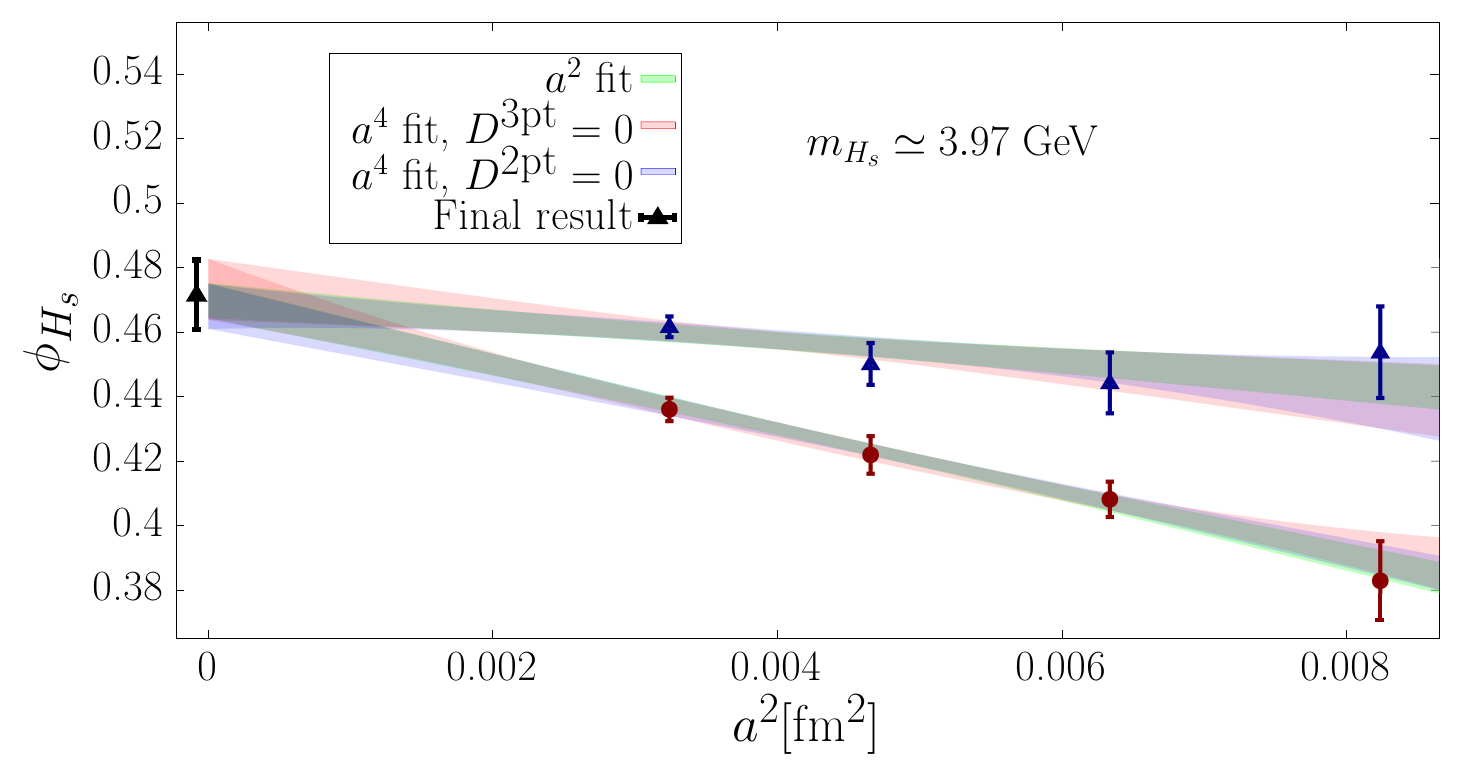}
\caption{\small\it Continuum limit extrapolation of $\phi_{H_{s}}$ for the five different simulated values of the heavy-strange meson mass $m_{H_{s}}$. For each mass we show the continuum limit fits obtained employing all four lattice spacings. The blue and red data points correspond respectively to the estimator $\phi^{\rm 2pt}$ and $\phi^{\rm 3pt}$. The green, red, and blue bands correspond respectively to the pure $a^{2}$ fit, and to the $a^{4}$ fit with $D^{\rm 3pt}=0$ and $D^{\rm 2pt}=0$. Finally, the black data point at $a^{2} \simeq 0$ correspond to our final determination, which is obtained combining the six different fits we performed using the AIC. \label{fig:fBs_cont_lim}}
\end{figure}

The second estimator of $f_{H_{s}}$ that we use is obtained exploiting the fact that in the zero photon-momentum limit $\bs{k} = 0$, the spatial part $H_{A}^{ii}(\bs{0},\bs{0})$ of the axial hadronic tensor is equal to $if_{H_s}$ if we choose fictitious values $q'_{s}$ and $q'_{h}$ for the electric charges of the strange and heavy quark entering $J^{\mu}_{\rm em}$ in Eq.~(\ref{eq:1PS_em}), in such a way that $q'_{s}+q'_{h}=1$ (see Appendix C of Refs.~\cite{Desiderio:2020oej} for more details on this point). In the following we denote the determination of $f_{H_{s}}$ from Eqs.~(\ref{eq:A1})-(\ref{eq:A4}) as $f_{H_{s}}^{\rm 2pt}$, and the one from $H_{A}^{ii}(\bs{0},\bs{0})$ as $f_{H_{s}}^{\rm 3pt}$. The two estimates of the the decay constants only differ by cut-off effects and allow us to better constraint the result of the continuum fits. Our strategy is to perform the continuum limit extrapolation at fixed values of $m_{H_{s}}$, fitting simultaneously $f_{H_{s}}^{\rm 2pt}$ and $f_{H_{s}}^{\rm 3pt}$ using the following Ansatz
\begin{align}
\phi_{H_{s}}^{\rm 2pt} \equiv f_{H_{s}}^{\rm 2pt}\sqrt{m_{H_{s}}}  = A + B^{\rm 2pt} a^{2} + D^{\rm 2pt} a^{4},\qquad\quad \phi_{H_{s}}^{\rm 3pt} \equiv f_{H_{s}}^{\rm 3pt}\sqrt{m_{H_{s}}} = A + B^{\rm 3pt} a^{2} + D^{\rm 3pt} a^{4}~,  
\end{align}
where $A, B^{\rm 2pt/3pt}$ and $D^{\rm 2pt/3pt}$ are free fit parameters. Note that a common continuum value is enforced. In order to avoid overfitting, given the limited number of gauge ensembles that we employ, we do not include fits with 
both $D^{\rm 2pt}$ and $D^{\rm 3pt}$ as fit parameters, i.e. we set either $D^{\rm 2pt}$ or $D^{\rm 3pt}$ or both to zero.
The fits are performed minimizing a correlated $\chi^{2}$-function which takes into account the correlation between $\phi_{H_{s}}^{\rm 2pt}$ and $\phi_{H_{s}}^{\rm 3pt}$ evaluated on the same gauge ensemble. The results of the extrapolation to the continuum limit for the five simulated values of $m_{H_{s}}$ are illustrated in Figure\,\ref{fig:fBs_cont_lim}. We have performed a total of six different fits for each $m_{H_{s}}$, which differ depending on whether the ensemble with the coarsest lattice spacing is included or not, and on whether we 
include or neglect the terms proportional to $a^4$, i.e. whether we set $D^{\rm 2pt}$ and/or $D^{\rm 3pt}$ to zero. 
The results obtained from the different fits are then combined using the AIC, which has already been discussed in Section\,\ref{subsec:formfactorsnumerical}. To be conservative, we add linearly the systematic and statistical errors from AIC.   
Having extrapolated the decay constants $f_{H_{s}}$ for each of the five simulated values of $m_{H_{s}}$ to the continuum limit, in order to obtain $f_{B_{s}}$ we need to perform the extrapolation in the mass. To do so, we make use of the heavy-quark scaling relation
\begin{align}
\label{eq:fBs_HQET}
\phi(m_{H_{s}}) = \textrm{const.} \times \left(  1 + \frac{B}{m_{H_{s}}} + \mathcal{O}( m_{H_{s}}^{-2} )    \right)~.
\end{align}
As is well known, Eq.~(\ref{eq:fBs_HQET}) is valid exactly only in the effective theory at the bare level. Logarithmic corrections to Eq.\,(\ref{eq:fBs_HQET}) are generated by the non-zero anomalous dimension of the axial current in the HQET, as well as from its matching to the axial current in full QCD. 
Let $J_{\Gamma}(\mu') = Z_{J}^{-1}(\alpha_{s}(\mu))\,\bar{\ell}\Gamma h$ b, where $\Gamma$ is one of the Dirac matrices and $\ell$ and $h$ denote light and heavy quark fields in QCD, be a generic renormalized heavy-light current in QCD and let $\tilde{J}_{\Gamma}(\mu) = Z^{-1}_{\tilde{J}}(\alpha_{s}(\mu'))\,\bar{\ell}\Gamma h_{v}$ be its counterpart in the HQET. 
The relation between the two currents is given, at leading order in the heavy-quark mass $m_{h}$, by~\cite{Chetyrkin:2003vi} 
\begin{align}
\label{eq:matching}
J_{\Gamma}(\mu') = C_{\Gamma}(m_{h}, m_{h}) \exp\left\{  \int_{\alpha_{s}(m_{h})}^{\alpha_{s}(\mu')} \frac{\gamma_{J}(\alpha_{s})}{2\beta(\alpha_{s})}\frac{d\alpha_{s}}{\alpha_{s}} -  \int_{\alpha_{s}(m_{h})}^{\alpha_{s}(\mu')} \frac{\gamma_{\tilde{J}}(\alpha_{s})}{2\beta(\alpha_{s})}\frac{d\alpha_{s}}{\alpha_{s}}         \right\}~\tilde{J}_{\Gamma}(\mu')~,
\end{align}
where $\beta(\alpha_{s})$ is the QCD $\beta$-function, and $\gamma_{J(\tilde{J})}$ is the anomalous dimension of the current $J(\tilde{J})$. In the case of the QCD axial current one has $\gamma_{J}=0$.  $C_{\Gamma}(m_{h},m_{h})$ is the matching coefficient and is obtained by imposing the equality between the renormalized proper vertices of the quark-bilinear in question, evaluated in QCD and in the HQET at the heavy quark scale $m_{h}$. Its two-loop expression for the axial current has been computed in Ref.\,\cite{Broadhurst:1994se} and is given by
\begin{align}
C_{\gamma^{0}\gamma^{5}}(m_{h},m_{h}) =  1 - \frac{2}{3}\frac{\alpha_{s}(m_{h})}{\pi} -2.95\left(\frac{\alpha_{s}(m_{h})}{\pi}\right)^{2} ~.
\end{align}
The anomalous dimension $\gamma_{\tilde J}$ in the HQET (which does not depend on the specific $\Gamma$ considered) has been computed at three-loops in Ref.~\cite{Chetyrkin:2003vi}. In summary, the relation in Eq.~(\ref{eq:fBs_HQET}) gets modified by the matching and by the running of $\tilde{J}_{\Gamma}(\mu')
$ into
\begin{align}
\label{eq:fit_func_fBs}
\phi(m_{H_{s}}) = C_{\gamma^{0}\gamma^{5}}(m_{h}, m_{h}) \exp\left\{   \int_{0}^{\alpha_{s}(m_{h})} \frac{\gamma_{\tilde{J}}(\alpha_{s})}{2\beta(\alpha_{s})}\frac{d\alpha_{s}}{\alpha_{s}}         \right\} \times \textrm{const}^\prime \times \left(  1 + \frac{B}{m_{H_{s}}} + \mathcal{O}( m_{H_{s}}^{-2} )    \right)~,
\end{align}
and we have reabsorbed the $\mu'$ dependence in Eq.\,(\ref{eq:matching}) into the constant factor, const$^\prime$. A delicate point of the analysis is the determination of the heavy quark mass $m_{h}$ of the HQET, which should be identified with the pole mass. Notoriously, the pole mass is affected by renormalon ambiguities and its perturbative expansion in terms of the $\msbar$ mass $m^{\msbar}_{h}(m^{\msbar}_{h})$ is asymptotically divergent. To avoid the use of the pole mass, we follow two different strategies and use, in place of $m_{h}$, either the heavy-strange meson mass $m_{H_{s}}$ ($m_{H_{s}} - m_{h} \simeq \mathcal{O}(\Lambda_{QCD})$), or the minimal-renormalon-subtracted mass advocated in Ref.~\cite{Brambilla:2017hcq} (see also Ref.~\cite{Gambino:2017vkx} for another alternative to the use of the pole mass). 

\begin{figure}
\includegraphics[scale=0.5]{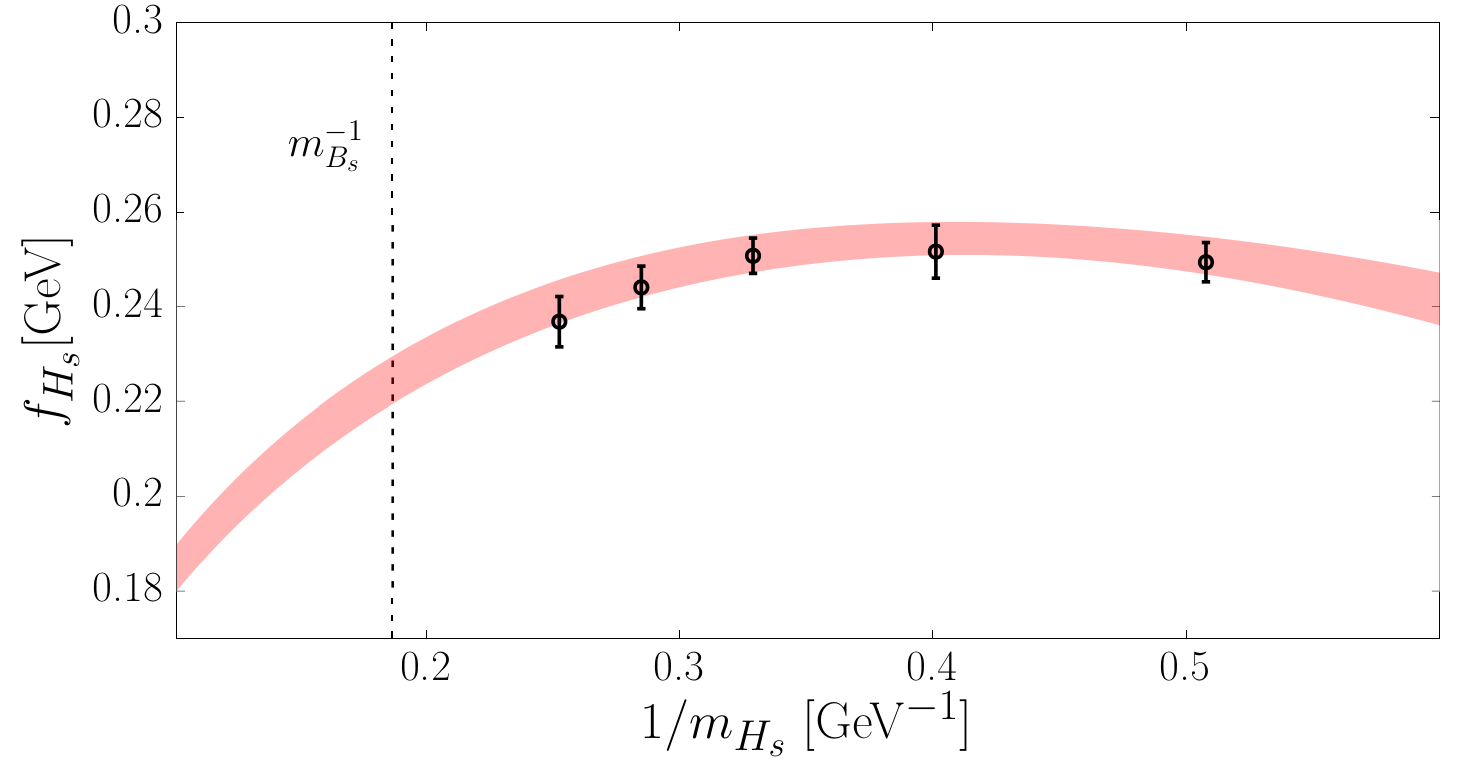}
\caption{\small\it Extrapolation of the decay constant $f_{H_{s}}$ to the physical $B_{s}$ meson mass $m_{B_{s}}\simeq 5.367~{\rm GeV}$ using the Ansatz in Eq.~(\ref{eq:fit_func_fBs}) and with $m_{h}$ replaced by $m_{H_{s}}$. The red curve corresponds to the best-fit function, while the dashed vertical line corresponds to the inverse mass of the $B_{s}$ meson. The reduced $\chi^{2}$ of the fit is smaller than unit. Our final determination of $f_{B_{s}}$ is given in Eq.~(\ref{eq:final_fBs}). \label{fig:extr_fBs}}
\end{figure}

The result of the extrapolation obtained replacing $m_{h}$ with $m_{H_{s}}$ in Eq.~(\ref{eq:fit_func_fBs}), and with the constant term and the $B$ coefficient considered as free fit parameters, is shown in Figure~\ref{fig:extr_fBs}.
The fit takes fully into account the correlation between the values of the decay constant obtained at the different values of $m_{H_{s}}$. The resulting $\chi^{2}/dof$ of the fit is very good and smaller than unity. By repeating the fit employing the minimal-renormalon-subtracted mass, we find that the value of $f_{B_{s}}$ changes by less than $0.3\sigma$, and therefore we do not add any additional systematic error. We obtain the value
\begin{align}
\label{eq:final_fBs}
f_{B_{s}} = 224.5\,(5.0)\,{\rm MeV}\,,
\end{align}
which agrees with the $N_{f}=2+1+1$ FLAG average $f_{B_{s}}^{\rm FLAG} = 230.3~(1.3)~{\rm MeV}$ at the level of $1.1\sigma$. Our final uncertainty is however much larger, which reflects the fact that our calculation is not tailored to be a precise determination of $f_{B_{s}}$. Our main interest is in the calculation of the  local form factors contributing to the $B_{s}\to \mu^{+}\mu^{-}\gamma$ decay amplitude. 
The main limitation preventing a more precise determination of $f_{B_{s}}$ comes from the systematic uncertainty associated to the continuum limit extrapolation, performed here with a rather limited number of lattice spacings. The new gauge ensembles that the Extended Twisted Mass (ETM) Collaboration will produce at smaller values of the lattice spacing, will allow for significantly reduced uncertainties on $f_{B_{s}}$.

\end{document}